\documentclass[aps,prx,reprint,twocolumns,notitlepage,longbibliography]{revtex4-1}
\usepackage{amssymb,bm}
\usepackage{amsmath,eucal}
\usepackage[normalem]{ulem}
\usepackage[usenames,dvipsnames]{color}
\usepackage{graphicx}
\usepackage{units}

\newcommand*{\TRANS}{^{\mkern-1.5mu\mathsf{T}}}
\newcommand*{\VEC}[1]{\boldsymbol{#1}}
\newcommand*{\TENSOR}[1]{\mathsf{#1}}

\newcommand*{\kB}{k_{\mathrm{B}}}

\renewcommand{\d}{\mathrm{d}}	
\newcommand{\D}{\mathcal{D}}

\newcommand*{\Da}{D_{\mathrm{a}}}
\newcommand*{\ta}{\tau_{\mathrm{a}}}
\newcommand*{\traj}[1]{\underline{#1}}
\newcommand*{\Traj}[1]{\overline{#1}}
\newcommand*{\trajR}[1]{\underline{\tilde{#1}}}
\newcommand*{\TrajR}[1]{\overline{\tilde{#1}}}
\newcommand*{\pp}[1]{\mathfrak{p}[{#1}]}
\newcommand*{\ppR}[1]{\tilde{\mathfrak{p}}[{#1}]}

\newcommand*{\I}{\mathcal{I}}
\newcommand*{\IR}{\tilde{\mathcal{I}}}

\begin{document}

\title{Irreversibility in active matter systems:
Fluctuation theorem and mutual information}
\author{Lennart Dabelow}
\affiliation{Fakult\"at f\"ur Physik, Universit\"at Bielefeld, 33615 Bielefeld, Germany}
\author{Stefano Bo}
\author{Ralf Eichhorn}
\affiliation{Nordita, Royal Institute of Technology and Stockholm University,
Roslagstullsbacken 23, SE-106 91 Stockholm, Sweden}
\begin{abstract}
We consider a Brownian particle which, in addition to being in contact with a thermal bath,
is driven by
fluctuating forces which stem from active processes in the system,
such as self-propulsion or collisions with other active particles.
These active
fluctuations do not fulfill a fluctuation-dissipation relation and therefore play the
role of a non-equilibrium environment,
which keeps the system permanently out of thermal equilibrium
even in the absence of external forces.
We investigate how the out-of-equilibrium character of the active matter system
and the associated irreversibility is reflected in
the trajectories of the Brownian particle.
Specifically, we analyze
the log-ratio of path probabilities for
observing a certain particle trajectory forward in time versus
observing its time-reversed twin trajectory.
For passive Brownian motion,
it is well-known that this path probability ratio quantifies
irreversibility in terms of entropy production.
For active Brownian motion,
we show that in addition to the usual entropy produced in the thermal environment
the path probability ratio contains a contribution to irreversibility from
mutual information ``production'' between the
particle trajectory and the history of the non-equilibrium environment.
The resulting irreversibility measure fulfills an integral fluctuation theorem
and a second-law like relation.
When deriving and discussing these relations,
we keep in mind that the active fluctuations
can occur either due to
a suspension of active particles pushing around a passive colloid or
due to active self-propulsion of the particle itself;
we point out the similarities and differences between these two situations.
We obtain explicit expressions for active fluctuations modeled
by an Ornstein-Uhlenbeck process.
Finally, we illustrate our general results by analyzing a Brownian particle
which is trapped in a static or moving harmonic potential.
\end{abstract}
\date{\today}
\maketitle

\section{Introduction and main results}
\subsection{Introduction}
Active particle systems consist of individual entities (``particles'') which have the ability to
perform motion by consuming energy from the environment and converting
it into a self-propulsion drive
\cite{Ramaswamy:2010tma,romanczuk2012active,Cates:2012dtw,marchetti2013hydrodynamics,
Bechinger:2016api,ramaswamy2017active,fodor2018statistical}.
Prototypical examples are
collections of macro-organisms, such as
animal herds, schools of fish, flocks of birds, or ant colonies
\cite{parrish1997animal,cavagna2014bird,cavagna2017physics},
and suspensions of biological microorganisms
or artificial microswimmers, such as bacteria and colloidal particles with catalytic surfaces
\cite{romanczuk2012active,Cates:2012dtw,elgeti2015physics,Bechinger:2016api,patteson2016active}.
Systems of this kind exhibit a variety of intriguing properties, e.g.\
clustering and swarming
\cite{Ben-Jacob:2000cso,Peruani:2006nco,Fily:2012aps,maggi2015multidimensional,Marconi:2015tas,paoluzzi2016critical},
bacterial turbulence \cite{Dombrowski:2004sca}, or motility-induced phase separation
\cite{Cates:2015mip,Farage:2015eii},
to name but a few.

\begin{figure}[t]
	\includegraphics[width=0.45\columnwidth]{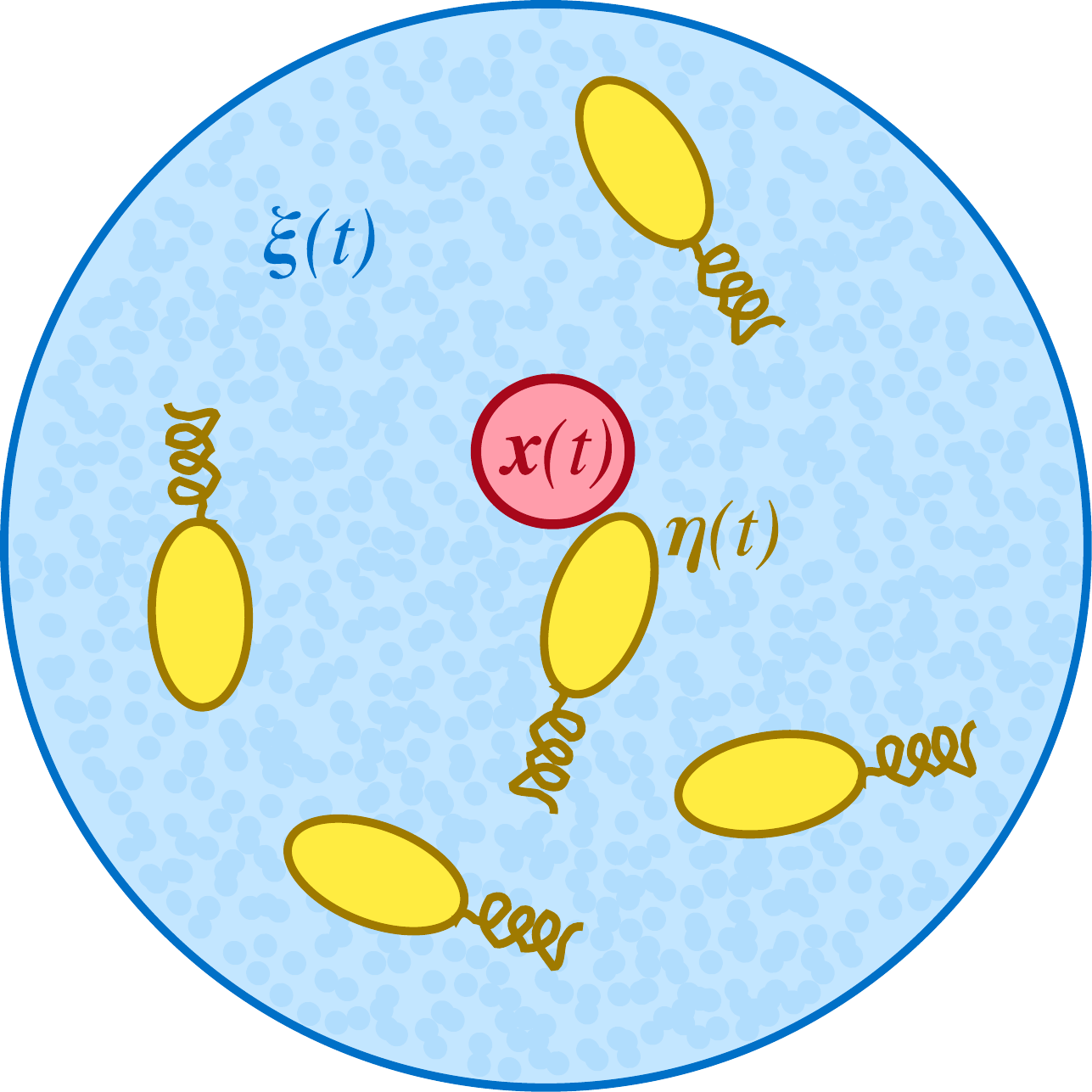}
	\hspace*{0.5cm}
	\includegraphics[width=0.45\columnwidth]{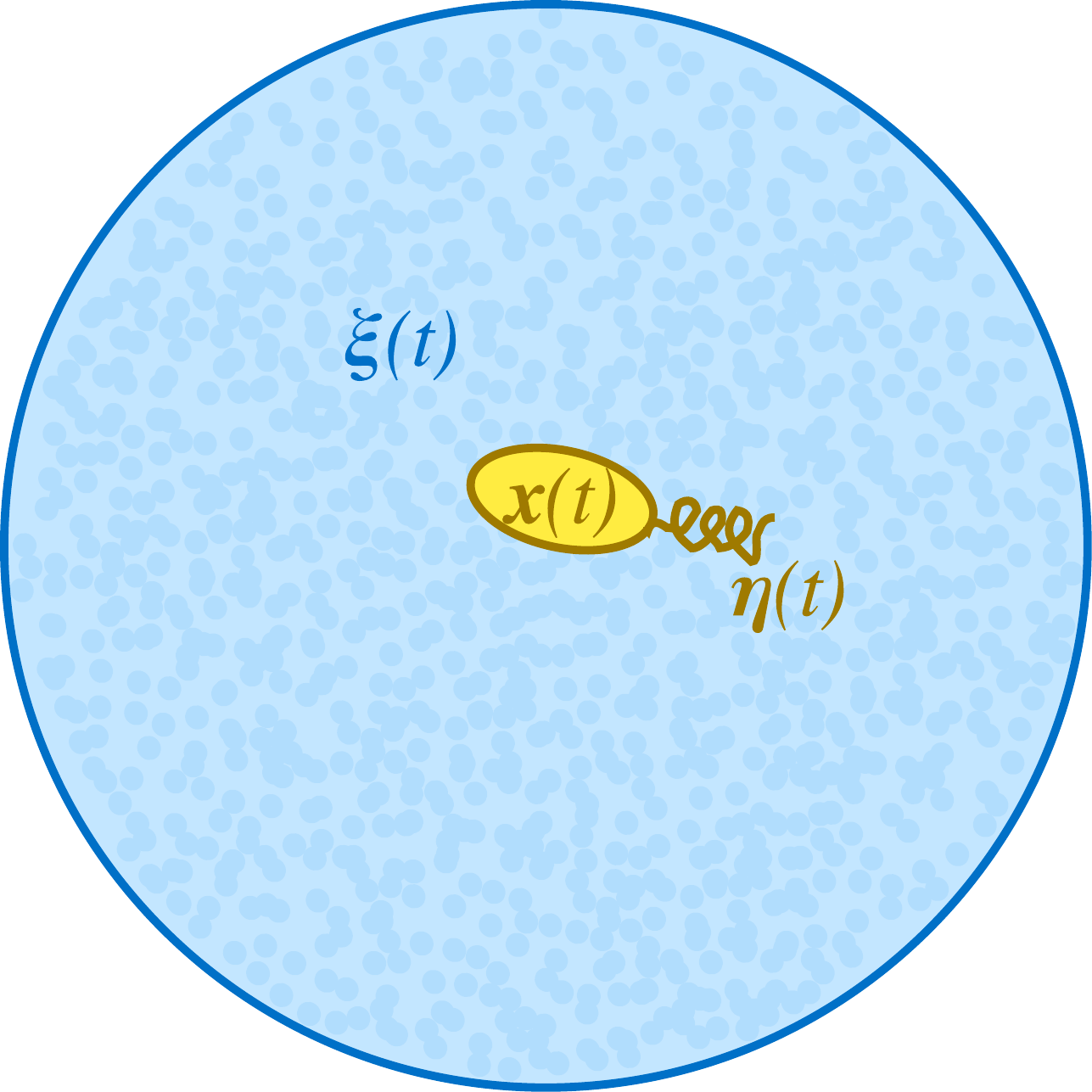}
\caption{
Illustration of the two different kinds of active matter systems.
Left: Passive Brownian particle (red) in a thermal bath (blue) with
a suspension of active, self-propelled particles (yellow);
collisions of the active particles with the passive one are modeled
by a fluctuating force $\VEC{\eta}(t)$.
Right: Self-propelled particle (yellow) in a thermal bath (blue);
the fluctuating self-propulsion velocity is modeled to result from
a fluctuation force $\VEC{\eta}(t)$.
In both cases thermal fluctuations are described by unbiased,
delta-correlated Gaussian white noise sources $\VEC{\xi}(t)$,
and the position of the particle of interest is denoted by $\VEC{x}(t)$,
see Eqs.~\eqref{eq:LE}-\eqref{eq:OUP} in the main text.
}
\label{fig:1}
\end{figure}

Microorganisms and mircoswimmers are usually
dispersed in an aqueous solution at room temperature and therefore
experience thermal fluctuations which give rise to a diffusive component
in their self-propelled swimming motion.
In addition, the self-propulsion mechanism is typically
noisy in itself \cite{romanczuk2012active}, for instance,
due to environmental factors or intrinsic stochasticity
of the mechanisms creating self-propulsion.
These ``active fluctuations'' exhibit two essential features.
First, a certain
persistence in the direction of driving
over length and time scales comparable to
observational scales. Second, an inherent non-equilibrium character
as a consequence of permanently converting and dissipating energy
in order to fuel self-propulsion.
Interestingly, similar active fluctuations with
the same characteristics can be observed in a complementary class
of active matter systems, namely a passive colloidal ``tracer'' particle which
is suspended in an aqueous solution of active swimmers.
The collisions with the active particles in the environment entail
directional persistence and non-equilibrium features in the
motion of the passive tracer particle \cite{Maggi:2014gee,Argun:2016nbs,maggi2017memory}.
The sketch in Fig.~\ref{fig:1} illustrates these two types of
active particle systems.

Despite the inherent non-equilibrium properties of
active matter systems they appear to bear striking similarities to equilibrium systems
\cite{Tailleur:2008smo,speck2014effective,takatori2014swim,Farage:2015eii},
for instance, the dynamics of individual active particles
at large scales often looks like passive Brownian diffusion \cite{Cates:2012dtw}
(i.e., at scales beyond at least the ``persistence length'' of the particle motion).
In connection with the ongoing attempts to
describe active matter systems by thermodynamic (-like) theories
this observation raises the important question
``How far from equilibrium is active matter?"
\cite{Fodor:2016hff,nardini2017entropy}.
More specifically, the questions are:
in which respects do emergent
properties of active systems resemble the thermodynamics of
thermal equilibrium systems, in which respects do they not,
and how do these deviations manifest themselves
in observables describing the thermodynamic character of the active matter
system, like, e.g, the entropy production?

Such questions should be answered \emph{independently} of
the specific microscopic processes creating the active driving,
because, while these processes maintain the active particle system
out of equilibrium, they contain
little or no information on
whether the emergent system behavior appears equilibrium-like or not.
It is, for instance, of little relevance
for the system properties emerging on the coarse-grained
scale of particle trajectories (e.g., the ``diffusion'' of active particles)
and their similarities or dissimilarities to thermal equilibrium,
if an active
particle is self-propelled by chemical processes occurring at its
surface or by a beating flagellum.
In this spirit,
we here address the above questions from a fundamental non-equilibrium statistical
mechanics viewpoint
by developing a trajectory-wise thermodynamic description
of active particle systems as a natural generalization of 
stochastic energetics \cite{Sekimoto:1998lea,Sekimoto:StochasticEnergetics}
and thermodynamics
\cite{Seifert:2008stp,Jarzynski:2011eai,Seifert:2012stf,van2015ensemble,seifert2018stochastic}
of passive particles in a purely thermal environment.
In particular,
we study the breakdown of time-reversal symmetry in the motion 
of particles which are driven by thermal and active fluctuations.
We quantify the associated irreversibility of individual
particle trajectories as a measure for non-equilibrium
by comparing the probability
of a specific particle trajectory to occur forward in time
to that of observing its time-reversed counterpart.
We link the corresponding probability ratio to functionals over
the forward trajectory,
and provide a thermodynamic interpretation of the
resulting extensive quantities in terms of entropy production
and mutual information.

In the case of passive particles in contact with a single, purely thermal bath,
the path probability ratio is a well-established fundamental concept in
stochastic thermodynamics for
assessing irreversibility
\cite{Seifert:2008stp,Jarzynski:2011eai,Seifert:2012stf,van2015ensemble}.
It is known to provide relations between
thermodynamic quantities, such as entropy production, work, or heat, and dynamical
properties encoded in path probability ratios.
Many such relations have been found over the last two decades
\cite{Crooks:1999epf,Seifert:2005epa,Chetrite:2008frf,Jarzynski:2011eai,Seifert:2012stf},
which, in their integrated forms, typically yield refinements
of the second law of thermodynamics \cite{Jarzynski:1997nef,Seifert:2005epa,Jarzynski:2011eai}.
In particular, the fluctuation theorem for the total entropy production
of a passive particle in a thermal environment
reinforces the fundamental interpretation of entropy as a measure of irreversibility
(we give a brief summary of the results most relevant to our present work 
in Sec.~\ref{sec:Da0}).
In the presence of active fluctuations, however,
the identification of dissipated heat and entropy production is less
straightforward, and is connected to the problem of
how to calculate and interpret the path probability ratio
in a physically meaningful way \cite{pietzonka2017entropy,Puglisi:2017crf}.

In the present work, we analyze the path probability ratio
based on particle trajectories, i.e.\ from the evolution of the
particle positions in time, without explicitly modeling the
(system-specific) microscopic processes behind the active fluctuations.
This allows for deriving general results on the non-equilibrium
character and thermodynamic content
of the dynamical behavior emerging in active matter systems
independently of the specific processes generating the active fluctuations,
but with the caveat that dissipation and irreversibility occurring
in these processes cannot be assessed \cite{pietzonka2017entropy}.
The derived quantitative irreversibility and thermodynamic measures are
easily accessible in standard experiments, e.g.\ from recording
particle trajectories using video microscopy.
In order to account for the active fluctuations, which
either stem from an active bath the (passive) particle is
dispersed in or from active self-propulsion
(see Fig.~\ref{fig:1} and
Section \ref{sec:EE} for specific examples of both cases),
without resolving the microscopic processes that
govern the interactions between active and passive particles
or drive self-propulsion,
we follow the common approach to include stochastic ``active forces''
in the equations of motion
of the individual particle of interest
(see, e.g., \cite{romanczuk2012active,Bechinger:2016api}
and references therein).
When calculating the path probabilities, we
treat these active non-equilibrium forces
in the same way as thermal equilibrium noise,
namely as a ``bath'' the particle is exposed to
with unknown microscopic details, but known statistical properties
(which are correlated in time and break detailed balance
\cite{Note1}%
).
Our main results following this approach and their implications are
briefly summarized in the next section.

\subsection{Brief summary of the main results}
Our central findings
are the following:

(i) Using a path-integral approach,
we calculate the probability density
$\mathfrak{p}[\traj{\VEC{x}}|\VEC{x}_0]$ of particle trajectories
$\Traj{\VEC{x}}=\{\VEC{x}(t)\}_{t=0}^\tau
=\VEC{x}_0\cup\{\VEC{x}(t)\}_{t>0}^\tau=\VEC{x}_0\cup\traj{\VEC{x}}$
as a result of the statistical properties of thermal and active
fluctuations simultaneously affecting the dynamics of the particle
[Sec.~\ref{sec:pp}, Eqs.~\eqref{eq:pp} and \eqref{eq:GammaEtc}].
Here, the coordinate $\VEC{x}(t)$ may either refer to the
position of a passive tracer particle in an active bath, or to the
position of an active, self-propelled particle.
When calculating these path weights, we consider the general situation
in which the particle is also subject to
deterministic external (or interaction) forces
$\VEC{f}_t=\VEC{f}(\VEC{x}(t),t)$,
which may explicitly depend on time via externally controlled
driving protocols.
We summarize the mathematical details of the derivation
as well as various known limiting cases in Appendices
\ref{app:integrateEta}-\ref{app:limits}.

(ii) We then find [Sec.~\ref{sec:FT}, Eqs.~\eqref{eq:ppratioDeltaSigma}]
that the log-ratio of the probabilities
for particle trajectories to occur forward in time versus backward in time
(time-backward trajectories are marked by a tilde symbol)
can be expressed as a functional
along the forward path with a non-local ``memory kernel''
$\Gamma_\tau(t,t')$:
\begin{subequations}
\label{eq:ppratioDeltaSigmaIntro}
\begin{equation}
\label{eq:ppratioIntro}
\frac{\ppR{\trajR{\VEC{x}}|\tilde{\VEC{x}}_0}}{\pp{\traj{\VEC{x}}|\VEC{x}_0}}
= e^{-\Delta \Sigma[\Traj{\VEC{x}}]/\kB}
\, ,
\end{equation}
with
\begin{equation}
\label{eq:DeltaSigmaIntro}
\Delta\Sigma[\Traj{\VEC{x}}]
= \frac{1}{T} \int_0^\tau \!\! \mathrm{d}t \int_0^\tau \!\! \mathrm{d}t^\prime \;
	\dot{\VEC{x}}_t\TRANS \VEC{f}_{t^\prime} \left[ \delta(t-t^\prime) - \frac{\Da}{D} \Gamma_\tau(t, t^\prime) \right]
\, .
\end{equation}
\end{subequations}
The functional $\Delta\Sigma$ therefore quantifies
irreversibility in our active matter systems.
With the explicit form of $\Gamma_\tau(t,t')$ given in Eq.~\eqref{eq:GammaEtc},
it can be calculated for any particle trajectory $\VEC{x}(t)$.
Information on the
thermal and active baths enters into
$\Delta\Sigma$ and $\Gamma_\tau(t,t')$ only via their statistical parameters,
like the thermal diffusion coefficient $D$,
thermal bath temperature $T$, active diffusion coefficient $\Da$,
and active correlation time $\ta$ (implicit in $\Gamma_\tau$).
It is, in particular, not necessary to make any assumptions on the behavior
of the active fluctuations under time-reversal to arrive at the above result
for $\Delta\Sigma$.

Combined with the change in system entropy
$\Delta S_{\mathrm{sys}}= -\kB \ln [p(\VEC{x}_\tau,\tau)/p(\VEC{x}_0,0)]$,
which compares system probabilities at the beginning ($t=0$) and end
($t=\tau$) of the trajectory \cite{Seifert:2005epa,Seifert:2012stf},
$\Delta\Sigma+\Delta S_{\mathrm{sys}}$
fulfill an integral fluctuation theorem
[Sec.~\ref{sec:FT}, Eq.~\eqref{eq:IFT}], valid for any duration $\tau$ of the particle
trajectory,
\begin{equation}
\label{eq:IFTIntro}
\left\langle e^{-(\Delta\Sigma + \Delta S_{\mathrm{sys}})/\kB} \right\rangle = 1
\, ;
\end{equation}
the average
$\langle \cdot \rangle$ is over all possible trajectories
advancing from $t=0$ to $t=\tau$.
As a consequence, they
obey a second-law like relation [Eq.~\eqref{eq:2ndLaw}],
\begin{equation}
\label{eq:2ndLawIntro}
\left\langle \Delta\Sigma[\Traj{\VEC{x}}]+ \Delta S_{\mathrm{sys}}(\VEC{x}_0,\VEC{x}_\tau) \right\rangle \geq 0
\, .
\end{equation}
In case of the equal sign, the active matter system appears reversible on
the level of particle trajectories $\VEC{x}(t)$.
From \eqref{eq:DeltaSigmaIntro} we see that in the stationary state (with $\Delta S_{\mathrm{sys}}=0$)
the trajectories for ``free active diffusion''
without external force, $\VEC{f}_t \equiv 0$,
always look reversible and thus equilibrium-like.
In the presence of external forces, we have $\left\langle \Delta\Sigma + \Delta S_{\mathrm{sys}} \right\rangle > 0$
in general, where $\left\langle \Delta\Sigma + \Delta S_{\mathrm{sys}} \right\rangle$ becomes larger
for more irreversible particle dynamics, indicating that
the system appears to be further away from an equilibrium-like behavior.

The path probability ratio
${\ppR{\trajR{\VEC{x}}|\tilde{\VEC{x}}_0}}/{\pp{\traj{\VEC{x}}|\VEC{x}_0}}$
is an important fundamental concept in stochastic thermodynamics,
because it contains
information on the thermodynamics of the system,
such as second-law like bounds on entropy production
as a consequence of the fluctuation theorems, features of heat,
work and efficiency distributions in systems that represent
heat engines, or violations of the fluctuation-response relation
due to non-equilibrium drivings.
In the present work we derive its most
immediate implications, namely fluctuation theorems and
second-law like relations.
Further insights about our active matter system
that may be derived in future work from the
explicit knowledge of the path probability ratio
are briefly discussed in the conclusions,
Section~\ref{sec:CD}.

(iii) By keeping track of the specific realization of the active fluctuations
which participated in generating the particle trajectory,
we can identify the two individual contributions to $\Delta\Sigma$
stemming from the thermal environment and the active bath.
The thermal part is the usual entropy
$\Delta S_{\mathrm{tot}}[\Traj{\VEC{x}}|\Traj{\VEC{\eta}}]$
produced in the thermal environment and in the system
along the particle trajectory (for the given active
noise realization,
which we denote $\Traj{\VEC{\eta}}=\{\VEC{\eta}(t)\}_{t=0}^\tau$).
The part associated with the active bath
is given by the
difference $\Delta\I[\Traj{\VEC{x}},\Traj{\VEC{\eta}}]$
in the amount of correlations (measured in terms of path-wise mutual
information), that are built up between the particle trajectory
and the active noise in the time-forward direction
as compared to the time-backward direction (Sec.~\ref{sec:MI}).
This splitting provides the thermodynamic interpretation
\begin{equation}
\label{eq:splitIntro}
\Delta\Sigma[\Traj{\VEC{x}}]+ \Delta S_{\mathrm{sys}}(\VEC{x}_0,\VEC{x}_\tau) =
\Delta S_{\mathrm{tot}}[\Traj{\VEC{x}}|\Traj{\VEC{\eta}}] - \kB \Delta\I[\Traj{\VEC{x}},\Traj{\VEC{\eta}}]
\end{equation}
for our irreversibility measure $\Delta\Sigma$ derived from
the path probability ratio of particle trajectories, see Eq.~\eqref{eq:ppratioDeltaSigmaIntro}.
It is valid for any active matter model which includes the
active fluctuations as ``active forces'' in the particle equations of motion,
no matter if we assume the active fluctuations to be
even [Eq.~\eqref{eq:DeltaSigma+}]
or odd [Eq.~\eqref{eq:DeltaSigma-}]
under time-reversal (denoted later in the text by $\pm$ sub- or superscripts).
The interpretation \eqref{eq:splitIntro} generalizes the trajectory-wise
stochastic thermodynamics of passive Brownian particles to active matter systems.
It shows that entropy production in the thermal bath, which fully quantifies the path
probability ratio for passive Brownian motion, is complemented by the difference
in mutual information accumulated with the active (non-equilibrium) bath in
the time-forward versus the time-backward direction.
From the fluctuation theorem
for $\Delta\Sigma$ we directly obtain an integral fluctuation
theorem and a second-law relation for the mutual information difference $\Delta\I$
[Eqs.~\eqref{eq:IFT+}, \eqref{eq:IFT-} and \eqref{eq:2nd+}, \eqref{eq:2nd-}].

We illustrate all these findings in Sec.~\ref{sec:HP} by discussing
the example of a colloidal
particle subject to thermal and active fluctuations,
which is trapped in a (static or moving) harmonic potential.
For this simple linear system all relevant quantities can be calculated
explicitly. We give the most important mathematical details
in Appendix \ref{app:HP}.

We finally remark that the explicit expression 
obtained under (i) and (ii) for $\Gamma_\tau(t,t')$
[see \eqref{eq:GammaEtc}]
is valid for a Gaussian
Ornstein-Uhlenbeck process
\cite{Gardiner:HandbookOfStochasticMethods,vanKampen1992stochastic}
as a model for the active fluctuations,
while result (iii)
is valid for general types of active
fluctuating forces $\VEC{\eta}$ with arbitrary (but well-defined)
statistical properties.
The Ornstein-Uhlenbeck process has become quite popular
and successful in describing active fluctuations
\cite{Fily:2012aps,Farage:2015eii, 
Maggi:2014gee,Argun:2016nbs,maggi2017memory,chaki2018entropy, 
Fodor:2016hff,marconi2017heat,mandal2017entropy,Puglisi:2017crf,
koumakis2014directed,szamel2014self,szamel2015glassy,maggi2015multidimensional,Flenner:2016tng,paoluzzi2016critical,
Marconi:2016vdi,szamel2017evaluating,sandford2017pressure,caprini2018linear,fodor2018statistical, 
berthier2013non,Marconi:2015tas,shankar2018hidden}, 
because it constitutes a minimal model for persistency in the active forcings
due to its exponentially decaying correlations with finite correlation time,
and it can easily be set up to break detailed balance
by a ``mismatched'' damping term
which does not validate the fluctuation-dissipation relation
\cite{Kubo:1966tfd,marconi2008fluctuation}.
Moreover, it is able to
describe both situations of interest mentioned above,
namely passive motion in a bath of active swimmers
\cite{Maggi:2014gee,Argun:2016nbs,maggi2017memory,chaki2018entropy}
and active motion driven by self-propulsion
\cite{Fily:2012aps,Farage:2015eii, 
Fodor:2016hff,marconi2017heat,mandal2017entropy,Puglisi:2017crf,
koumakis2014directed,szamel2014self,szamel2015glassy,maggi2015multidimensional,Flenner:2016tng,paoluzzi2016critical,Marconi:2016vdi,szamel2017evaluating,sandford2017pressure,caprini2018linear,fodor2018statistical, 
berthier2013non,Marconi:2015tas,shankar2018hidden}. 
The details of the model class we are considering are given in the next section.

\section{Model}
\label{sec:model}
We consider a colloidal particle at position $\VEC{x}$
in $d=1,2$, or $3$ dimensions, which is suspended in an
aqueous solution at thermal equilibrium with temperature $T$.
The particle diffuses under the influence of
deterministic external forces, in general consisting of
potential forces $-\nabla U(\VEC{x},t)$ (with the potential $U(\VEC{x},t)$)
and non-conservative force components $\VEC{F}(\VEC{x},t)$.
In addition, it experiences fluctuating driving forces due to permanent energy
conversion from active processes in the environment or the particle itself. 
The specific examples we have in mind are a passive tracer
in an active non-thermal bath (composed, e.g., of bacteria in aqueous solution,
see left panel of Fig.~\ref{fig:1})
\cite{Maggi:2014gee,Argun:2016nbs,maggi2017memory,chaki2018entropy},
or a self-propelled particle (e.g., a  colloidal microswimmer or bacterium,
see right panel of Fig.~\ref{fig:1})
\cite{Fily:2012aps,Farage:2015eii, 
Fodor:2016hff,marconi2017heat,mandal2017entropy,Puglisi:2017crf,
koumakis2014directed,szamel2014self,szamel2015glassy,maggi2015multidimensional,Flenner:2016tng,paoluzzi2016critical,Marconi:2016vdi,szamel2017evaluating,sandford2017pressure,caprini2018linear,fodor2018statistical, 
berthier2013non,Marconi:2015tas,shankar2018hidden}. 

Neglecting inertia effects \cite{Purcell:1977lal}, we model the overdamped
Brownian motion of the colloidal particle by the Langevin equation
\cite{Gardiner:HandbookOfStochasticMethods,vanKampen1992stochastic,snook2006langevin,mazo2002brownian}
\begin{equation}
\label{eq:LE}
\dot{\VEC{x}}(t) = \VEC{v}(\VEC{x}(t),t) + \sqrt{2\Da}\,\VEC{\eta}(t) + \sqrt{2D}\,\VEC{\xi}(t)
\, .
\end{equation}
The deterministic forces are collected in
\begin{subequations}
\label{eq:vf}
\begin{equation}
\label{eq:v}
\VEC{v}(\VEC{x},t) = \frac{1}{\gamma}\VEC{f}(\VEC{x},t)
\, ,
\end{equation}
where $\gamma$ is the hydrodynamic friction coefficient of the particle
and
\begin{equation}
\label{eq:f}
\VEC{f}(\VEC{x},t) = -\nabla U(\VEC{x},t) + \VEC{F}(\VEC{x},t)
\, .
\end{equation}
\end{subequations}
Thermal fluctuations are described by unbiased Gaussian white noise sources
$\VEC{\xi}(t)$ with mutually independent,
delta-correlated components $\xi_i(t)$, i.e.\
$\langle \xi_i(t)\xi_j(t') \rangle = \delta_{ij}\delta(t-t')$,
where the angular brackets denote the average over many realizations
of the noise and $i,j \in \{1,\ldots,d\}$.
The strength of the thermal fluctuations is given by the
particle's diffusion coefficient $D$, which is connected to the
temperature $T$ and the friction $\gamma$ by the
fluctuation-dissipation relation $D=\kB T/\gamma$
\cite{Einstein:1905udv,Nyquist:1928tao,Callen:1951iag,Kubo:1966tfd}
($\kB$ is Boltzmann's constant),
as a consequence of the equilibrium properties of the thermal bath.

The term $\sqrt{2\Da}\VEC{\eta}(t)$ in \eqref{eq:LE}
represents the active force components,
with  $\VEC\eta(t) = (\eta_1(t),\ldots,\eta_d(t))$
being unbiased, mutually independent noise processes,
and $\Da$ being an effective ``active diffusion'' characterizing
the strength of the active fluctuations.
The model \eqref{eq:LE} does not contain ``active friction'',
i.e.\ an integral term over $\dot{\VEC{x}}(t)$ with a friction kernel
modelling the damping effects associated with the
active forcing.
It thus represents the situation of low activity or particle density,
where active fluctuations do not dominate over thermal ones and such
friction effects are negligibly small \cite{maggi2017memory,chen2007fluctuations}.
In spite of this approximation,
the model \eqref{eq:LE} has been applied successfully to
describe various active particle systems
\cite{Fily:2012aps,Farage:2015eii,Maggi:2014gee,Argun:2016nbs,maggi2017memory,chaki2018entropy,Marconi:2015tas,shankar2018hidden},
and has become quite popular in an even more simplified variant
which neglects thermal fluctuations ($D=0$)
\cite{Fodor:2016hff,marconi2017heat,mandal2017entropy,Puglisi:2017crf,
koumakis2014directed,szamel2014self,szamel2015glassy,maggi2015multidimensional,Flenner:2016tng,paoluzzi2016critical,Marconi:2016vdi,szamel2017evaluating,sandford2017pressure,caprini2018linear,fodor2018statistical}. 
However, for thermodynamic consistency
it is necessary to consider both noise sources \cite{Puglisi:2017crf},
especially when assessing the flow of heat and entropy,
which is in general produced in the
thermal environment as well as in the processes fuelling the active
fluctuations \cite{gaspard2017communication,pietzonka2017entropy,speck2018active}.

As active fluctuations are the result of perpetual
energy conversion (e.g., by the bacteria in the bath, or by the
propulsion mechanism of the particle), their
salient feature is that they
do not fulfill a fluctuation-dissipation relation.
In our model \eqref{eq:LE}, this is particularly obvious as
active friction effects are absent by the assumption of being negligibly small;
in general, the active friction kernel would not match the
active noise correlations \cite{zamponi2005fluctuation,ohkuma2007fluctuation}.
In that sense, the active fluctuations $\VEC{\eta}(t)$
characterize a non-thermal environment or bath for the
Brownian particle with intrinsic non-equilibrium properties.
We emphasize that $\VEC{\eta}(t)$ is not a quantity directly measurable,
but rather embodies the net effects of the active components in the environment
or the self-propulsion mechanism of the particle,
similar to the white noise $\VEC{\xi}(t)$ representing an
effective description of the innumerable
collisions with the fluid molecules in the aqueous solution.

Although
several results in the present paper will be generally valid for any
noise process $\VEC{\eta}(t)$ (with finite moments),
our main aim is to study the specific situation of
exponentially correlated, Gaussian non-equilibrium fluctuations
\cite{koumakis2014directed,szamel2014self,marconi2016velocity,marconi2017heat,
Fodor:2016hff,Flenner:2016tng,mandal2017entropy,Puglisi:2017crf},
generated from an explicitly solvable Ornstein-Uhlenbeck process,
\begin{equation}
\label{eq:OUP}
\dot{\VEC{\eta}}(t) = -\frac{1}{\ta}\VEC{\eta}(t) + \frac{1}{\ta} \VEC{\zeta}(t)
\, ,
\end{equation}
i.e.\ so-called Gaussian colored noise \cite{hanggi1994colored}.
Here, 
$\VEC\zeta(t)=(\zeta_1(t),\ldots,\zeta_d(t))$ are mutually independent,
unbiased, delta-correlated Gaussian noise sources, just like $\VEC{\xi}(t)$,
but completely unrelated to them.
The characteristic time $\ta$
quantifies the correlation time of the process ($i,j \in \{1,\ldots,d\}$),
\begin{equation}
\label{eq:corrEta}
\langle \eta_i(t) \eta_j(t') \rangle = \frac{\delta_{ij}}{2\ta} e^{-|t-t'|/\ta}
\, ,
\end{equation}
as can be verified easily from the explicit steady-state solution of
\eqref{eq:OUP}.
It is thus a measure for the persistence of the active
fluctuations.

The model \eqref{eq:LE}, \eqref{eq:vf} describes a single
Brownian particle
in simultaneous contact with a thermal bath and
an active environment as a source of non-equilibrium fluctuations $\VEC{\eta}(t)$.
Our main aim in this paper is to investigate in detail the
role these non-equilibrium fluctuations play for the stochastic
energetics \cite{Sekimoto:StochasticEnergetics} and
thermodynamics \cite{Seifert:2012stf} of the Brownian particle.
While we adopt the single particle picture for simplicity,
all our results hold for
multiple, interacting Brownian particles as well.
In this case, the symbol $\VEC{x}$ in \eqref{eq:LE}, \eqref{eq:vf}
denotes a super-vector
collecting the positions $\VEC{x}_i$ ($i=1,2,\ldots,N$) of all $N$ particles,
i.e.\ $\VEC{x}=(\VEC{x}_1,\VEC{x}_2,\ldots,\VEC{x}_N)$,
and similarly for the forces, velocities, and so on.
The only requirement is that the particles are identical
in the sense that they have the same coupling coefficients
$D$ and $\Da$, and that
the active fluctuations $\VEC{\eta}_i(t)$
of the individual constituents are independent, but share
identical statistical properties.

\section{The ideal thermal bath: $D_a=0$}
\label{sec:Da0}
In order to set up the framework and further establish notation,
we start with
briefly recalling the well-known case of a Brownian particle in
sole contact with a thermal equilibrium reservoir,
\begin{equation}
\label{eq:LEDa0}
\dot{\VEC{x}}(t) = \VEC{v}(\VEC{x}(t),t) + \sqrt{2D}\,\VEC{\xi}(t)
\, ,
\end{equation}
where the deterministic driving $\VEC{v}(\VEC{x}(t),t)$
is defined as in \eqref{eq:vf}.
The Brownian particle can be prevented from equilibrating with
the thermal bath
by time-variations in the potential $U(\VEC{x},t)$ or by 
non-conservative external forces $\VEC{F}(\VEC{x},t)$.

\subsection{Energetics}
Following Sekimoto \cite{Sekimoto:1998lea,Sekimoto:StochasticEnergetics},
the heat $\delta Q$ that the particle exchanges with the thermal bath while
moving over an infinitesimal distance $\mathrm{d}\VEC{x}(t)$ during a time-step
$\mathrm{d}t$ from $t$ to $t+\mathrm{d}t$ is quantified as
the energy the thermal bath transfers to the particle along this displacement
due to friction $-\gamma\dot{\VEC{x}}(t)$ and fluctuations
$\sqrt{2\kB T\gamma}\,\VEC{\xi}(t)$, i.e.\
\begin{equation}
\label{eq:Qdef}
\delta Q(t) = \left( -\gamma\dot{\VEC{x}}(t)+\sqrt{2\kB T\gamma}\,\VEC{\xi}(t) \right) \cdot \mathrm{d}\VEC{x}(t)
\, ,
\end{equation}
where the product needs to be interpreted in Stratonovich sense
\cite{Note2}.
With the definition \eqref{eq:Qdef}, heat is counted as positive if
received by the particle and as negative when dumped into the environment
(this sign convention is thus the same as Sekimoto's original one
\cite{Sekimoto:1998lea,Sekimoto:StochasticEnergetics}).
From the equation of motion \eqref{eq:LEDa0}, we immediately see that
the heat exchange can be equivalently written as
\begin{equation}
\label{eq:QDa0}
\delta Q(t) = - \left( -\nabla U(\VEC{x}(t),t) + \VEC{F}(\VEC{x},t) \right) \cdot \mathrm{d}\VEC{x}(t)
\, .
\end{equation}
The change of the particle's ``internal'' energy $\mathrm{d}U(t)$
over the same
displacement $\mathrm{d}\VEC{x}(t)$ is given by the total differential
\begin{equation}
\label{eq:U}
\mathrm{d}U(t) = \nabla U(\VEC{x}(t),t) \cdot \mathrm{d}\VEC{x}(t) + \frac{\partial U(\VEC{x}(t),t)}{\partial t}\,\mathrm{d}t
\, .
\end{equation}
Hence, if the potential does not vary over time,
the only contribution to $\mathrm{d}U(t)$ comes from the change in the
potential energy associated with the displacement $\mathrm{d}\VEC{x}$.

On the other hand, even if the particle does not move within $\mathrm{d}t$,
its ``internal'' energy can still change due to variations of the potential
landscape by an externally applied, time-dependent protocol.
Prototype examples
are intensity- or position-variations of optical tweezers,
which serve as a trap for
the colloidal particle \cite{gomez2010steady,schmiedl2007optimal,Argun:2016nbs,ciliberto2017experiments,mcgloin2016colloidal}.
Being imposed and controlled externally, these contributions are
interpreted as work performed on the particle.
A second source of external forces which may contribute to such
work are the non-conservative components $\VEC{F}(\VEC{x},t)$ in \eqref{eq:LEDa0}
(originally not considered by Sekimoto \cite{Sekimoto:1998lea},
but systematically analyzed later by Speck et al.~\cite{speck2008role}).
The total work applied on the particle by external forces is
therefore given by
\begin{equation}
\label{eq:W}
\delta W(t) = \frac{\partial U(\VEC{x}(t),t)}{\partial t}\,\mathrm{d}t + \VEC{F}(\VEC{x}(t),t) \cdot \mathrm{d}\VEC{x}(t)
\, .
\end{equation}

Combining Eqs.\ \eqref{eq:QDa0}, \eqref{eq:U}, and \eqref{eq:W} we
obtain the first law
\begin{equation}
dU = \delta Q + \delta W
\end{equation}
for the energy balance over infinitesimal displacements $\mathrm{d}\VEC{x}(t)$,
valid at any point in time $t$.
After integration along a specific trajectory $\{\VEC{x}(t)\}_{t=0}^\tau$ of duration $\tau$,
which starts at $\VEC{x}(0)=\VEC{x}_0$ and ends at some $\VEC{x}(\tau)=\VEC{x}_\tau$
(which is different for every realization of the thermal
noise  $\VEC{\xi}(t)$ in \eqref{eq:LEDa0}, even if $\VEC{x}(0)=\VEC{x}_0$ is kept fixed),
we find the first law at the trajectory level \cite{van2015ensemble}
\begin{equation}
\Delta U[\Traj{\VEC{x}}] = U(\VEC{x}_\tau,\tau) - U(\VEC{x}_0,0) = Q[\Traj{\VEC{x}}] + W[\Traj{\VEC{x}}]
\, .
\end{equation}
Here, $\Delta U[\Traj{\VEC{x}}]$, $Q[\Traj{\VEC{x}}]$, and $W[\Traj{\VEC{x}}]$
denote the integrals of the expressions
\eqref{eq:QDa0} [or, equivalently, \eqref{eq:Qdef}], \eqref{eq:U}, and \eqref{eq:W}, respectively,
along the trajectory $\Traj{\VEC{x}}=\{\VEC{x}(t)\}_{t=0}^\tau$; the
notation $[\Traj{\VEC{x}}]$ explicitly indicates the dependence on that trajectory.

\subsection{Path probability and entropy}
\label{sec:ppDeltaSDa0}
The next step towards a thermodynamics characterization of the Brownian
motion \eqref{eq:LEDa0} is to introduce an entropy change or
entropy production associated with individual trajectories \cite{Seifert:2005epa}.
From the viewpoint of irreversibility, such an entropy concept has
been defined as the log-ratio of the probability densities for observing
a certain particle trajectory and its time-reversed twin
\cite{andrieux2007entropy}.
We express the probability density of a particle trajectory
by the standard Onsager-Machlup path integral
\cite{Onsager:1953fai,Machlup:1953fai,Chernyak:2006pia,cugliandolo2017rules},
\begin{equation}
\label{eq:ppDa0}
\pp{\traj{\VEC{x}}|\VEC{x}_0}
\propto
\exp\left\{
-\int_0^\tau \!\!\!\mathrm{d}t \, \left[ 
	\frac{(\dot{\VEC{x}}_t-\VEC{v}_t)^2}{4D}
	+ \frac{\VEC{\nabla}\cdot\VEC{v}_t}{2}
\right]
\right\}
\, ,
\end{equation}
where we condition on a fixed starting position $\VEC{x}_0$,
and, accordingly, introduce the notation $\traj{\VEC{x}}=\{\VEC{x}(t)\}_{t>0}^\tau$
to denote a trajectory which starts
at fixed position $\VEC{x}(0)=\VEC{x}_0$.
In contrast to $\Traj{\VEC{x}}$ from above
the set of points $\traj{\VEC{x}}$ 
does therefore not contain the initial point $\VEC{x}_0$
(i.e.\ $\Traj{\VEC{x}}=\VEC{x}_0 \cup \traj{\VEC{x}}$).
The path probability \eqref{eq:ppDa0} is to be understood as a product of transition
probabilities in the limit of infinitesimal time-step size, using a mid-point discretization rule.
The divergence of $\VEC{v}$ in the second term represents the path-dependent part of normalization
\cite{Chernyak:2006pia,cugliandolo2017rules},
while all remaining normalization
factors are path-independent constants and thus omitted.
For convenience, we have introduced the short-hand notation
$\VEC{x}_t = \VEC{x}(t)$ and $\VEC{v}_t=\VEC{v}(\VEC{x}(t),t)$.

We now consider the time-reversed version $\tilde{\VEC{x}}(t)$
of this trajectory, which is traced out backward from the final point
$\VEC{x}(\tau)=\VEC{x}_\tau$ to the initial point $\VEC{x}_0$
when advancing time,
\begin{equation}
\label{eq:xrev}
\tilde{\VEC{x}}(t) = \VEC{x}(\tau-t)
\, ,
\end{equation}
and ask how likely it is that $\trajR{{\VEC{x}}}=\{\tilde{\VEC{x}}(t)\}_{t>0}^\tau$
is generated by the same Langevin equation \eqref{eq:LEDa0} as $\traj{\VEC{x}}$,
with the same deterministic forces $\VEC{v}$ acting at identical positions along
the path. The latter requirement implies that in case of explicitly time-dependent
external forces the force protocol has to be time-inverted,
i.e.\ $\VEC{v}(\VEC{x},t)$ is replaced by $\VEC{v}(\VEC{x},\tau-t)$ in \eqref{eq:LEDa0}
to construct the Langevin equation for the time-reserved path
\cite{Note3}.
From that Langevin equation we can deduce the probability
$\ppR{\trajR{\VEC{x}}|\tilde{\VEC{x}}_0}$
for observing the backward trajectory $\trajR{\VEC{x}}$,
conditioned on its initial position $\tilde{\VEC{x}}_0=\VEC{x}(\tau)=\VEC{x}_\tau$,
in analogy to \eqref{eq:ppDa0}.
Using \eqref{eq:xrev}, we can then express $\ppR{\trajR{\VEC{x}}|\tilde{\VEC{x}}_0}$
in terms of the forward path $\traj{\VEC{x}}$,
so that we find for the path probability ratio
\begin{equation}
\label{eq:ppratioDa0}
\frac{\ppR{\trajR{\VEC{x}}|\tilde{\VEC{x}}_0}}{\pp{\traj{\VEC{x}}|\VEC{x}_0}}
= e^{-\Delta S[\Traj{\VEC{x}}]/\kB}
\, ,
\end{equation}
with the quantity $\Delta S[\Traj{\VEC{x}}]$ being a functional of the forward path only,
\begin{align}
\label{eq:DeltaSDa0}
\Delta S[\Traj{\VEC{x}}] 
& = \frac{1}{T} \int_0^\tau \mathrm{d}t \, \dot{\VEC{x}}(t)\cdot\VEC{f}(\VEC{x}(t),t)
\nonumber\\
& = \frac{1}{T} \int_0^\tau \VEC{f}(\VEC{x}(t),t)\cdot\mathrm{d}\VEC{x}(t)
\, .
\end{align}
As a quantitative measure of irreversibility,
$\Delta S[\Traj{\VEC{x}}]$ is identified with the entropy
production along the path $\traj{\VEC{x}}$ with
given initial position $\VEC{x}_0$.

For an infinitesimal displacement $\mathrm{d}\VEC{x}(t)$ the corresponding
entropy change reads
\begin{equation}
\label{eq:deltaSDa0}
\delta S(t) = \frac{1}{T}\,\VEC{f}(\VEC{x}(t),t)\cdot\mathrm{d}\VEC{x}(t)
= -\frac{\delta Q(t)}{T}
\, .
\end{equation}
The last equality follows from comparison with \eqref{eq:QDa0}
[see also \eqref{eq:vf}] and states that the entropy production along
$\mathrm{d}\VEC{x}(t)$ is given by the heat $-\delta Q(t)$
dissipated into the environment
during that step divided by the bath temperature. For that reason,
$\delta S(t)$, and $\Delta S[\Traj{\VEC{x}}]$, are more accurately called
\textit{entropy production in the environment}.

The entropy of the Brownian particle itself (i.e.\ the entropy associated with
the system degrees of freedom $\VEC{x}$) is defined as the state function
\cite{Seifert:2005epa,Seifert:2012stf}
\begin{equation}
S_{\mathrm{sys}}
(\VEC x, t)
 = -\kB \ln p(\VEC{x},t)
\, ,
\end{equation}
where $p(\VEC{x},t)$ is the time-dependent solution of the
Fokker-Planck equation
\cite{Gardiner:HandbookOfStochasticMethods,vanKampen1992stochastic,mazo2002brownian}
associated with \eqref{eq:LEDa0},
for the same initial distribution $p_0(\VEC{x}_0)=p(\VEC{x}_0,0)$
from which the initial value $\VEC{x}_0$
for the path $\traj{\VEC{x}}$ is drawn.
The change in system entropy along the trajectory is therefore
given by
\begin{equation}
\label{eq:DeltaSsysDa0}
\Delta S_{\mathrm{sys}}(\VEC{x}_0,\VEC{x}_\tau) = -\kB \ln p(\VEC{x}_\tau,\tau) + \kB \ln p(\VEC{x}_0,0)
\, .
\end{equation}

Combining $\Delta S[\Traj{\VEC{x}}]$ and $\Delta S_{\mathrm{sys}}(\VEC{x}_0,\VEC{x}_\tau)$,
we define the total entropy production along a trajectory $\Traj{\VEC{x}}$ \cite{Seifert:2005epa},
\begin{equation}
\label{eq:DeltaStotDa0}
\Delta S_{\mathrm{tot}}[\Traj{\VEC{x}}] = 
\Delta S[\Traj{\VEC{x}}] + \Delta S_{\mathrm{sys}}(\VEC{x}_0,\VEC{x}_\tau)
\, .
\end{equation}
It fulfills the integral fluctuation theorem
\cite{Seifert:2008stp,Seifert:2012stf}
\begin{equation}
\label{eq:IFTDa0}
\left\langle e^{-\Delta S_{\mathrm{tot}}/\kB} \right\rangle = 1
\end{equation}
as a direct consequence of
\eqref{eq:ppratioDa0} [see also \eqref{eq:DeltaSsysDa0}].
The average in \eqref{eq:IFTDa0} is over all trajectories
with a given, but arbitrary distribution $p_0(\VEC{x}_0)$
of initial values $\VEC{x}_0$ \cite{Seifert:2005epa}.

\section{The non-equilibrium environment}
\label{sec:noneqBath}
We now focus on the full model \eqref{eq:LE}, \eqref{eq:vf}, \eqref{eq:OUP}
for Brownian motion subject to active fluctuations $\VEC{\eta}(t)$.
Our main goal is to develop a trajectory-wise thermodynamic description
as a natural generalization of 
stochastic energetics and thermodynamics in
a purely thermal environment (see previous section).
We thus treat the active forces $\VEC{\eta}(t)$ in the same
way as the thermal noise $\VEC{\xi}(t)$,
namely as a source of fluctuations whose
specific realizations are not accessible but whose statistical properties
determine the probability $\pp{\traj{\VEC{x}}|\VEC{x}_0}$
for observing a certain particle trajectory $\traj{\VEC{x}}$.
We are interested in how the non-equilibrium characteristics of the active fluctuations
affect the irreversibility measure encoded in the ratio between forward and
backward path probabilities, and how this measure is connected to
the energetics of the active Brownian motion.

\subsection{Energetics and entropy}
\label{sec:EE}
Comparing \eqref{eq:LE} with \eqref{eq:LEDa0}, we may conclude that the
stochastic energetics associated with \eqref{eq:LE} can be obtained from the
energetics for \eqref{eq:LEDa0}
by adjusting the total forces acting on the Brownian particle,
i.e.\ by replacing $\VEC{v}(\VEC{x},t)$ with
$\VEC{v}(\VEC{x},t) + \sqrt{2D_a}\VEC{\eta}(t)$.
However, there is another, maybe less obvious way of turning 
\eqref{eq:LEDa0} into the model \eqref{eq:LE} with active
fluctuations, namely by substituting
$\dot{\VEC{x}}$ with $\dot{\VEC{x}}-\sqrt{2D_a}\VEC{\eta}(t)$.
In the following we argue that
these two approaches correspond to two different physical situations
(depicted in Fig.~\ref{fig:1}),
with different trajectory-wise energy balances.

\subsubsection{Active bath}
\label{sec:energeticsAB}
Examples for an ``active bath'' are
the intracellular matrix of a living cell
\cite{mackintosh2010active,guo2014probing,ahmed2015active,needleman2017active,bernheim2018living,
ben2011effective,fodor2015activity}
or an aqueous suspension containing swimming bacteria
\cite{Bechinger:2016api,Argun:2016nbs}.
These active environments then drive
a passive Brownian particle, e.g.\ a colloid,
via interactions or ``collisions'', which are
included in the equation of motion for the passive tracer
as effective
non-equilibrium fluctuations $\VEC{\eta}(t)$.
In this case, the fluctuations $\VEC{\eta}(t)$ can
indeed be interpreted as additional time-dependent forces 
from sources external to the Brownian particle.
Hence, the total external force acting on the Brownian particle at time $t$
is given by $\gamma\left(\VEC{v}(\VEC{x}(t),t) + \sqrt{2D_a}\VEC{\eta}(t)\right)$.
This modification of the external force does obviously not affect the basic definition
\eqref{eq:Qdef} of heat exchanged with the thermal bath.
Using the force balance expressed in the Langevin equation \eqref{eq:LE}
to replace $-\gamma\dot{\VEC{x}}(t)+\gamma\sqrt{2D}\,\VEC{\xi}(t)$ we obtain
\begin{equation}
\label{eq:deltaQ+}
\delta Q_+(t) = - \Big( \VEC{f}(\VEC{x}(t),t) + \sqrt{2D_a}\gamma\,\VEC{\eta}(t) \Big)
	\cdot \mathrm{d}\VEC{x}(t)
\, .
\end{equation}
The active fluctuations $\VEC{\eta}(t)$ formally play the role of additional
non-conservative force components, and thus affect the heat which is dissipated into the
thermal environment in order to balance all acting external forces.
However, they cannot be controlled to perform work on the particle
due to their inherent fluctuating character as an active bath, such
that the definition \eqref{eq:W} of the work remains unchanged.
Indeed, this standard definition of work from stochastic thermodynamics
has been used to analyze a microscopic
heat engine operating between two active baths \cite{Krishnamurthy:2016ams}.
Finally, the change of ``internal'' energy \eqref{eq:U} over
a time-interval $\mathrm{d}t$ is determined by the potential
$U(\VEC{x},t)$ only,
and thus is not altered by the presence of the active fluctuations $\VEC{\eta}(t)$
either.

Combining \eqref{eq:U}, \eqref{eq:W} and \eqref{eq:deltaQ+} we find
the energy balance (first law)
\begin{equation}
dU = \delta Q_+ + \delta W + \delta A_+
\end{equation}
for Brownian motion in an active bath.
Here, we have introduced the energy exchanged with the active bath,
\begin{equation}
\label{eq:deltaA+}
\delta A_+ = \sqrt{2\Da}\gamma\,\VEC{\eta}(t)\cdot\mathrm{d}\VEC{x}(t)
\, .
\end{equation}
It might be best interpreted as ``heat''
in the sense of Sekimoto's general definition, that any energy
exchange with unknown or inaccessible degrees of freedom may be identified
as ``heat'' \cite{Sekimoto:StochasticEnergetics}.
In our setup, the active fluctuations $\VEC{\eta}(t)$ represent an effective
description of the forces from the active environment, and are thus in general not
directly measurable in an experiment.

Based on the heat
the Brownian particle exchanges with the thermal environment,
we can identify the entropy production in the thermal environment as
\begin{equation}
\label{eq:deltaS+}
\begin{split}
\delta S_+(t)
& = -\frac{\delta Q_+(t)}{T}
\\
& = \frac{1}{T} \left( \VEC{f}(\VEC{x}(t),t)+\sqrt{2\Da}\gamma\,\VEC{\eta}(t) \right) \cdot\mathrm{d}\VEC{x}(t)
\, ,
\end{split}
\end{equation}
in analogy to \eqref{eq:deltaSDa0}.
We refrain, however, from defining an entropy production in
the active bath.
Such an environment,
itself being in a non-equilibrium
state due to continuous dissipation of energy,
does not possess a well-defined entropy,
so that the ``heat'' \label{eq:deltaQa} dissipated into
the active bath cannot be associated with a change of bath entropy.

\subsubsection{Self-propulsion}
\label{sec:energeticsSP}
Self-propulsion occurs if a particle is able to locally convert
external fuel on its own into directed motion; such fuel may, e.g., be
nutrients for a bacterium or some kind of chemical components
like hydrogen peroxide for a colloidal Janus particle;
see Fig.~1 in the Review \cite{Bechinger:2016api} for an extensive
collection of self-propelled (Brownian) particles.
This self-propulsion entails that
incessant particle motion can occur already without any external forces
being applied. It is represented by
the active fluctuations $\VEC{\eta}(t)$
in the equations of motion \eqref{eq:LE} for the self-propelled Brownian particle.
For $\VEC{f}(\VEC{x},t)\equiv 0$, and
without thermal fluctuations, $\sqrt{2D}\,\VEC{\xi}(t)\equiv 0$,
the momentary particle velocity $\dot{\VEC{x}}(t)$ is exactly equal to
the active velocity $\sqrt{2\Da}\VEC{\eta}(t)$.
In that sense, self-propulsion is force-free.
More precisely, the driving force, which is created locally by
the particle for self-propulsion, is compensated according to
\textit{actio est reactio}, such that
the total force acting on
a fluid volume comprising the particle and its
active self-propulsion mechanism is zero.
The corresponding dissipation (and entropy production) inside such a fluid volume,
which results from the conversion of energy or fuel
to generate the self-propulsion drive, cannot be
quantified by our effective description
of the active propulsion as fluctuating forces $\VEC{\eta}$,
because it does not contain any information on the underlying microscopic
processes.
In order to quantify such entropy production,
a specific model for
the self-propulsion mechanism is required 
\cite{gaspard2017communication,pietzonka2017entropy,speck2018active}.
In other words, our ``coarse-grained'' description \eqref{eq:OUP} does not allow
us to assess how much energy the conversion process behind the
self-propulsion drive dissipates.

On the other hand, if the force-free motion of the self-propelled
particle is disturbed by the presence of external forces,
$\VEC{f}(\VEC{x},t) \neq 0$,
or thermal fluctuations $\sqrt{2D}\gamma\,\VEC{\xi}(t)$,
its actual velocity deviates from the
self-propulsion velocity $\sqrt{2\Da}\VEC{\eta}(t)$,
resulting in hydrodynamic friction
$-\gamma \left[\dot{\VEC{x}}(t) -\sqrt{2D_a}\VEC{\eta}(t) \right]$ between the
fluid volume comprising the particle and its
active self-propulsion mechanism
on the one hand and the surrounding fluid environment on the other hand.
Together with the thermal fluctuations
$\sqrt{2D}\gamma\,\VEC{\xi}(t)$ these friction forces
are the forces exchanged with the thermal environment,
entailing an energy transfer between
particle and thermal bath.
It is this dissipation associated with the deviations
of the particle trajectory from the self-propulsion path
which we can quantify.
The corresponding heat exchange with the thermal bath for a displacement $\mathrm{d}\VEC{x}(t)$
taking place over a time-interval $\mathrm{d}t$ at time $t$ is given by
$\left(-\gamma \left[\dot{\VEC{x}}(t) -\sqrt{2D_a}\VEC{\eta}(t) \right] + \sqrt{2D}\gamma\,\VEC{\xi}(t) \right)\cdot
\left( \mathrm{d}\VEC{x}(t) - \sqrt{2\Da}\,\VEC{\eta}(t)\,\mathrm{d}t \right)$,
or, using that hydrodynamic friction and thermal fluctuations are
exactly balanced by the external force $\VEC{f}$ [see \eqref{eq:LE}],
\begin{equation}
\label{eq:deltaQ-}
\delta Q_-(t) = -\VEC{f}(\VEC{x}(t),t)
\cdot
\left( \mathrm{d}\VEC{x}(t) - \sqrt{2\Da}\,\VEC{\eta}(t)\,\mathrm{d}t \right)
\, .
\end{equation}
This definition of heat is equivalent to the one suggested in
\cite{speck2018active} for active particles,
which are self-propelled by chemical conversions.

The work $\delta W(t)$, on the other hand, performed on or by the particle
during the time-step $\mathrm{d}t$, which can be controlled or harvested by
an external agent, is exactly the same as without active propulsion,
given in \eqref{eq:W},
because from the operational viewpoint of the external agent it is irrelevant
what kind of mechanisms propel the particle.
Accordingly, it has been used in \cite{speck2016stochastic} for connecting
work in active particle systems to pressure,
and in \cite{martin2018extracting}
for measuring the work performed by a colloidal heat engine with an active particle as
the working medium.
The definition \eqref{eq:U}, finally, for the change in ``internal'' energy $\mathrm{d}U(t)$
is independent of how particle motion is driven as well, and thus remains unaffected
by our interpretation of $\sqrt{2\Da}\,\VEC{\eta}(t)$ as active propulsion.

Combining \eqref{eq:U}, \eqref{eq:W} and \eqref{eq:deltaQ-},
we find the first law-like relation
\begin{equation}
\delta U = \delta Q_- + \delta W + \delta A_-
\end{equation}
for active, self-propelled Brownian motion
described by the Langevin equation \eqref{eq:LE}.
Here, we balance the different energetic contributions by
introducing
\begin{equation}
\label{eq:deltaA-}
\delta A_-(t) =
-\VEC{f}(\VEC{x}(t),t) \cdot\sqrt{2\Da}\,\VEC{\eta}(t)\,\mathrm{d}t
\, .
\end{equation}
This quantity represents the contribution to the heat exchange
with the thermal bath, which is contained
only in the active component $\sqrt{2\Da}\,\VEC{\eta}(t)\,\mathrm{d}t$ of the
full particle displacement $\mathrm{d}\VEC{x}(t)$,
see \eqref{eq:deltaQ-}.
We can therefore interpret it as the ``heat'' transferred
from the active fluctuations to the thermal bath via the
Brownian particle. Again, this is a quantity which can in general
not be measured in an experiment, since it is produced
by the inaccessible active fluctuations $\VEC{\eta}(t)$.

Finally, we can relate the heat exchange
$\delta Q_-$ with the thermal
bath to dissipation and define a corresponding entropy production
in the environment,
\begin{equation}
\label{eq:deltaS-}
\begin{split}
\delta S_-(t) 
& = -\frac{\delta Q_-(t)}{T}
\\
& = \frac{1}{T} \, \VEC{f}(\VEC{x}(t),t) \cdot
\left( \mathrm{d}\VEC{x}(t) - \sqrt{2\Da}\,\VEC{\eta}(t)\,\mathrm{d}t \right)
\, .
\end{split}
\end{equation}

\subsection{Path probability}
\label{sec:pp}
We now calculate the probability $\pp{\traj{\VEC{x}}|\VEC{x}_0}$ for
observing a certain path $\traj{\VEC{x}}=\{\VEC{x}(t)\}_{t>0}^\tau$
starting at $\VEC{x}_0$,
generated under the combined influence of thermal and active fluctuations.
We treat these two noise sources on equal terms,
namely as fluctuating forces with unknown specific realizations but known statistical
properties.
Due to the memory of the active noise $\VEC{\eta}(t)$, the system~\eqref{eq:LE} is non-Markovian.
Therefore, the standard Onsager-Machlup path integral
\cite{Onsager:1953fai,Machlup:1953fai,Chernyak:2006pia} cannot be applied directly to
obtain $\pp{\traj{\VEC{x}}|\VEC{x}_0}$.
However, for the Ornstein-Uhlenbeck model \eqref{eq:OUP} of $\VEC{\eta}(t)$
the combined set of variables $(\VEC{x},\VEC{\eta})$ is Markovian and we can
easily write down the path probability
$\pp{\traj{\VEC{x}},\traj{\VEC{\eta}}|\VEC{x}_0,\VEC{\eta}_0}$
for the joint trajectory $(\traj{\VEC{x}},\traj{\VEC{\eta}})=\{(\VEC{x}(t),\VEC{\eta}(t))\}_{t>0}^\tau$,
conditioned on an initial configuration $(\VEC{x}_0,\VEC{\eta}_0)$:
\begin{multline}
\label{eq:ppxeta}
\pp{\traj{\VEC{x}},\traj{\VEC{\eta}}|\VEC{x}_0,\VEC{\eta}_0}
\propto
\exp \left\{ -\int_{0}^\tau \!\!\!\mathrm{d}t
\, \left[ 
	\frac{(\dot{\VEC{x}}_t-\VEC{v}_t-\sqrt{2\Da}\,\VEC{\eta}_t)^2}{4D}
\right.\right.
\\[0.5ex]
\left.\left.
	+ \frac{(\ta \dot{\VEC{\eta}}_t+\VEC{\eta}_t)^2}{2}
	+ \frac{\VEC{\nabla}\cdot\VEC{v}_t}{2}
\right]
\right\}
\, .
\end{multline}
To calculate the path weight for the particle trajectories $\pp{\traj{\VEC{x}}|\VEC{x}_0}$,
we have to integrate out the active noise history,
\begin{equation}
\label{eq:ppintegrateEta}
\pp{\traj{\VEC{x}}|\VEC{x}_0} = \int \D\traj{\VEC{\eta}}\,\mathrm{d}\VEC{\eta}_0 \;
\pp{\traj{\VEC{x}},\traj{\VEC{\eta}}|\VEC{x}_0,\VEC{\eta}_0} \, p_0(\VEC{\eta}_0|\VEC{x}_0)
\, ,
\end{equation}
where $p_0(\VEC{\eta}_0|\VEC{x}_0)$ characterizes the initial distribution of the
active fluctuations at $t=0$.

Since the variables $\VEC{\eta}(t)$ represent an effective description of the active
fluctuations which are not experimentally accessible,
it is in general not possible to set up a specific initial state for $\VEC{\eta}(t)$.
As a consequence there are basically two physically reasonable choices for
$p_0(\VEC{\eta}_0|\VEC{x}_0)$.

On the one hand, we can assume that the active bath has reached its stationary
state, before we immerse the Brownian particle, such that the bath's initial distribution is
\textit{independent} of $\VEC{x}_0$ and given by the stationary distribution of $\VEC{\eta}(t)$, i.e.\
$p_0(\VEC{\eta}_0|\VEC{x}_0)=p_{\mathrm{s}}(\VEC{\eta}_0)$.
For the Ornstein-Uhlenbeck process \eqref{eq:OUP}, the stationary distribution reads
\cite{Risken:TheFokkerPlanckEquation,Gardiner:HandbookOfStochasticMethods}
\begin{equation}
\label{eq:pseta}
p_{\mathrm{s}}(\VEC{\eta}_0) = \sqrt{\frac{\ta}{\pi}} \, e^{-\ta \VEC{\eta}_0^2}
\, .
\end{equation}
At $t=0$ we place the Brownian particle into the fluid with an initial distribution
$p_0(\VEC{x}_0)$ of particle positions which can be prepared arbitrarily,
and start measuring immediately.

On the other hand, we may let the Brownian particle 
adapt to the active and thermal environments before performing measurements,
and assume that the system $(\VEC{x}_0,\VEC{\eta}_0)$
is in a \textit{joint} steady state $p_{\mathrm{s}}(\VEC{x}_0,\VEC{\eta}_0)=p_0(\VEC{\eta}_0|\VEC{x}_0)p_0(\VEC{x}_0)$
at $t=0$. In that case, control over the distribution $p_0(\VEC{x}_0)$ of
initial particle positions is limited, as it is influenced by the active fluctuations.
The form of $p_{\mathrm{s}}(\VEC{x}_0,\VEC{\eta}_0)$ depends on the particular set-up,
i.e.\ the specific choices for $U(\VEC{x},t)$ and $\VEC{F}(\VEC{x},t)$ in \eqref{eq:vf}.

In the following, we will perform the path integration \eqref{eq:ppintegrateEta} for
the first option, starting from independent initial conditions
$p_0(\VEC{\eta}_0|\VEC{x}_0)=p_{\mathrm{s}}(\VEC{\eta}_0)$;
the calculation for the second option 
can be carried out along the same lines, if the joint steady state
$p_{\mathrm{s}}(\VEC{x}_0,\VEC{\eta}_0)$ is Gaussian in the active noise $\VEC{\eta}_0$
\cite{activeGyrator}.
Plugging \eqref{eq:ppxeta} and \eqref{eq:pseta} into \eqref{eq:ppintegrateEta},
and performing a partial integration of the term proportional to
$\dot{\VEC{\eta}}_t^2$ in the exponent, we obtain
\begin{multline}
\label{eq:pp1}
\pp{\traj{\VEC{x}}|\VEC{x}_0}
\propto
\exp \left\{ \int_0^\tau \!\!\mathrm{d}t
	\left[ -\frac{\left( \dot{\VEC{x}}_t - \VEC{v}_t \right)^2}{4D} - \frac{\VEC{\nabla} \cdot \VEC{v}_t}{2} \right]
\right\}
\\[0.5ex]
\times \! \int \D\Traj{\VEC{\eta}} \;
\exp \left\{ \int_0^\tau \!\! \mathrm{d}t
	\frac{\sqrt{2\Da}}{2D} \VEC{\eta}_t\TRANS \left( \dot{\VEC{x}}_t - \VEC{v}_t \right) \right.
\\[0.5ex]
\left. \mbox{}- \frac{1}{2} \int_0^\tau \!\! \mathrm{d}t \int_0^\tau \!\! \mathrm{d} t^\prime \,
	\VEC{\eta}_t\TRANS \hat{V}_\tau(t, t^\prime) \VEC{\eta}_{t^\prime} \right\}
\, ,
\end{multline}
where we have used the abbreviation $\Traj{\VEC{\eta}}=\VEC{\eta}_0 \cup \traj{\VEC{\eta}}$
for the full path including the initial point $\VEC{\eta}_0$, in order to write
$\D\traj{\VEC{\eta}}\,\mathrm{d}\VEC{\eta}_0=\D\Traj{\VEC{\eta}}$.
The differential operator
\begin{subequations}
\label{eq:Vfull}
\begin{equation}
\label{eq:V}
\hat{V}_\tau(t,t') = 
\delta(t-t') \left[ \hat{V}(t) + \hat{V}_0(t) + \hat{V}_\tau(t) \right]
\end{equation}
consists of an ordinary component,
\begin{equation}
\hat{V}(t) = -\ta^2 \partial_t^2 + \left( 1 + \Da/D \right)
\, ,
\end{equation}
and two boundary components
\begin{gather}
\label{eq:V0}
\hat{V}_0(t) = \delta(t) \left( \ta - \ta^2 \partial_t \right)
\, ,
\\
\hat{V}_\tau(t) = \delta(t-\tau) \left( \ta + \ta^2 \partial_t \right)
\, ,
\end{gather}
\end{subequations}
which include the boundary terms picked up by the partial
integration of $\dot{\VEC{\eta}}_t^2$,
and from the initial distribution~\eqref{eq:pseta} of $\VEC\eta_0$.
The subscript $\tau$ in \eqref{eq:V} indicates that the operator $\hat{V}_\tau$
is acting on trajectories of duration $\tau$.

Since the path integral over the active noise histories $\VEC{\eta}(t)$ is Gaussian,
we can perform it exactly \cite{Zinn-Justin:QFTAndCriticalPhenomena}.
We find
\begin{widetext}
\begin{equation}
\label{eq:pp}
\pp{\traj{\VEC{x}}|\VEC{x}_0}
\propto
\exp \left\{ -\frac{1}{4D} \int_{0}^\tau \!\! \mathrm{d}t  \int_{0}^\tau \!\! \mathrm{d}t^\prime
	\left( \dot{\VEC{x}}_t - \VEC{v}_t \right)\TRANS
		\left[ \delta(t - t^\prime) - \frac{\Da}{D} \Gamma_\tau(t,t^\prime) \right]
	\left( \dot{\VEC{x}}_{t^\prime} - \VEC{v}_{t^\prime} \right)
- \frac{1}{2} \int_0^\tau \!\! \mathrm{d}t \, \VEC{\nabla} \cdot \VEC{v}_t
\right\}
\, ,
\end{equation}
\end{widetext}
where $\Gamma_\tau(t,t^\prime)$ denotes the operator inverse or Green's function of
$\hat{V}_\tau(t, t^\prime)$ in the sense that
\begin{equation}
\label{eq:defGamma}
\int_0^\tau \!\! \d t^\prime \; \hat{V}_\tau(t,t^\prime) \, \Gamma_\tau(t^\prime, t^{\prime\prime})
= \delta(t - t^{\prime\prime})
\, .
\end{equation}
Roughly speaking, this Gaussian integration can be understood by
thinking of $\hat{V}_\tau$ and $\Gamma_\tau$
as matrices with continuous indices $t$, $t^\prime$.
The path integral~\eqref{eq:pp1} is then a continuum generalization of an
ordinary Gaussian integral for finite-dimensional matrices,
and can be performed by ``completing the square''.
We provide a rigorous derivation of \eqref{eq:pp} and~\eqref{eq:defGamma}
in Appendix~\ref{app:integrateEta}.

In order to obtain the explicit form of the Green's function $\Gamma_\tau(t,t')$,
we need to solve the integro-differential equation \eqref{eq:defGamma}.
In our case, the operator $\hat{V}_\tau(t,t')$ is proportional to $\delta(t-t')$
[see \eqref{eq:V}]
and thus has a ``diagonal'' structure, such 
that \eqref{eq:defGamma} 
turns into an ordinary linear differential equation.
We can solve it by following standard methods
\cite{BenderOrszag:AdvMathMethods,StakgoldHolst:GreensFunctionsAndBVPs},
details are given in Appendix~\ref{app:GreenFunctionConstruction}.
We obtain
\begin{subequations}
\label{eq:GammaEtc}
\begin{widetext}
\begin{equation}
\label{eq:Gamma}
\Gamma_\tau(t, t^\prime)
= \left( \frac{1}{2 \ta^2 \lambda} \right)
	\frac{\kappa_+^2 \, e^{-\lambda |t-t^\prime|}
		+ \kappa_-^2 \, e^{-\lambda (2\tau - |t - t^\prime|)}
		- \kappa_+ \kappa_- \left[ e^{-\lambda (t+t^\prime)} + e^{-\lambda (2\tau - t - t^\prime)} \right]}
	{\kappa_+^2 - \kappa_-^2 \, e^{-2 \lambda \tau}}
\, ,
\end{equation}
\end{widetext}
with
\begin{equation}
\label{eq:lambda}
\lambda = \frac{1}{\ta} \sqrt{1 + \Da/D}
\, ,
\end{equation}
and
\begin{equation}
\kappa_\pm = 1 \pm \lambda \ta = 1 \pm \sqrt{1+\Da/D}
\, .
\end{equation}
\end{subequations}

With this expression for $\Gamma_\tau(t,t')$,
\eqref{eq:pp} represents the exact path probability density for the dynamics
of the Brownian particle \eqref{eq:LE}, under the influence of
active Ornstein-Uhlenbeck fluctuations \eqref{eq:OUP}.
This is our first main result.
We see that the active fluctuations with their colored noise character
lead to correlations in the path weight via the memory kernel $\Gamma_\tau(t, t^\prime)$.
They relate trajectory points at different times by an exponential weight factor
similar to the active noise correlation function \eqref{eq:corrEta},
but with a correlation time which is a factor of $(1+\Da/D)^{-1/2}$ smaller.
We emphasize again that we assumed independent initial conditions
for the particle's position $\VEC{x}_0$
and the active bath variables $\VEC{\eta}_0$,
$p_0(\VEC{\eta}_0|\VEC{x}_0)=p_{\mathrm{s}}(\VEC{\eta}_0)$
[see also the discussion around Eq.~\eqref{eq:pseta}].
A different choice for the initial distribution, for instance the joint
stationary state for $\VEC{x}_0$ and $\VEC{\eta}_0$, would result in a modified
$\Gamma_\tau(t, t^\prime)$, whose precise form in general depends on the specific
implementation of the deterministic forces $\VEC{f}$.
The correlations in the path weight we measure via the memory
kernel $\Gamma_\tau(t, t^\prime)$ are thus influenced by our choice of the time instance
at which we start observing the particle trajectory. In that sense,
the system ``remembers its past'' even prior to the initial time point $t=0$,
because of the finite correlation time in the active fluctuations. 

We finally remark that our general expression \eqref{eq:pp} with \eqref{eq:Gamma}
reduces to known results in the three limiting cases
$\Da \to 0$ (passive particle),
$\ta \to 0$ (white active noise), and
$D \to 0$ (no thermal bath);
details of the calculations can be found in
Appendix~\ref{app:limits}.
Without active fluctuations ($D_a \to 0$), we trivially recover
the standard Onsager-Machlup expression
\eqref{eq:ppDa0} for passive Brownian motion.
In the white noise limit for $\VEC{\eta}(t)$ ($\ta \to 0$),
the equation of motion \eqref{eq:LE} involves two independent Gaussian white noise
processes with vanishing means and variances $2D$ and $2\Da$, respectively.
Their sum is itself a white noise source with zero mean, but variance $2(D+\Da)$.
Accordingly, as $\ta \to 0$, we obtain from \eqref{eq:pp}
an Onsager-Machlup path weight of the form \eqref{eq:ppDa0},
but with the diffusion coefficient $D$ being replaced by $D + \Da$.
In the third limiting case of vanishing thermal fluctuations,
we are left with a pure colored noise path weight \cite{McKane:1990pia,Hanggi:1989pis},
\begin{multline}
\label{eq:ppD0}
\pp{\traj{\VEC{x}}|\VEC{x}_0}_{D \to 0}
\propto \exp \Bigg\{ - \int_0^\tau \!\! \mathrm{d}t \,
	\frac{\left[ \left( 1 + \ta \partial_t \right) \left( \dot{\VEC{x}}_t - \VEC{v}_t \right) \right]^2}{4\Da}
\\
	\mbox{}- \frac{\VEC{\nabla} \cdot \VEC{v}_t}{2} \Bigg\} \;
	p_{\mathrm{s}}\!\left( \frac{\dot{\VEC{x}}_0 - \VEC{v}_0}{\sqrt{2\Da}} \right)
\, ,
\end{multline}
where $p_{\text{s}}$ denotes the steady-state distribution
\eqref{eq:pseta} of the colored noise.

\subsection{Fluctuation theorem}
\label{sec:FT}
With the explicit form \eqref{eq:pp} of the path probability, we can now
derive an exact
expression for the log-ratio of path probabilities
which relates time-forward and \nobreakdash-backward paths.
Following exactly the same line of reasoning as described in Sec.~\ref{sec:ppDeltaSDa0}
for the case of a passive Brownian particle,
we consider the ratio of probabilities for observing a specific trajectory
$\traj{\VEC{x}}$ and its time-reversed twin $\trajR{\VEC{x}}$,
created under a time-reversed protocol $\VEC{v}(\VEC{x},\tau-t)$ \cite{Note3}.
With the definition \eqref{eq:xrev} of the time-reversed trajectory,
we can express its probability in terms of the forward path.
Using the property $\Gamma_\tau(\tau-t,\tau-t')=\Gamma_\tau(t,t')$ of
the memory kernel [see Eq.~\eqref{eq:Gamma}],
we then obtain the path probability ratio
\begin{subequations}
\label{eq:ppratioDeltaSigma}
\begin{equation}
\label{eq:ppratio}
\frac{\ppR{\trajR{\VEC{x}}|\tilde{\VEC{x}}_0}}{\pp{\traj{\VEC{x}}|\VEC{x}_0}}
= e^{-\Delta \Sigma[\Traj{\VEC{x}}]/\kB}
\, ,
\end{equation}
with
\begin{equation}
\label{eq:DeltaSigma}
\Delta\Sigma[\Traj{\VEC{x}}]
= \frac{1}{T} \int_0^\tau \!\! \mathrm{d}t \int_0^\tau \!\! \mathrm{d}t^\prime \;
	\dot{\VEC{x}}_t\TRANS \VEC{f}_{t^\prime} \left[ \delta(t-t^\prime) - \frac{\Da}{D} \Gamma_\tau(t, t^\prime) \right]
\, .
\end{equation}
\end{subequations}
As a stochastic integral, this is to be understood in the
Stratonovich sense.
Note that the procedure of time inversion does not involve the active fluctuations
$\VEC{\eta}(t)$ in any way, and thus does not require any
assumptions on their properties under time reversal. In fact, the probability density
$\pp{\traj{\VEC{x}}|\VEC{x}_0}$ for the trajectories of the Brownian particle
is a result of integrating over \textit{all} possible realizations $\VEC{\eta}(t)$
of the active fluctuations, containing any pair of conceivable time-forward and time-backward twins
with their natural weight of occurrence [see also Eq.~\eqref{eq:ppxeta}].
Hence, for the probability ratio of particle trajectories \eqref{eq:pp}
the behavior of the
active fluctuations $\VEC{\eta}(t)$ under time inversion is irrelevant.

As described in Sec.~\ref{sec:ppDeltaSDa0}
a relation like \eqref{eq:ppratio} based on path probability
ratios entails an integral
fluctuation theorem, if the entropy production in
the system
$\Delta S_{\mathrm{sys}}(\VEC{x}_0,\VEC{x}_\tau) = -\kB \ln p(\VEC{x}_\tau,\tau) + \kB \ln p(\VEC{x}_0,0)$
[see Eq.~\eqref{eq:DeltaSsysDa0}]
is taken into consideration.
Explicitly, we find
\begin{equation}
\label{eq:ppfullratio}
\frac{\ppR{\TrajR{\VEC{x}}}}{\pp{\Traj{\VEC{x}}}}
= e^{-\left( \Delta \Sigma[\Traj{\VEC{x}}]+ \Delta S_{\mathrm{sys}}(\VEC{x}_0,\VEC{x}_\tau) \right)/\kB}
\, ,
\end{equation}
and therefore
\begin{equation}
\label{eq:IFT}
\left\langle e^{-(\Delta\Sigma + \Delta S_{\mathrm{sys}})/\kB} \right\rangle = 1
\, .
\end{equation}
By Jensen's inequality we conclude
\begin{equation}
\label{eq:2ndLaw}
\left\langle \Delta\Sigma + \Delta S_{\mathrm{sys}} \right\rangle \geq 0
\, .
\end{equation}
The fluctuation theorem \eqref{eq:IFT} and the bound \eqref{eq:2ndLaw}
are a direct consequence of the definition of
$\Delta\Sigma+\Delta S_{\mathrm{sys}}$
as the logarithm of a path probability ratio \eqref{eq:ppfullratio}.
Obviously, formally equivalent relations are valid
for any quantity defined in this way.
Here, these relations \eqref{eq:ppfullratio}-\eqref{eq:2ndLaw}
become useful because our analysis equips us
with an explicit expression for $\Delta\Sigma$ as a functional
of the particle trajectory in forward time.

In \eqref{eq:2ndLaw} equality is achieved
if and only if the dynamics is symmetric under time reversal.
The setting considered here, however, is generally not symmetric under time-reversal
because of our choice of particle position and active
fluctuations being independent initially.
The approach to a correlated (stationary) state is irreversible, such
that we inevitably pick up transient contributions to $\Delta\Sigma+\Delta S_{\mathrm{sys}}$,
which are strictly positive on average, even if the external forces $\VEC{f}$
are time independent and conservative.

For the same reason, namely the build-up of correlations between
the particle trajectory $\VEC{x}(t)$ and the active fluctuation $\VEC{\eta}(t)$,
the quantity $\Delta\Sigma$ is non-additive,
i.e.\ $\Delta\Sigma[\Traj{\VEC{x}}] \neq
\Delta\Sigma[\{\VEC{x}(t)\}_{t=0}^{t'}] + \Delta\Sigma[\{\VEC{x}(t)\}_{t=t'}^{\tau}]$
in general, for any intermediate time $0 < t^\prime < \tau$.
In order to establish such an additivity property, we would have to
take into account the correlations between $\VEC{x}(t)$ and $\VEC{\eta}(t)$
in the initial distribution $p_{t^\prime}(\VEC{\eta}_{t^\prime}|\VEC{x}_{t^\prime})$
for the path integral~\eqref{eq:ppintegrateEta} of the second part of the trajectory
$\VEC{x}(t)$ with $t^\prime < t < \tau$
that have been build up until the time point $t^\prime$.
However, it is not possible to specify this conditional distribution in the general situation
considered here allowing for arbitrary forces and force protocols $\VEC{f}=\VEC{f}(\VEC{x},t)$,
see \eqref{eq:f} and the remark below \eqref{eq:pseta}.

The path probability ratio \eqref{eq:ppratio},
relating the time reversibility of trajectories to the extensive
quantity $\Delta\Sigma[\Traj{\VEC{x}}]$, together with its corresponding integral fluctuation
theorem \eqref{eq:IFT}, constitute our second main result.
The interpretation of $\Delta\Sigma$ and its fluctuation theorems
in physical terms is, however, not as
straightforward as in the case of a passive Brownian particle,
in which we could identify the logarithm of the path probability ratio as
the heat dissipated into the thermal environment divided by the bath temperature
\cite{Seifert:2005epa,Seifert:2012stf}, see Eq.~\eqref{eq:deltaSDa0}.
Although we cannot make such a simple identification for the
Brownian particle driven by active fluctuations, it is still possible
to connect $\Delta\Sigma$ to physically meaningful quantities,
as we will show in the following.

As a first step in this direction, we observe a formal similarity between
$\Delta S$, given in Eq.~\eqref{eq:DeltaSDa0}, and $\Delta\Sigma$, given in Eq.~\eqref{eq:DeltaSigma}.
Defining the non-local ``force''
\begin{multline}
\VEC{\varphi}_\tau[\Traj{\VEC{x}},t]
= \VEC{f}(\VEC{x}(t),t)
\\
\mbox{}- \frac{\Da}{D}\int_0^\tau \mathrm{d}t' \, \VEC{f}(\VEC{x}(t'),t') \, \Gamma_\tau(t,t')
\, ,
\end{multline}
we can bring \eqref{eq:DeltaSigma} into the form
\begin{align}
\label{eq:DeltaSigmaphi}
\Delta \Sigma[\Traj{\VEC{x}}] 
& = \frac{1}{T} \int_0^\tau \mathrm{d}t \, \dot{\VEC{x}}(t)\cdot\VEC{\varphi}_\tau[\Traj{\VEC{x}},t]
\nonumber\\
& = \frac{1}{T} \int_0^\tau \mathrm{d}\VEC{x}(t) \cdot \VEC{\varphi}_\tau[\Traj{\VEC{x}},t]
\, ,
\end{align}
in obvious analogy to \eqref{eq:DeltaSDa0},
but with the essential difference that the ``force'' $\VEC{\varphi}_\tau[\Traj{\VEC{x}},t]$
at time $0 \leq t \leq \tau$ depends not only on $\VEC{x}(t)$ but rather
on the full trajectory $\Traj{\VEC{x}}=\{\VEC{x}(t)\}_{t=0}^\tau$ via the memory
term $\int_0^\tau \mathrm{d}t' \, \VEC{f}(\VEC{x}(t'),t') \Gamma_\tau(t,t')$.
Similar ``memory forces'' have already been found to affect irreversibility by
contributing to dissipation in Langevin systems with colored noise,
which does not obey the fluctuation-dissipation relation \cite{puglisi2009irreversible}.

From \eqref{eq:DeltaSigmaphi} we can read off a production rate for $\Sigma$,
\begin{subequations}
\begin{equation}
\label{eq:SigmaRateInstant}
\sigma_\tau(t) = \dot{\VEC{x}}(t) \cdot \frac{1}{T} \, \VEC{\varphi}_\tau[\Traj{\VEC{x}},t]
\, ,
\end{equation}
or
\begin{equation}
\label{eq:deltaSigma}
\delta\Sigma_\tau(t) = \mathrm{d}{\VEC{x}}(t) \cdot \frac{1}{T} \, \VEC{\varphi}_\tau[\Traj{\VEC{x}},t]
\, .
\end{equation}
\end{subequations}
Even though this expression for the $\Sigma$-production is
analogous to the actual entropy productions in the environment as
identified in \eqref{eq:deltaS+} or \eqref{eq:deltaS-}
[see also \eqref{eq:deltaSDa0}],
and even contains a term 
$\left[\mathrm{d}{\VEC{x}}(t) \cdot \VEC{f}(\VEC{x}(t),t)\right]/T$,
which quantifies dissipation due to the external force $\VEC{f}$,
its physical meaning beyond formally defined ``memory forces'' is
unclear.
In the following, we will argue that
a valid interpretation is provided by the
mutual information between the active fluctuations $\VEC{\eta}(t)$
and the particle trajectory $\VEC{x}(t)$.

\section{Mutual information}
\label{sec:MI}
The path-wise mutual information \cite{parrondo2015thermodynamics}
between the particle trajectory $\Traj{\VEC{x}}$ (starting at $\VEC{x}_0$)
and a realization $\Traj{\VEC{\eta}}$
of the active fluctuations (with $\VEC{\eta}(0)=\VEC{\eta}_0$) is given as
\begin{equation}
\label{eq:Idef}
\I[\Traj{\VEC{x}},\Traj{\VEC{\eta}}]
= \ln \frac{\pp{\Traj{\VEC{x}},\Traj{\VEC{\eta}}}}
		     {\pp{\Traj{\VEC{x}}}\,\pp{\Traj{\VEC{\eta}}}}
= \ln \frac{\pp{\Traj{\VEC{x}}|\Traj{\VEC{\eta}}}}
		     {\pp{\Traj{\VEC{x}}}}
\, ,
\end{equation}
where
$\pp{\Traj{\VEC{x}},\Traj{\VEC{\eta}}}=\pp{\traj{\VEC{x}},\traj{\VEC{\eta}}|\VEC{x}_0,\VEC{\eta}_0}\,p(\VEC{x}_0,\VEC{\eta}_0)$,
$\pp{\Traj{\VEC{x}}}=\pp{\traj{\VEC{x}}|\VEC{x}_0}\,p(\VEC{x}_0)$,
and $\pp{\Traj{\VEC{\eta}}}=\pp{\traj{\VEC{\eta}}|\VEC{\eta}_0}\,p(\VEC{\eta}_0)$
include the initial densities
$p(\VEC{x}_0,\VEC{\eta}_0)$, $p(\VEC{x}_0)=\int\mathrm{d}\VEC{\eta}_0\,p(\VEC{x}_0,\VEC{\eta}_0)$,
and $p(\VEC{\eta}_0)=\int\mathrm{d}\VEC{x}_0\,p(\VEC{x}_0,\VEC{\eta}_0)$
of the Brownian particle and the active fluctuations
(see also the discussion in Sec.~\ref{sec:pp}).
This path-wise mutual information quantifies the reduction in uncertainty about the path $\Traj{\VEC{x}}$ when
we know the realization $\Traj{\VEC{\eta}}$ and vice versa, and can therefore,
loosely speaking, be seen as a measure of the correlations between $\Traj{\VEC{x}}$
and $\Traj{\VEC{\eta}}$. Note that it can become
negative if
$\pp{\Traj{\VEC{x}}|\Traj{\VEC{\eta}}}<\pp{\Traj{\VEC{x}}}$,
while its average is always positive.

Likewise, the path-wise mutual information between the time-reversed trajectory
$\TrajR{\VEC{x}}$ from \eqref{eq:xrev} and a suitably chosen time-reversed
realization $\TrajR{\VEC{\eta}}$ of the active fluctuations is
\begin{equation}
\IR[\TrajR{\VEC{x}},\TrajR{\VEC{\eta}}]
= \ln \frac{\ppR{\TrajR{\VEC{x}}|\TrajR{\VEC{\eta}}}}
		   {\ppR{\TrajR{\VEC{x}}}}
\, .
\end{equation}
For their difference
\begin{subequations}
\label{eq:DeltaI}
\begin{equation}
\label{eq:DeltaIdef}
\Delta\I[\Traj{\VEC{x}},\Traj{\VEC{\eta}}]
= \I[\Traj{\VEC{x}},\Traj{\VEC{\eta}}]
	- \IR[\TrajR{\VEC{x}},\TrajR{\VEC{\eta}}]
\end{equation}
we thus find
\begin{equation}
\label{eq:DeltaIpp}
\Delta\I[\Traj{\VEC{x}},\Traj{\VEC{\eta}}] =
  \ln \frac{\ppR{\TrajR{\VEC{x}}}}{\pp{\Traj{\VEC{x}}}}
- \ln \frac{\ppR{\TrajR{\VEC{x}}|\TrajR{\VEC{\eta}}}}
		   {\pp{\Traj{\VEC{x}}|\Traj{\VEC{\eta}}}}
\, .
\end{equation}
\end{subequations}
This expression represents the path-wise
mutual information difference
between a combined forward process $(\Traj{\VEC{x}},\Traj{\VEC{\eta}})$
and its backward twin.
If $\Delta\I[\Traj{\VEC{x}},\Traj{\VEC{\eta}}]$ is positive,
the path-wise mutual information
$\I[\Traj{\VEC{x}},\Traj{\VEC{\eta}}]$ along the combined forward path is larger
than $\IR[\TrajR{\VEC{x}},\TrajR{\VEC{\eta}}]$ for the time-reversed path
[see \eqref{eq:DeltaIdef}],
implying that correlations between the particle trajectory $\Traj{\VEC{x}}$
and the active fluctuation $\Traj{\VEC{\eta}}$
are stronger in the time-forward direction. Intuitively, we may thus 
say that they are more likely
to occur together than their time-reversed twins, making the combined
forward process $(\Traj{\VEC{x}},\Traj{\VEC{\eta}})$ the more ``natural'' one
of the two processes in terms of path-wise mutual information.

To make the connection to irreversibility more rigorous,
we rewrite in \eqref{eq:DeltaIpp} the path-wise mutual information
difference $\Delta\I[\Traj{\VEC{x}},\Traj{\VEC{\eta}}]$ explicitly
as path probability ratios between time-forward and time-backward processes.
We now see
that $\Delta\I$, via the term $-\ln\!\left(\ppR{\TrajR{\VEC{x}}}/\pp{\Traj{\VEC{x}}}\right)$,
is directly related to the
irreversibility measure $\Delta\Sigma$
for the actively driven Brownian particle
and its change in system entropy $\Delta S_{\mathrm{sys}}$.
The additional log-ratio
$\ln\!\left( \ppR{\TrajR{\VEC{x}}|\TrajR{\VEC{\eta}}}/\pp{\Traj{\VEC{x}}|\Traj{\VEC{\eta}}} \right)$
involves the probability
of the forward path $\Traj{\VEC{x}}$
being generated by the specific realization $\Traj{\VEC{\eta}}$
of the active fluctuations for which
we measure the mutual information content with $\Traj{\VEC{x}}$ in $\Delta\I$,
and, likewise, the path probability
of the time-reversed twin being generated by the time-reversed fluctuation.
We can rewrite this term by
splitting off
the contributions from the initial densities,
\begin{equation}
\label{eq:pp|eta}
\ln \frac{\ppR{\TrajR{\VEC{x}}|\TrajR{\VEC{\eta}}}}
		 {\pp{\Traj{\VEC{x}}|\Traj{\VEC{\eta}}}}
=
\ln \frac{\ppR{\trajR{\VEC{x}}|\tilde{\VEC{x}}_0,\TrajR{\VEC{\eta}}}}{\pp{\traj{\VEC{x}}|\VEC{x}_0,\Traj{\VEC{\eta}}}}
+ \ln \frac{p(\tilde{\VEC{x}}_0|\TrajR{\VEC{\eta}})}{p(\VEC{x}_0|\VEC{\eta}_0)}
\, .
\end{equation}
We here keep the possibility that the initial particle position is
conditioned on the initial state of the active fluctuations,
$p(\VEC{x}_0|\Traj{\VEC{\eta}})=p(\VEC{x}_0|\VEC{\eta}_0)$,
which is more general than the situation
$p(\VEC{x}_0|\Traj{\VEC{\eta}})=p(\VEC{x}_0)$ for which
we calculated $\Delta\Sigma$ in Section~\ref{sec:pp}
[see also the discussion around Eq.\ \eqref{eq:pseta}].
According to \eqref{eq:xrev}, the time-reversed initial position is given
by the final point of the forward path, $\tilde{\VEC{x}}_0=\VEC{x}_\tau$.
It therefore depends on the complete history of the active
fluctuations,
which is captured equivalently by $\TrajR{\VEC{\eta}}$ or $\Traj{\VEC{\eta}}$,
for any reasonable choice of the time-reversed fluctuations in terms
of the forward realization [see also Eq.\ \eqref{eq:etarev} below].
With $p(\tilde{\VEC{x}}_0|\TrajR{\VEC{\eta}})=p(\VEC{x}_\tau|\Traj{\VEC{\eta}})$
we can identify the boundary term in \eqref{eq:pp|eta} as the
change in system entropy
\begin{equation}
\label{eq:DeltaSsys|eta}
\Delta S_{\mathrm{sys}}^{|\Traj{\VEC{\eta}}}[\Traj{\VEC{x}}|\Traj{\VEC{\eta}}] =
-\kB \ln p(\VEC{x}_\tau|\Traj{\VEC{\eta}}) + \kB \ln p(\VEC{x}_0|\VEC{\eta}_0)
\, ,
\end{equation}
which occurs along the trajectory $\Traj{\VEC{x}}$ under
the specific realization $\Traj{\VEC{\eta}}$ of the active fluctuations
(we explicitly indicate the conditioning on the realization of the active
fluctuation by adding `$|\Traj{\VEC{\eta}}\,$' as a superscript).

To quantify the first term in \eqref{eq:pp|eta},
we exploit that a prescribed active fluctuation
$\Traj{\VEC{\eta}}=\{\VEC{\eta}(t)\}_{t=0}^\tau$
acts like an additional driving ``force'' in \eqref{eq:LE}.
Without needing to know any further details about $\VEC{\eta}(t)$
(e.g., of how it is generated),
we can thus derive the conditioned forward path probability
$\pp{\traj{\VEC{x}}|\VEC{x}_0,\Traj{\VEC{\eta}}}$
directly from \eqref{eq:LE} as a standard
Onsager-Machlup path integral,
\begin{multline}
\pp{\traj{\VEC{x}}|\VEC{x}_0,\Traj{\VEC{\eta}}}
\propto
\exp \left\{ -\int_{0}^\tau \!\!\!\mathrm{d}t
\, \frac{(\dot{\VEC{x}}_t-\VEC{v}_t-\sqrt{2\Da}\,\VEC{\eta}_t)^2}{4D} \right.
\\
\left.
\mbox{}- \frac{\VEC{\nabla} \cdot \VEC{v}_t}{2} \right\}
\, .
\end{multline}
In order to evaluate the time-reversed counterpart, we have to specify the behavior of
the active fluctuation signal $\VEC{\eta}(t)$ under inversion of the direction of time.
There are essentially two choices,
\begin{equation}
\label{eq:etarev}
\tilde{\VEC{\eta}}_\pm(t) = \pm \VEC{\eta}(\tau-t)
\, ,
\end{equation}
corresponding to $\VEC{\eta}(t)$ being an even or odd process.
While both these options a priori appear to be equally valid, they
are connected to different interpretations of the active fluctuations.
We argue below (beginning of Section \ref{sec:MIAB}) that for
passive tracer particles which are suspended in an active bath
the active fluctuations $\VEC{\eta}(t)$ bear the character of an
external force and therefore should be considered even
under time-reversal.
In contrast, for an active particle with self-propulsion
(see beginning of Section \ref{sec:MISP})
the active fluctuations $\VEC{\eta}(t)$ are internal to the particle and
thus should change sign when time is reversed (odd under
time-reversal), otherwise
the particle trajectory would not be reversed under
inversion of the direction of time
in a setup in which only 
self-propulsion is acting (no thermal noise and
no external forces)
\cite{Note4}.

\subsection{Active bath}
\label{sec:MIAB}
If we interpret $\VEC{\eta}(t)$ as fluctuations coming from an active bath,
they are external to the particle and as such are similar to externally
applied forces. The question of irreversibility under a
prescribed realization $\Traj{\VEC{\eta}}$ is then
related to the question of how likely the backward particle trajectory
$\tilde{\VEC{x}}(t)$ [see \eqref{eq:xrev}]
is to be observed when exactly the same forces act on the particle at
identical positions during the forward and backward motion.
Hence, this case is described by active fluctuations which
are even under time-reversal, i.e.\ by the plus-sign in \eqref{eq:etarev}.

Using $\tilde{\VEC{\eta}}_+(t) = +\VEC{\eta}(\tau-t)$ as the time-reversed fluctuation,
we find
\begin{subequations}
\label{eq:DeltaS+}
\begin{equation}
\label{eq:DeltaS+pp}
\ln \frac{\ppR{\trajR{\VEC{x}}|\tilde{\VEC{x}}_0,\TrajR{\VEC{\eta}}_+}}
		   {\pp{\traj{\VEC{x}}|\VEC{x}_0,\Traj{\VEC{\eta}}}}
= - \Delta S_+[\Traj{\VEC{x}}|\Traj{\VEC{\eta}}]/\kB
\end{equation}
for the path probability ratio, where
$\Delta S_+[\Traj{\VEC{x}}|\Traj{\VEC{\eta}}]$ is given
by the entropy production in the thermal environment from \eqref{eq:deltaS+},
\begin{align}
\Delta S_+[\Traj{\VEC{x}}|\Traj{\VEC{\eta}}]
& = \int_0^\tau \mathrm{d}t \; \delta S_+(t)
\nonumber \\
& = \frac{1}{T} \int_0^\tau \mathrm{d}\VEC{x}(t) \cdot 
\left( \VEC{f}(\VEC{x}(t),t)+\sqrt{2\Da}\gamma\,\VEC{\eta}(t) \right)
\, .
\end{align}
\end{subequations}
Together with \eqref{eq:DeltaSsys|eta}, we thus obtain
the total entropy production
along the particle trajectory $\Traj{\VEC{x}}$ which is generated by a given realization
$\Traj{\VEC{\eta}}$ of the active fluctuations, assumed to be even
under time-inversion.
It is defined in analogy to \eqref{eq:DeltaSsysDa0} as
\begin{equation}
\label{eq:DeltaStot+}
\Delta S_{\mathrm{tot}}^+[\Traj{\VEC{x}}|\Traj{\VEC{\eta}}] = 
\Delta S_+[\Traj{\VEC{x}}|\Traj{\VEC{\eta}}] + \Delta S_{\mathrm{sys}}^{|\Traj{\VEC{\eta}}}[\Traj{\VEC{x}}|\Traj{\VEC{\eta}}]
\, .
\end{equation}
By construction
[see \eqref{eq:DeltaSsys|eta}, \eqref{eq:DeltaS+pp} and also \eqref{eq:pp|eta}],
$\Delta S_{\mathrm{tot}}^+[\Traj{\VEC{x}}|\Traj{\VEC{\eta}}]$
is identical to the log-ratio
$-\kB \ln\left( \ppR{\TrajR{\VEC{x}}|\TrajR{\VEC{\eta}}_+}/\pp{\Traj{\VEC{x}}|\Traj{\VEC{\eta}}} \right)$.

Combining this result with \eqref{eq:pp|eta}, \eqref{eq:ppfullratio}
and \eqref{eq:DeltaI}, we then infer
that $\Delta\Sigma[\VEC{x}]+\Delta S_{\mathrm{sys}}(\VEC{x}_0,\VEC{x}_\tau)$
is a combination of the conditioned entropy production and 
the difference in mutual information between
time-forward and time-backward processes,
\begin{equation}
\label{eq:DeltaSigma+}
\Delta\Sigma[\Traj{\VEC{x}}]+\Delta S_{\mathrm{sys}}(\VEC{x}_0,\VEC{x}_\tau)
= \Delta S_{\mathrm{tot}}^+[\Traj{\VEC{x}}|\Traj{\VEC{\eta}}]
- \kB \, \Delta\I_+[\Traj{\VEC{x}},\Traj{\VEC{\eta}}]
\, ,
\end{equation}
where $\Delta\I_+[\Traj{\VEC{x}},\Traj{\VEC{\eta}}]$ is given by \eqref{eq:DeltaI}
with $\TrajR{\VEC{\eta}}=\TrajR{\VEC{\eta}}_+=\{\tilde{\VEC{\eta}}_+(t)\}_{t=0}^\tau$.
Note that the individual terms on the right-hand side both depend on
the active noise $\Traj{\VEC{\eta}}$
with a dependency that is compensated exactly, because the
left-hand side is independent of $\Traj{\VEC{\eta}}$.

As an immediate consequence of \eqref{eq:IFT}, the total conditioned entropy production
together with the mutual information difference
fulfill the integral fluctuation theorem
\begin{equation}
\label{eq:IFT+}
\left\langle e^{-(\Delta S_{\mathrm{tot}}^+/\kB - \Delta\I_+)} \right\rangle_{\Traj{\VEC{x}}} = 1
\, ,
\end{equation}
and the second-law like relation
\begin{subequations}
\label{eq:2nd+}
\begin{equation}
\label{eq:2nd+eta}
\left\langle \Delta S_{\mathrm{tot}}^+[\Traj{\VEC{x}}|\Traj{\VEC{\eta}}] \right\rangle_{\Traj{\VEC{x}}}
	- \kB \left\langle \Delta\I_+[\Traj{\VEC{x}},\Traj{\VEC{\eta}}] \right\rangle_{\Traj{\VEC{x}}} \geq 0
\, .
\end{equation}
These results
\eqref{eq:IFT+} and \eqref{eq:2nd+} are valid for any realization $\Traj{\VEC{\eta}}$
of the active fluctuations,
because the original average $\langle\cdot\rangle$ in \eqref{eq:IFT}
is just over the distribution of all
particle trajectories $\Traj{\VEC{x}}$.
We accentuate this fact here by the subscript $\Traj{\VEC{x}}$
at the averages $\langle\cdot\rangle_{\Traj{\VEC{x}}}$
to distinguish them from the standard meaning of the brackets
$\langle\cdot\rangle$ as an average over \emph{all} stochastic quantities present.
Since the terms in \eqref{eq:2nd+eta} taken together are independent of
$\Traj{\VEC{\eta}}$ [see Eq.~\eqref{eq:DeltaSigma+}],
we can even average over the joint distribution
$\pp{\Traj{\VEC{x}},\Traj{\VEC{\eta}}}$ to obtain a further
second-law like relation
\begin{equation}
\label{eq:2nd+av}
\left\langle \Delta S_{\mathrm{tot}}^+ \right\rangle
	- \kB \left\langle \Delta\I_+ \right\rangle \geq 0
\, .
\end{equation}
\end{subequations}

We emphasize that
\eqref{eq:DeltaSigma+}, \eqref{eq:IFT+} and \eqref{eq:2nd+},
being a direct consequence of the structure of the equation of motion \eqref{eq:LE},
are completely independent of the specific
model behind the active fluctuations.
Moreover, they are formally similar to the fluctuation theorem with information exchange
derived by Sagawa and Ueda \cite{sagawa2012fluctuation} for a completely different physical setup.
Sagawa and Ueda considered the difference between initial and final correlations
of a system of interest with an ``information reservoir'', which provides correlations as a resource
of entropy changes.

For the Ornstein-Uhlenbeck representation \eqref{eq:OUP}, we can use our
result \eqref{eq:DeltaSigma} for $\Delta\Sigma[\Traj{\VEC{x}}]$
to derive an explicit expression for the mutual information difference,
\begin{align}
&\hspace*{-5ex}
\Delta\I_+[\Traj{\VEC{x}},\Traj{\VEC{\eta}}]
+ \left[ \ln\frac{p(\VEC{x}_\tau|\Traj{\VEC{\eta}})}{p(\VEC{x}_\tau)} - \ln\frac{p(\VEC{x}_0|\VEC{\eta}_0)}{p(\VEC{x}_0)} \right]
\nonumber \\
& = \frac{1}{\kB T} \left[
	\sqrt{2\Da} \int_0^\tau \!\! \mathrm{d}t \; \dot{\VEC{x}}_t\TRANS \VEC{\eta}_{t}
	\right.
\nonumber \\
& \qquad \mbox{}+ \left.
	 \frac{\Da}{D} \int_0^\tau \!\! \mathrm{d}t \int_0^\tau \!\! \mathrm{d}t^\prime \;
	\dot{\VEC{x}}_t\TRANS \VEC{f}_{t^\prime} \Gamma_\tau(t, t^\prime) \right]
\nonumber \\
& = \frac{1}{\kB T} \bigg[
	\Delta A_+[\Traj{\VEC{x}},\Traj{\VEC{\eta}}]
\nonumber \\
& \qquad \mbox{}+
	\frac{\Da}{D} \int_0^\tau \!\! \mathrm{d}t \int_0^\tau \!\! \mathrm{d}t^\prime \;
	\dot{\VEC{x}}_t\TRANS \VEC{f}_{t^\prime} \Gamma_\tau(t, t^\prime) \bigg]
\, .
\label{eq:DeltaI+}
\end{align}
where in the last equality we have used
$\Delta A_+[\Traj{\VEC{x}},\Traj{\VEC{\eta}}]=\int_0^\tau \!\! \mathrm{d}t \, \delta A_+(t)$
with the ``heat'' $\delta A_+(t)$ exchanged with the active bath as identified in
\eqref{eq:deltaA+}.

\subsection{Self-propulsion}
\label{sec:MISP}
If the active fluctuations represent an effective model for a self-propulsion
mechanism of the particle, we choose $\VEC{\eta}(t)$ to be odd under time-reversal
for a proper assessment of irreversibility, i.e.\ we choose the minus-sign in
\eqref{eq:etarev}.
Otherwise we would relate irreversibility to the likelihood that thermal
fluctuations make the particle
move a certain path backward \textit{against} its internal self-propulsion,
rather than to the probability that a self-propelled particle traces out a given
trajectory backward in time driven by the same propulsion forces
\cite{Note4}.

Calculating the path probability ratio 
${\pp{\trajR{\VEC{x}}|\tilde{\VEC{x}}_0,\TrajR{\VEC{\eta}}}}/
{\pp{\traj{\VEC{x}}|\VEC{x}_0,\Traj{\VEC{\eta}}}}$
for odd active fluctuations,
$\tilde{\VEC{\eta}}(t)=\tilde{\VEC{\eta}}_-(t)=-\VEC{\eta}(\tau-t)$,
we obtain
\begin{subequations}
\label{eq:DeltaS-}
\begin{equation}
\label{eq:DeltaS-pp}
\ln \frac{\ppR{\trajR{\VEC{x}}|\tilde{\VEC{x}}_0,\TrajR{\VEC{\eta}}_-}}
		 {\pp{\traj{\VEC{x}}|\VEC{x}_0,\Traj{\VEC{\eta}}}}
= -\Delta S_-[\Traj{\VEC{x}}|\Traj{\VEC{\eta}}]/\kB
\, .
\end{equation}
The quantity $\Delta S_-[\Traj{\VEC{x}}|\Traj{\VEC{\eta}}]$
is given by
\begin{align}
\Delta S_-[\Traj{\VEC{x}}|\Traj{\VEC{\eta}}]
& = \int_0^\tau \!\mathrm{d}t \; \delta S_-(t) 
\nonumber \\
& = \frac{1}{T} \int_0^\tau \!
\left( \mathrm{d}\VEC{x}(t) - \sqrt{2\Da}\,\VEC{\eta}(t)\,\mathrm{d}t \right)
\cdot \VEC{f}(\VEC{x}(t),t)
\, .
\end{align}
\end{subequations}
It thus represents
the entropy production in the environment along the trajectory $\Traj{\VEC{x}}$,
because $\delta S_-(t)$ results from the heat $-\delta Q_-(t)$ dissipated into the thermal bath
during a displacement of the active self-propelled particle,
which deviates from the
``force-free'' path that would be carved out by the self-propulsion drive alone;
see also \eqref{eq:deltaS-} and the discussion above Eq.~\eqref{eq:deltaQ-}.
In analogy to \eqref{eq:DeltaSsysDa0},
we finally combine $\Delta S_-[\Traj{\VEC{x}}|\Traj{\VEC{\eta}}]$
with \eqref{eq:DeltaSsys|eta} to define the total entropy production
under a given realization $\Traj{\VEC{\eta}}$ of the active propulsion as
\begin{equation}
\label{eq:DeltaStot-}
\Delta S_{\mathrm{tot}}^-[\Traj{\VEC{x}}|\Traj{\VEC{\eta}}] = 
\Delta S_-[\Traj{\VEC{x}}|\Traj{\VEC{\eta}}] + \Delta S_{\mathrm{sys}}^{|\Traj{\VEC{\eta}}}[\Traj{\VEC{x}}|\Traj{\VEC{\eta}}]
\, .
\end{equation}
Note that with \eqref{eq:DeltaSsys|eta} and \eqref{eq:DeltaS-pp}
[see also \eqref{eq:pp|eta}],
we can write this total entropy production as the log-ratio
$-\kB \ln\left( \ppR{\TrajR{\VEC{x}}|\TrajR{\VEC{\eta}}_-}/\pp{\Traj{\VEC{x}}|\Traj{\VEC{\eta}}} \right)$
of path weights.

Hence, we find again that $\Delta\Sigma[\Traj{\VEC{x}}]+\Delta S_{\mathrm{sys}}(\VEC{x}_0,\VEC{x}_\tau)$ consists of
conditional total entropy production and mutual
information difference
[compare Eqs.\ \eqref{eq:DeltaI}, \eqref{eq:ppfullratio} and \eqref{eq:DeltaStot-}],
\begin{equation}
\label{eq:DeltaSigma-}
\Delta\Sigma[\Traj{\VEC{x}}]+\Delta S_{\mathrm{sys}}(\VEC{x}_0,\VEC{x}_\tau)
= \Delta S_{\mathrm{tot}}^-[\Traj{\VEC{x}}|\Traj{\VEC{\eta}}]
- \kB \, \Delta\I_-[\Traj{\VEC{x}},\Traj{\VEC{\eta}}]
\, .
\end{equation}
This is exactly the same interpretation of
$\Delta\Sigma[\Traj{\VEC{x}}]+\Delta S_{\mathrm{sys}}(\VEC{x}_0,\VEC{x}_\tau)$
as we found earlier for the case of an active bath [c.f.\ \eqref{eq:DeltaSigma+}],
but with different expressions for the entropy production and
the mutual information difference.
The mutual information in the backward process is now measured
with respect to the sign-inverted active fluctuation process
$\tilde{\VEC{\eta}}_-(t)=-\VEC{\eta}(\tau-t)$, i.e.\
$\Delta\I_-[\Traj{\VEC{x}},\Traj{\VEC{\eta}}]$ in \eqref{eq:DeltaSigma-}
is given by \eqref{eq:DeltaI} with
$\TrajR{\VEC{\eta}}=\TrajR{\VEC{\eta}}_-=\{\tilde{\VEC{\eta}}_-(t)\}_{t=0}^\tau$.
As in the case of an active bath \eqref{eq:DeltaSigma+},
the dependence of the individual terms $\Delta S_{\mathrm{tot}}^-$ and $\Delta\I_-$
in \eqref{eq:DeltaSigma-} 
on the active forcing $\Traj{\VEC{\eta}}$ cancel
exactly to result in the $\Traj{\VEC{\eta}}$-independent expression 
$\Delta\Sigma[\Traj{\VEC{x}}]+\Delta S_{\mathrm{sys}}(\VEC{x}_0,\VEC{x}_\tau)$.

From \eqref{eq:IFT}
we immediately obtain the integral fluctuation theorem
\begin{equation}
\label{eq:IFT-}
\left\langle e^{-(\Delta S_{\mathrm{tot}}^-/\kB - \Delta\I_-)} \right\rangle_{\Traj{\VEC{x}}} = 1
\, ,
\end{equation}
and the second-law like relation
\begin{subequations}
\label{eq:2nd-}
\begin{equation}
\label{eq:2nd-eta}
\left\langle \Delta S_{\mathrm{tot}}^-[\Traj{\VEC{x}}|\Traj{\VEC{\eta}}] \right\rangle_{\Traj{\VEC{x}}}
	- \kB \left\langle \Delta\I_-[\Traj{\VEC{x}},\Traj{\VEC{\eta}}] \right\rangle_{\Traj{\VEC{x}}} \geq 0
\, 
\end{equation}
for active self-propulsion.Like in \eqref{eq:IFT+} we use the subscript $\Traj{\VEC{x}}$ to explicitly indicate
that the averages are over the distribution $\pp{\Traj{\VEC{x}}}$ alone and that the resulting
quantities are still functionals of the active fluctuations and
are valid for any realization $\Traj{\VEC{\eta}}$ of the self-propelling forces.
After extending to the average over $\pp{\Traj{\VEC{x}},\Traj{\VEC{\eta}}}$
to include these active fluctuations we end up with the bound
\begin{equation}
\label{eq:2nd-av}
\left\langle \Delta S_{\mathrm{tot}}^- \right\rangle - \kB \left\langle \Delta\I_- \right\rangle \geq 0
\, .
\end{equation}
\end{subequations}

Again, these findings 
\eqref{eq:DeltaSigma-}, \eqref{eq:IFT-} and \eqref{eq:2nd-}
are independent of the specific model for the active fluctuations,
because they are a direct consequence of the particle's equation of
motion \eqref{eq:LE}.
Moreover, they are formally similar to the Sagawa-Ueda fluctuation theorem
with information \cite{sagawa2012fluctuation}.
If the active propulsion $\VEC{\eta}(t)$ is modeled by
\eqref{eq:OUP}, we can write the mutual information difference
in the explicit form
\begin{align}
&\hspace*{-5ex}
\Delta\I_-[\Traj{\VEC{x}},\Traj{\VEC{\eta}}]
+ \left[ \ln\frac{p(\VEC{x}_\tau|\Traj{\VEC{\eta}})}{p(\VEC{x}_\tau)} - \ln\frac{p(\VEC{x}_0|\VEC{\eta}_0)}{p(\VEC{x}_0)} \right]
\nonumber \\
& = \frac{1}{\kB T} \left[
	-\sqrt{2\Da} \int_0^\tau \!\! \mathrm{d}t \; \VEC{\eta}_{t}\TRANS \VEC{f}_t
	\right.
\nonumber \\
& \qquad \mbox{}+ \left.
	 \frac{\Da}{D} \int_0^\tau \!\! \mathrm{d}t \int_0^\tau \!\! \mathrm{d}t^\prime \;
	\dot{\VEC{x}}_t\TRANS \VEC{f}_{t^\prime} \Gamma_\tau(t, t^\prime) \right]
\nonumber \\
& = \frac{1}{\kB T} \bigg[
	\Delta A_-[\Traj{\VEC{x}},\Traj{\VEC{\eta}}]
\nonumber \\
& \qquad \mbox{}+
	\frac{\Da}{D} \int_0^\tau \!\! \mathrm{d}t \int_0^\tau \!\! \mathrm{d}t^\prime \;
	\dot{\VEC{x}}_t\TRANS \VEC{f}_{t^\prime} \Gamma_\tau(t, t^\prime) \bigg]
\, ,
\label{eq:DeltaI-}
\end{align}
by using \eqref{eq:DeltaSigma} and by defining
$\Delta A_-[\Traj{\VEC{x}},\Traj{\VEC{\eta}}]=\int_0^\tau \!\! \mathrm{d}t \, \delta A_-(t)$
[see \eqref{eq:deltaA-}],
as the total ``heat'' transferred from the active propulsion to the thermal bath along the trajectory
$\Traj{\VEC{x}}$.

\subsection{Relation to information theory}
The second-law like relations \eqref{eq:2nd+} and \eqref{eq:2nd-}
involve the averages 
$\left\langle \Delta\I_+ \right\rangle$ and
$\left\langle \Delta\I_- \right\rangle$
of the mutual information difference.
As we will show in the following, these averages are closely
related to central concepts in information theory,
entailing additional bounds
and an interpretation
in terms of hypothesis testing.

Due to the unbiased Gaussian character of the active fluctuations,
the probabilities for
observing a time-forward and its time-reversed realization
are the same, i.e.\
$\ppR{\TrajR{\VEC{\eta}}}=\pp{\Traj{\VEC{\eta}}}$.
We can therefore rewrite \eqref{eq:DeltaIpp} as
\begin{equation}
\Delta\I_\pm[\Traj{\VEC{x}},\Traj{\VEC{\eta}}] =
  \ln \frac{\ppR{\TrajR{\VEC{x}}}}{\pp{\Traj{\VEC{x}}}}
- \ln \frac{\ppR{\TrajR{\VEC{x}},\TrajR{\VEC{\eta}}_\pm}}
		   {\pp{\Traj{\VEC{x}},\Traj{\VEC{\eta}}}}
\, ,
\end{equation}
where we have inserted our two options
$\TrajR{\VEC{\eta}}=\TrajR{\VEC{\eta}}_\pm$
from \eqref{eq:etarev} for the time-reversed
active fluctuation.
Performing the average over $(\Traj{\VEC{x}},\Traj{\VEC{\eta}})$
with density
$\pp{\Traj{\VEC{x}},\Traj{\VEC{\eta}}}$,
we then find
\begin{equation}
\label{eq:KL}
\left\langle \Delta\I_\pm[\Traj{\VEC{x}},\Traj{\VEC{\eta}}] \right\rangle =
D_{\mathrm{KL}}(\pp{\Traj{\VEC{x}},\Traj{\VEC{\eta}}}||\ppR{\TrajR{\VEC{x}},\TrajR{\VEC{\eta}}_\pm}) -
D_{\mathrm{KL}}(\pp{\Traj{\VEC{x}}}||\ppR{\TrajR{\VEC{x}}})
\, .
\end{equation}
Here, we use the definition
\begin{equation}
D_{\mathrm{KL}}(\pp{\Traj{\VEC{y}}}||\ppR{\TrajR{\VEC{y}}})
= \int \D\Traj{\VEC{y}} \; \pp{\Traj{\VEC{y}}} \, \ln\frac{\pp{\Traj{\VEC{y}}}}{\ppR{\TrajR{\VEC{y}}}}
\end{equation}
of the Kullback-Leibler divergence
between a probability density $\pp{\Traj{\VEC{y}}}$
for a process $\Traj{\VEC{y}}$ and another density $\ppR{\TrajR{\VEC{y}}}$
for a second process $\TrajR{\VEC{y}}$;
in our case, the two processes are related by time-reversal.
The Kullback-Leibler divergence is a standard concept in information theory to
measure how distinct two probability densities are. It is non-negative
and equals zero if and only if the two probabilities are identical
\cite{kullback1951information,lexa2004useful}.

The result~(\ref{eq:KL}) shows that
the average mutual information difference
$\left\langle \Delta\I_\pm[\Traj{\VEC{x}},\Traj{\VEC{\eta}}] \right\rangle$
is the difference between the Kullback-Leibler divergence of 
the particle trajectory $\Traj{\VEC{x}}$ relative to its time-reversed twin $\TrajR{\VEC{x}}$
and the Kullback-Leibler divergence of the combined path
$(\Traj{\VEC{x}},\Traj{\VEC{\eta}})$,
relative to the time-reversed realization $(\TrajR{\VEC{x}},\TrajR{\VEC{\eta}}_\pm)$.
We can therefore interpret it
to measure how much harder it is to discriminate between time-forward and
time-backward realizations if only the particle trajectory is known, rather
than the full dynamics including the active noise realization $\Traj{\VEC{\eta}}$.
Since $\Traj{\VEC{x}}$ can be seen as a ``coarse-graining projection''
of $(\Traj{\VEC{x}},\Traj{\VEC{\eta}})$, we expect the discrimination to become harder,
i.e.\ $D_{\mathrm{KL}}(\pp{\Traj{\VEC{x}}}||\ppR{\TrajR{\VEC{x}}})$ to become smaller
compared to
$D_{\mathrm{KL}}(\pp{\Traj{\VEC{x}},\Traj{\VEC{\eta}}}||\ppR{\TrajR{\VEC{x}},\TrajR{\VEC{\eta}}_\pm})$
(see \cite{bo2017multiple} for a similar discussion).
This intuition is corroborated by the so-called data processing inequality
\cite{kullback1951information,lexa2004useful}, which, applied to our situation, proves
\begin{equation}
\label{eq:DKL>}
D_{\mathrm{KL}}(\pp{\Traj{\VEC{x}},\Traj{\VEC{\eta}}}||\ppR{\TrajR{\VEC{x}},\TrajR{\VEC{\eta}}_\pm})
\geq
D_{\mathrm{KL}}(\pp{\Traj{\VEC{x}}}||\ppR{\TrajR{\VEC{x}}})
\, .
\end{equation}
As a direct consequence, we find from \eqref{eq:KL} the bound
\begin{equation}
\label{eq:DeltaIbound}
\left\langle \Delta\I_\pm[\Traj{\VEC{x}},\Traj{\VEC{\eta}}] \right\rangle \geq 0
\end{equation}
on the average mutual information difference.
Exploiting \eqref{eq:2nd+av} and \eqref{eq:2nd-av}, respectively,
we infer that $\langle\Delta S_{\mathrm{tot}}^\pm\rangle \geq 0$.
This follows also from the fact that the $\Delta S_{\mathrm{tot}}^\pm$
are given as log-ratios of
path probabilities,
which are conditioned on $\Traj{\VEC{\eta}}$
[see the discussion around \eqref{eq:DeltaS+} and \eqref{eq:DeltaS-}].
Accordingly, they obey an integral fluctuation theorem when averaged over
the conditioned density $\pp{\Traj{\VEC{x}}|\Traj{\VEC{\eta}}}$, 
but then also when averaged over
the full density $\pp{\Traj{\VEC{x}},\Traj{\VEC{\eta}}}$.

More interestingly, we can also use \eqref{eq:DeltaIbound} to equip
the second-law like relations 
for the total irreversibility measure $\Delta\Sigma+\Delta S_{\mathrm{sys}}$
with an upper bound. Taking the average of
\eqref{eq:DeltaSigma+} and \eqref{eq:DeltaSigma-}
over $(\Traj{\VEC{x}},\Traj{\VEC{\eta}})$, 
the non-negativity \eqref{eq:DeltaIbound} of $\left\langle \Delta\I_\pm \right\rangle$ implies
\begin{equation}
\label{eq:2ndX}
\left\langle \Delta S_{\mathrm{tot}}^\pm[\Traj{\VEC{x}}|\Traj{\VEC{\eta}}] \right\rangle
\geq
\left\langle \Delta\Sigma[\Traj{\VEC{x}}]+\Delta S_{\mathrm{sys}}(\VEC{x}_0,\VEC{x}_\tau) \right\rangle
\geq 0
\, .
\end{equation}
The total average irreversibility of a particle trajectory, measured as
$\langle\Delta\Sigma+\Delta S_{\mathrm{sys}}\rangle$,
is thus always smaller than (or equal to)
the mean entropy change
$\int\D\Traj{\VEC{x}}\,\pp{\Traj{\VEC{x}}|\Traj{\VEC{\eta}}}\Delta S_{\mathrm{tot}}^\pm[\Traj{\VEC{x}}|\Traj{\VEC{\eta}}]$,
which would occur for a given realization $\Traj{\VEC{\eta}}$ of the active fluctuations
if treated as a known (and measurable) external force contributing to the dissipation in the thermal environment,
averaged over the distribution $\pp{\Traj{\VEC{\eta}}}$ of all possible active fluctuations, i.e.\
$\int\D\Traj{\VEC{\eta}}\,\pp{\Traj{\VEC{\eta}}}\int\D\Traj{\VEC{x}}\,\pp{\Traj{\VEC{x}}|\Traj{\VEC{\eta}}}
\Delta S_{\mathrm{tot}}^\pm[\Traj{\VEC{x}}|\Traj{\VEC{\eta}}]=
\int\D\Traj{\VEC{\eta}}\int\D\Traj{\VEC{x}}\,\pp{\Traj{\VEC{x}},\Traj{\VEC{\eta}}}
\Delta S_{\mathrm{tot}}^\pm[\Traj{\VEC{x}}|\Traj{\VEC{\eta}}]=
\langle \Delta S_{\mathrm{tot}}^\pm[\Traj{\VEC{x}}|\Traj{\VEC{\eta}}] \rangle$.
For this bound to be valid it does not 
matter whether these forcings come from an active bath
and thus are considered even under time-reversal ('$+$' sign),
or from active self-propulsion, which is odd under time-reversal ('$-$' sign).
The difference between
$\langle \Delta S_{\mathrm{tot}}^\pm[\Traj{\VEC{x}}|\Traj{\VEC{\eta}}] \rangle$
and $\langle\Delta\Sigma+\Delta S_{\mathrm{sys}}\rangle$ is compensated by the build-up of mutual information.

We point out that the upper bound for the total average irreversibility in
\eqref{eq:2ndX} is stronger than the usual ``coarse-graining inequality''
obtained when integrating out ``hidden'' degrees of freedom (see, e.g., \cite{roldan2010estimating}).
In the present notation such an inequality would read
$\left\langle \Delta S_{\mathrm{tot}}[\Traj{\VEC{x}},\Traj{\VEC{\eta}}] \right\rangle
\geq
\left\langle \Delta\Sigma[\Traj{\VEC{x}}]+\Delta S_{\mathrm{sys}}(\VEC{x}_0,\VEC{x}_\tau) \right\rangle$.
Here, $\left\langle \Delta S_{\mathrm{tot}}[\Traj{\VEC{x}},\Traj{\VEC{\eta}}] \right\rangle$
would denote the total average entropy production in the thermal environment of the combined
processes $(\Traj{\VEC{x}},\Traj{\VEC{\eta}})$,
a quantity which is not well-defined, however, because we
cannot quantify the entropy production associated with the auxiliary process $\Traj{\VEC{\eta}}$.

Finally, we remark that the inequality \eqref{eq:DKL>}
has an interesting interpretation in the context of hypothesis testing
\cite{CoverThomas:ElementsOfInformationTheory,lexa2004useful},
when estimating the direction of the arrow of time in the system \cite{Jarzynski:2011eai,roldan2015decision}.
It states that
the probability of misclassifying a specific path as having been generated
by a forward dynamics when, in reality, it was generated by a backward one
decreases faster with the number of observations if
more detailed information on the dynamics of the system
is available by additionally
monitoring the realizations $\Traj{\VEC{\eta}}$
of the active fluctuations.
\subsection{Discussion}
The fluctuation theorems and second-law like relations
\eqref{eq:IFT+}, \eqref{eq:2nd+} and \eqref{eq:IFT-}, \eqref{eq:2nd-}
for an active bath and for active self-propulsion, respectively,
and their interpretation in terms of mutual information
differences are our third main result,
see also \eqref{eq:2ndX}.
For both cases, we find that the total irreversibility measure
$\Delta\Sigma[\Traj{\VEC{x}}]+\Delta S_{\mathrm{sys}}(\VEC{x}_0,\VEC{x}_\tau)$
consists of two contributions:
First, the ``usual'' entropy production $\Delta S_{\mathrm{tot}}^\pm$,
which we would measure for the motion of the Brownian particle under
a given realization of the active fluctuations
(as if the active fluctuations were just some additional known ``external'' driving force),
and, second, the difference in mutual information $\Delta\I_\pm$
accumulated between the particle trajectory and the active non-equilibrium environment
along the forward versus the backward path.
We note that even though the individual contributions in
$\Delta S_{\mathrm{tot}}^\pm - \kB \Delta\I_\pm$ depend on the specific realization
$\Traj{\VEC{\eta}}$ of the active fluctuations, their sum does not
[see \eqref{eq:DeltaSigma+} and \eqref{eq:DeltaSigma-}], i.e.\
entropy production and change in mutual information always compensate their
dependency on the realization of the active fluctuations.
However, if we want to have access to these individual contributions
$\Delta S_{\mathrm{tot}}^\pm$ and $\Delta\I_\pm$
directly, we would have to measure the active fluctuations $\Traj{\VEC{\eta}}$,
i.e.\ the direction and magnitude of the forces representing the activity
in the system.
In general, this may be a challenging task in typical
experiments with active Brownian particles,
as one would have to separate thermal from active fluctuations.
We can imagine, however, that at least partial information about
$\Traj{\VEC{\eta}}$ could be obtained experimentally.
For example, tracking the orientation of an active, self-propelled particle
(e.g., a bacterium with a flagellum)
it may be possible to infer the direction of $\VEC{\eta}$.
Moreover, we could think of an experimental setup in which the
active fluctuations $\Traj{\VEC{\eta}}$ are realized artificially by an
external colored noise source, like in \cite{Mestres:2014ron};
then all relevant quantities may be accessible.

The entropy productions $\Delta S_{\mathrm{tot}}^\pm$
for an active bath and for active self-propulsion, respectively,
as obtained from irreversibility arguments
in \eqref{eq:DeltaS+} and \eqref{eq:DeltaS-}
are consistent with the energetics derived 
in Secs.\ \ref{sec:energeticsAB} and \ref{sec:energeticsSP}
[compare with \eqref{eq:deltaQ+} and \eqref{eq:deltaQ-}].
They are thus
directly related to the heat dissipated into the \emph{thermal}
part of the environment.
Hence, the appearance of the path-wise mutual information difference
in the fluctuation theorem is a
consequence of the \emph{active}
fluctuations being present as a non-equilibrium bath in
addition to the usual thermal bath. Indeed, we can easily see
from \eqref{eq:DeltaI+} and \eqref{eq:DeltaI-} that
$\Delta\I_\pm$ vanish identically in the absence of active
fluctuations, $\Da=0$.
Accordingly, $\Delta S_\pm$ reduce to the standard entropy production
in the thermal environment in this limit
[see \eqref{eq:DeltaS+}, \eqref{eq:DeltaS-}, and \eqref{eq:DeltaSDa0}].

We emphasize that the path-wise mutual information quantifies how
the active fluctuations contribute to the
irreversibility of the particle trajectory, but does not
capture the unavoidable dissipation connected with maintaining
the active nature of the non-equilibrium environment itself.
In fact, our effective description \eqref{eq:OUP} of the active fluctuations
as a time-correlated Gaussian noise source does not have any
knowledge about the microscopic details generating this noise,
so that it can obviously not assess the associated dissipative processes.
On the one hand, we may therefore suspect that the appearance of
the path-wise mutual information is a consequence of this ``coarse-grained'' description
of the active environment, and might be replaced by a ``finer'' measure
once all the details are known \cite{gaspard2017communication,pietzonka2017entropy,speck2018active}.
On the other hand, we may argue that these microscopic details
behind the active fluctuations are irrelevant if we are only interested
in characterizing the irreversibility of the particle trajectory
and related thermodynamic-like properties of the system emerging
on this level of coarse-graining.
Then, knowledge of the statistical properties of the active fluctuations,
as provided by \eqref{eq:OUP}, is sufficient,
and the path-wise mutual information between particle trajectory and active
noise realization emerges as a natural and adequate
irreversibility measure.
This situation may be comparable to the one for entropy production in
a thermal bath: we quantify entropy production solely from the statistical
properties of the thermal noise without having to
rely on a detailed microscopic description of the bath.

\section{Example: Harmonic potential}
\label{sec:HP}
\begin{figure}[ht]
	\includegraphics[scale=1]{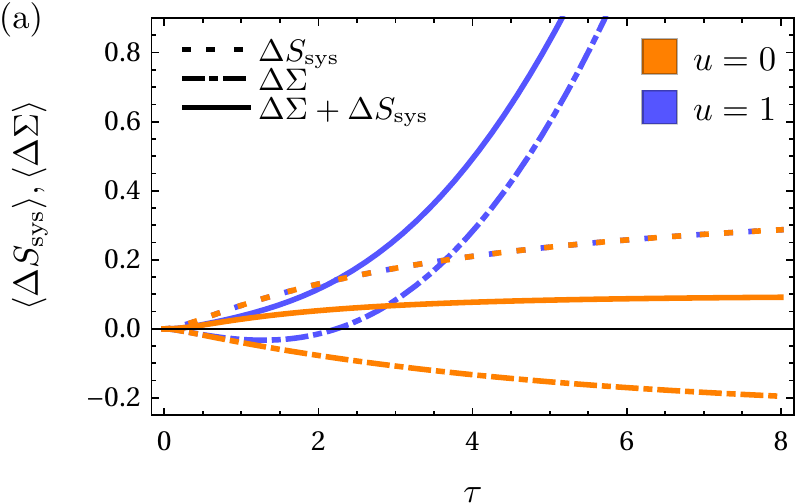}
	\\[2ex]
	\includegraphics[scale=1]{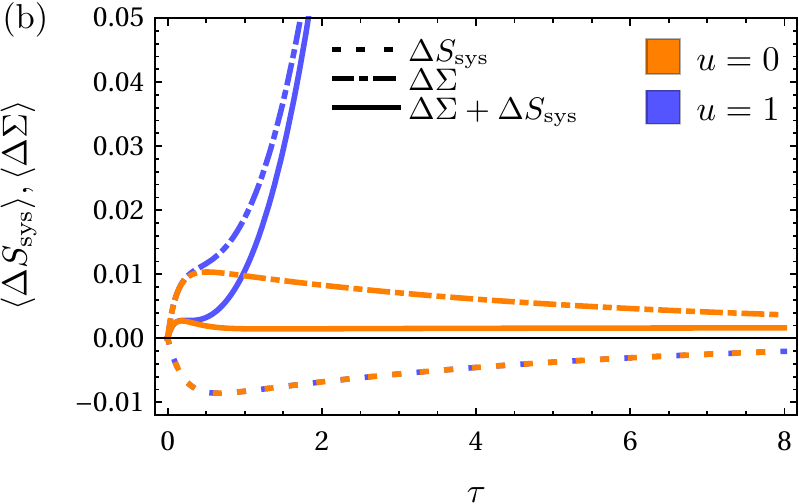}
\caption{Average contributions to irreversibility $\langle \Delta\Sigma \rangle$ and
average change in system entropy $\langle \Delta S_{\mathrm{sys}} \rangle$ as a function
of the observation time $\tau$. We compare the cases of a particle trapped in a static
($u=0$, orange lines) or moving ($u=1$, blue lines) harmonic potential having
an initial distribution which is Gaussian with variance
(a) $c_0^2 = \kB T / k = D\gamma/k$ and
(b) $c_0^2=\frac{1}{k}\left[ \kB T + \frac{\gamma\Da}{1+(k\ta/\gamma)} \right]$.
Parameter values are $k=0.1$, $\gamma=1$, $\ta = 0.2$, $D = \Da = 0.2$.
Note that the blue and orange lines for the change in system entropy
are on top of each other in both figures,
as $\langle \Delta S_{\mathrm{sys}} \rangle$ does not depend on $u$.
}
\label{fig:2}
\end{figure}
To illustrate our results, we consider the example of a one-dimensional
Brownian particle trapped in a harmonic potential
$U(x, t) = \frac{k}{2}(x-u t)^2$, whose center is either held at fixed position, $u=0$,
or else is displaced at constant velocity $u \neq 0$.
Such a setup can be readily implemented in experiment with state-of-the-art
optical tweezers, and in fact has been used to study various aspects of
stochastic thermodynamics and active matter, for instance in
\cite{Argun:2016nbs,chaki2018entropy,van2003stationary,van2004extended,schmiedl2007optimal,engel2009asymptotics}.

The Langevin equation of motion~\eqref{eq:LE} for this specific setup is linear,
\begin{equation}
\label{eq:LEex}
	\dot{x}(t) = -\frac{k}{\gamma} \left[ x(t) - u t \right] + \sqrt{2 D} \, \xi(t) + \sqrt{2 \Da} \, \eta(t) 
\, .
\end{equation}
We can therefore solve the associated Fokker-Planck equation analytically
\cite{VanKampen:1976teo,Gardiner:HandbookOfStochasticMethods,Risken:TheFokkerPlanckEquation}
to obtain the propagator in closed form.
From that we can explicitly compute the averages of the irreversibility measure $\Delta\Sigma$,
the mutual information difference $\Delta\I_\pm$, and the change in
system entropy $\Delta S_{\mathrm{sys}}$, for a Gaussian initial distribution $p_0(x_0)$
of particle positions with zero mean and variance $c_0^2$.
Although the calculations are straightforward
(we present some details in Appendix~\ref{app:HP}),
the resulting formulae for $\langle \Delta\Sigma \rangle$, $\langle \Delta S_{\mathrm{sys}} \rangle$
and $\langle \Delta\I_\pm \rangle$ are lengthy and bulky, so that we discuss them mostly
in graphical form (see Figs.~\ref{fig:2} and \ref{fig:4}) and give explicit
expressions only in some limiting cases
[see Eqs.~\eqref{eq:HPDeltaSsys}, \eqref{eq:HPDeltaSigma_u0thermalVar}, \eqref{eq:HPDeltaSigma_u0steadyVar} below].
For the initial variance $c_0^2$ two specific
cases are of particular physical relevance. First, $c_0^2=\kB T/k$,
corresponding to a Gaussian distribution that would be created as an equilibrium
state by thermal fluctuations only. Starting from this distribution,
the time evolution of
our irreversibility measures includes the transient relaxation
from the thermal state towards the steady state which develops due to the
presence of the active fluctuations.
Second, $c_0^2=\frac{1}{k}\left[ \kB T + \frac{\gamma\Da}{1+(k\ta/\gamma)} \right]$,
corresponding to a Gaussian distribution with a variance which is
exactly the same as the one the particle distribution has
when particle and active fluctuations are in their joint steady state.
In that case, the particle distributions at the beginning and end of the
process are identical, so that any contributions to the irreversibility
measure $\Delta\Sigma$ that are not associated with the displacement $u$ of
the trap are solely due to the build-up of correlations between
particle position and active degrees of freedom.
In the following, we will focus on the first alternative,
and briefly come back to the second alternative at the end of Sec.~\ref{sec:HPSigma}
and in Sec.~\ref{sec:HPIFT}.
\subsection{Irreversibility}
\label{sec:HPSigma}
In Fig.~\ref{fig:2} we compare $\langle \Delta\Sigma \rangle$, $\langle \Delta S_{\mathrm{sys}} \rangle$ and
$\langle \Delta\Sigma + \Delta S_{\mathrm{sys}} \rangle$ for $u=0$ and $u=1$ as a function of the duration
$\tau$ of the trajectories.
We find that the \emph{average} change in system entropy $\langle \Delta S_{\mathrm{sys}} \rangle$
is independent of the driving velocity $u$, because it only depends on the variance of the initial
and final Gaussian distributions, but not on their centers.
For long trajectories, i.e.\ large $\tau$, it approaches the constant value
\begin{equation}
\label{eq:HPDeltaSsys}
\lim_{\tau\to\infty}\langle\Delta S_{\mathrm{sys}}\rangle
	= \frac{\kB}{2} \ln \left[ 1+\frac{\Da}{D}\;\frac{1}{1+\frac{k\ta}{\gamma}} \right]
\, .
\end{equation}

In contrast, the irreversibility measures $\langle \Delta\Sigma \rangle$
for the static trap $u=0$ and
the moving trap $u=1$ are similar only during a short
transient phase, but then become qualitatively different.
In the static trap, $\langle \Delta\Sigma \rangle$ becomes constant
at large times $\tau$,
\begin{widetext}
\begin{equation}
\label{eq:HPDeltaSigma_u0thermalVar}
\lim_{\tau\to\infty}\langle\Delta\Sigma\rangle |_{u=0}
=  -\frac{\kB}{2} \; 
	\frac{\left( 1-\frac{k\ta}{\gamma} \right) + 2\frac{D}{\Da}
   		  \left( 1+\frac{k\ta}{\gamma} \right) \left( 1+\sqrt{1+\frac{\Da}{D}} \right)}
   		 {\left( 1+\sqrt{1+\frac{\Da}{D}} \right)^2
		  \left( \frac{k\ta}{\gamma} + \sqrt{1+\frac{\Da}{D}} \right)^2}
	\, \left( \frac{\Da}{D} \right)^2
\, ,
\end{equation}
\end{widetext}
in accordance with the system reaching a current-less,
equilibrium-like steady state.
In the moving trap, the particle is transported continuously by permanent
dissipation of energy, so that $\langle \Delta\Sigma \rangle$ increases with
the length $\tau$ of the trajectories.
For large $\tau$, the growth rate is given by the ensemble average of
the time-averaged production rate
$\sigma = \lim_{\tau\to\infty} \frac{1}{\tau} \int_0^\tau \d t \; \sigma_\tau(t)$,
where $\sigma_\tau(t)$ is defined in~\eqref{eq:SigmaRateInstant}.
Explicitly, we obtain
\begin{equation}
\label{eq:SigmaRate}
\langle \sigma \rangle
		= \lim_{\tau\to\infty} \frac{\langle  \Delta\Sigma \rangle}{\tau}
		= \kB \frac{u^2}{D + \Da}
\, .
\end{equation}
We thus see that the total ``irreversibility production'' 
$\langle \Delta\Sigma + \Delta S_{\mathrm{sys}} \rangle$ in the moving trap
grows extensively with the observation time as it would without active fluctuations, too.
The growth rate, however, is \emph{reduced} by the additional presence of active
fluctuations, as the combined environment can be interpreted to have
a higher ``effective temperature''.
Remarkably, the average growth rate \eqref{eq:SigmaRate} is independent of the
relevant system time-scales, i.e.\ the correlation time
$\ta$ of active fluctuations and the relaxation time $\gamma/k$
in the harmonic potential.

In the static trap $u=0$,
we find that no such extensive growth
occurs for $\langle \Delta\Sigma + \Delta S_{\mathrm{sys}} \rangle$.
Hence, the system reaches a steady state, which
appears equilibrium-like
from the viewpoint of particle trajectories
without access to the microscopic processes generating the active driving.
Nevertheless, we observe
$\lim_{\tau\to\infty} \langle \Delta\Sigma + \Delta S_{\mathrm{sys}} \rangle > 0$
regardless of the initial distribution $p_0(x_0)$.
For the case of a Gaussian with ``thermal variance''
$c_0^2=\kB T/k$ the relevant results
are given in Fig.~\ref{fig:2} and
Eqs.~\eqref{eq:HPDeltaSsys}, \eqref{eq:HPDeltaSigma_u0thermalVar}.
Both,
$\lim_{\tau\to\infty}\langle\Delta S_{\mathrm{sys}}\rangle$ and
$\lim_{\tau\to\infty}\langle\Delta\Sigma\rangle$,
depend on just two dimensionless parameters:
the ratio of the two noise intensities $\frac{\Da}{D}$ and the ratio of the
two system timescales $\frac{k\tau_a}{\gamma}$.
The limits of large and small correlation time and noise amplitudes
can be easily computed and present no difficulties.
As an interesting example, we consider the case
of vanishing correlation time of the active fluctuations,
\begin{multline}
\lim_{\tau_a\to 0}\lim_{\tau\to\infty}\langle\Delta\Sigma+\Delta S_{\mathrm{sys}}\rangle|_{u=0}
\\
= \frac{\kB}{2}\ln\left[1+\frac{\Da}{D} \right] -\frac{\kB}{2}\frac{\Da}{D+\Da}
\, .
\end{multline}
This is exactly the entropy, which would be produced by a
thermal white-noise process with diffusion constant $D+\Da$
relaxing from an initial Gaussian distribution with variance
$D\gamma/k=\kB T/k$ to its equilibrium state,
a Gaussian with variance $(D+\Da)\gamma/k$.

It is quite obvious that for a variance $c_0^2=\kB T/k$ of
the initial Gaussian distribution, there must be some change
in system entropy and a build-up of irreversibility,
because the Gaussian particle distribution approached at long times has a larger
variance $\frac{1}{k}\left[ \kB T + \frac{\gamma\Da}{1+k\ta/\gamma} \right]$
[see Eq.~\eqref{eq:Cinfty}].
But even if we start with the initial variance
$c_0^2=\frac{1}{k}\left[ \kB T + \frac{\gamma\Da}{1+(k\ta/\gamma)} \right]$,
we find a positive
$\langle\Delta\Sigma\rangle$ while reaching the steady state at large $\tau$,
\begin{equation}
\label{eq:HPDeltaSigma_u0steadyVar}
\lim_{\tau\to\infty}\langle\Delta\Sigma\rangle|_{u=0}
= \frac{\kB \, \frac{k\ta}{\gamma} \left( \frac{\Da}{D} \right)^2}
	   {\left( 1+\sqrt{1+\frac{\Da}{D}} \right)^2  \left( \frac{k\ta}{\gamma}+\sqrt{1+\frac{\Da}{D}} \right)^2}
\, .
\end{equation}
The origin of this positive contribution is our choice of \emph{independent}
initial conditions in form of a product density $p_0(x_0,\eta_0)=p_0(x_0)p_{\mathrm{s}}(\eta_0)$
[see also the discussion of Eq.~\eqref{eq:pseta}].
Hence, during the initial transient there must be
an irreversible build-up of correlations between
the particle and the active fluctuations,
becoming manifest in a non-zero total $\langle \Delta\Sigma + \Delta S_{\mathrm{sys}} \rangle$.
It turns out that the associated change in $\langle \Delta\Sigma + \Delta S_{\mathrm{sys}} \rangle$
is non-monotonic [see Fig.~\ref{fig:2}(b)],
indicating that the variance of the particle distribution
departs over some (transient) time period, even though
the initial variance $c_0^2=\frac{1}{k}\left[ \kB T + \frac{\gamma\Da}{1+(k\ta/\gamma)} \right]$
is identical to the final one.
In order to obtain a vanishing average
$\langle \Delta\Sigma + \Delta S_{\mathrm{sys}} \rangle$
we would have to start from a joint stationary state $p_{\mathrm{s}}(x_0,\eta_0)$
of particle positions and active fluctuations instead of a factorized one.
It can even be shown that forward and backward paths
are equally likely in that case (see \cite{activeGyrator}),
implying that $\Delta\Sigma[\Traj{x}] + \Delta S_{\mathrm{sys}}(x_0, x_\tau) = 0$
holds already on the level of individual trajectories.
It is unsolved, however, if the latter property is generic for stationary
states of trapped active particles (in the absence of symmetry-breaking forces),
or if it is specific to the harmonic
trap, to the Ornstein-Uhlenbeck realization
of the active fluctuations, or to dimensionality one.

We emphasize again that these correlations are the very reason
that the irreversibility measure $\Delta\Sigma$ is non-additive
[see also the discussion below Eq.~\eqref{eq:2ndLaw}].
The curves in Fig.~\ref{fig:2} apply only to trajectories
evolving over the complete time-interval $[0,\tau]$.
We cannot split a trajectory at an intermediate time,
calculate the individual $\Delta\Sigma$
for the two parts of the trajectory from \eqref{eq:DeltaSigma},
and then add them up to obtain $\Delta\Sigma$
for the full time interval, because the ``initial'' state for the
second part will inevitably depend on correlations between particle and active fluctuations
accumulated during the first part.
Such correlations are not taken into account in
\eqref{eq:DeltaSigma} which is based on
the assumption of an initial product state.

\subsection{Fluctuation theorem}
\label{sec:HPIFT}
\begin{figure}[ht]
	\includegraphics[scale=1]{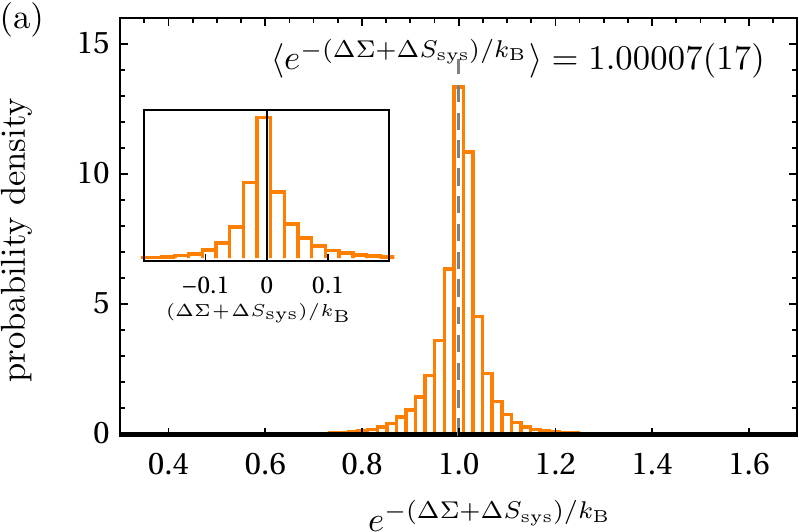}
	\\[2ex]
	\includegraphics[scale=1]{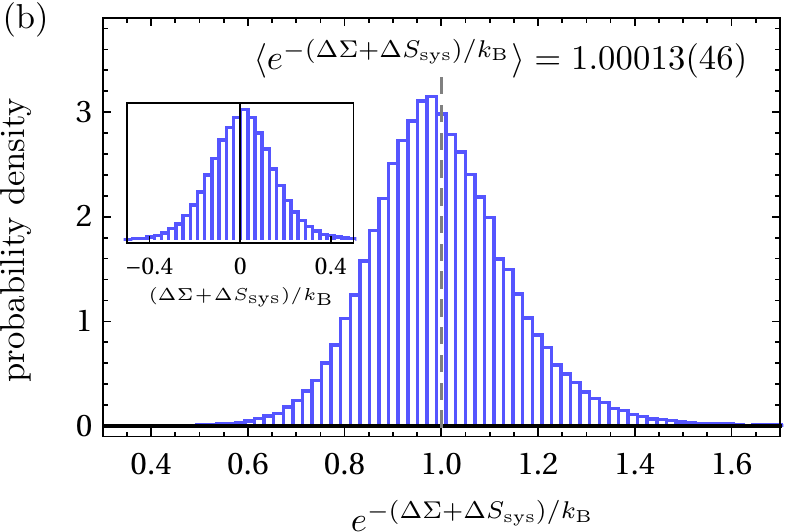}
\caption{Probability densities of the path probability ratio~\eqref{eq:ppfullratio}
for a harmonically trapped particle (a) in a static trap, $u=0$, and (b) in a moving trap, $u=1$.
The densities are obtained from numerically simulating $10^5$ sample trajectories of duration $\tau = 1$.
Parameters are $k=0.1$, $\gamma=1$, $\ta = 0.2$, $D = \Da = 0.2$,
$c_0^2=\frac{\gamma}{k} \left[D + \frac{\Da}{1+k \ta / \gamma} \right]$.
The insets show the corresponding densities of
$\Delta S_{\mathrm{tot}}=\Delta\Sigma+\Delta S_{\mathrm{sys}}$.
The average values 
$\langle \Delta\Sigma \rangle$, $\langle \Delta S_{\mathrm{sys}} \rangle$ and
$\langle \Delta\Sigma + \Delta S_{\mathrm{sys}} \rangle$ as
obtained from the numerical simulations are consistent with the corresponding
theoretical predictions:
(a) $u=0$, simulation:
$\langle \Delta\Sigma \rangle = 0.010(4)$,
$\langle \Delta S_{\mathrm{sys}} \rangle = -0.0090(13)$,
$\langle \Delta\Sigma + \Delta S_{\mathrm{sys}} \rangle = 0.00139(17)$.
$u=0$, theory:
$\langle \Delta\Sigma \rangle = 0.00975$,
$\langle \Delta S_{\mathrm{sys}} \rangle = -0.00828$,
$\langle \Delta\Sigma + \Delta S_{\mathrm{sys}} \rangle = 0.00148$.
(b) $u=1$, simulation:
$\langle \Delta\Sigma \rangle = 0.017(4)$,
$\langle \Delta S_{\mathrm{sys}} \rangle = -0.0073(13)$,
$\langle \Delta\Sigma + \Delta S_{\mathrm{sys}} \rangle = 0.0104(5)$.
$u=1$, theory:
$\langle \Delta\Sigma \rangle = 0.01883$,
$\langle \Delta S_{\mathrm{sys}} \rangle = -0.00828$,
$\langle \Delta\Sigma + \Delta S_{\mathrm{sys}} \rangle = 0.01056$. 
}
\label{fig:3}
\end{figure}
To illustrate the integral fluctuation theorem~\eqref{eq:IFT}
satisfied by $\Delta\Sigma + \Delta S_{\mathrm{sys}}$,
we show in Fig.~\ref{fig:3}
probability densities for the path probability ratio~\eqref{eq:ppfullratio}
for the same two situations of a static and a moving harmonic trapping potential
already analyzed in Fig.~\ref{fig:2}(b).
The probability densities are obtained from simulating $10^5$ trajectories of
length $\tau = 1$, well
within the transient regime of the system
evolution (see Fig.~\ref{fig:2}).
While the distribution for
$\exp[-(\Delta\Sigma + \Delta S_{\mathrm{sys}})/\kB]$
is almost symmetric about $1$ in the static trap $u=0$,
the most probable value lies visibly below $1$ for the
moving trap $u=1$, indicating that trajectories with a
positive $\Delta\Sigma + \Delta S_{\mathrm{sys}}$ are more
likely than those with a negative value (see insets in Fig.~\ref{fig:2}).
The sample mean for the path probability ratio lies well
within one standard deviation of $1$ in both cases,
in accordance with the exact result~\eqref{eq:IFT}.

\subsection{Mutual information}
\begin{figure}[ht]
	\includegraphics[scale=1]{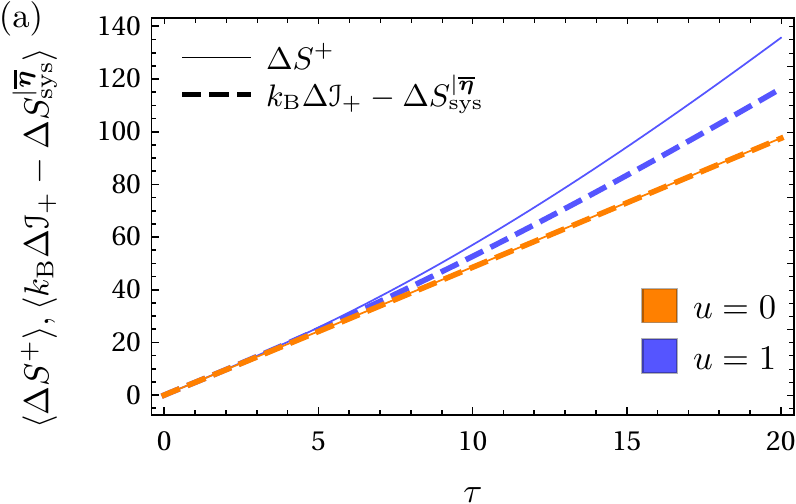}
	\\[2ex]
	\includegraphics[scale=1]{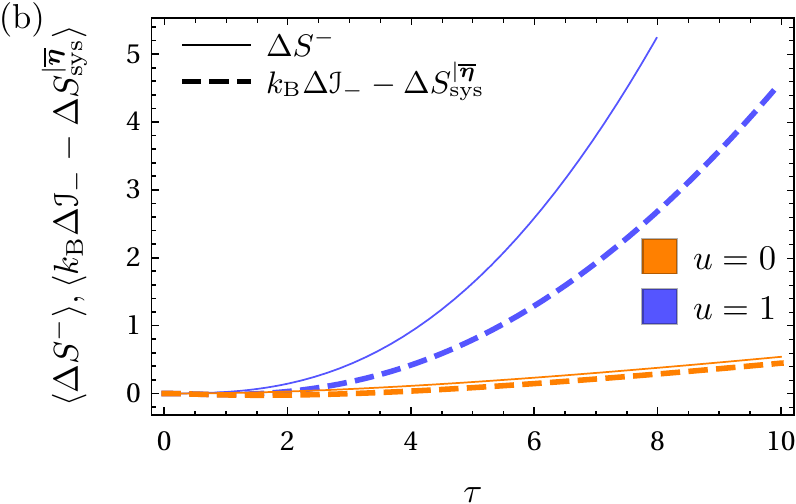}
\caption{Average mutual information and conditional entropy
production as a function of the observation time $\tau$ for a particle 
trapped in a static ($u=0$, orange lines) or moving ($u=1$, blue lines)
harmonic potential. Parameters are again $k=0.1$, $\gamma=1$, $\tau_a = 0.2$, $D = D_a = 0.2$.
(a) Interpretation of the active fluctuations as an active bath, i.e.\ they
are considered being even under time-reversal.
(b)	Interpretation of the active fluctuations as self-propulsion, i.e.\ they
are considered being odd under time-reversal.
Note that even for $u=0$ (orange lines) the curves for
$\langle \Delta S^+ \rangle$,
$\kB \langle \Delta\I_+ \rangle - \langle \Delta S_{\mathrm{sys}}^{|\Traj{\VEC{\eta}}} \rangle$ and
$\langle \Delta S^- \rangle$,
$\kB \langle \Delta\I_- \rangle - \langle \Delta S_{\mathrm{sys}}^{|\Traj{\VEC{\eta}}} \rangle$, respectively,
are not identical, but just appear very similar on the shown scale.
}
\label{fig:4}
\end{figure}
The quantities $\Delta\Sigma$ and $\Delta S_{\mathrm{sys}}$
analyzed in the previous two sections as a measure for the
irreversibility in the system evolution depend only
on the particle trajectories, but not on the realizations of the active fluctuations,
and thus are readily accessible in experiments and simulations.
The specific role of the active fluctuations, on the other hand, is
nicely captured by the splitting of irreversibility, i.e.\ the log-ratio
of path probabilities, into
total conditional entropy production $\Delta S_{\mathrm{tot}}^\pm$
and mutual information $\Delta\I_\pm$,
as described Sec.~\ref{sec:MIAB} for an active bath (`$+$' sign) and
in Sec.~\ref{sec:MISP} for self-propulsion (`$-$' sign).
Since for the present example of a particle trapped in a harmonic
potential we have analytical expressions at hand for the
combined propagator of particle position and active fluctuations
(see Appendix \ref{app:HP}),
we can calculate the conditional
entropy production from \eqref{eq:DeltaS+} and \eqref{eq:DeltaS-},
and the mutual information from
\eqref{eq:DeltaI+} and \eqref{eq:DeltaI-}
explicitly.
Only the quantity $\ln p(\VEC{x}_\tau|\Traj{\VEC{\eta}})$ is not easily
accessible because it is conditioned on the full realization $\Traj{\VEC{\eta}}$,
such that we add it to the change in mutual information in form 
of the change in (conditional) system entropy $\Delta S_{\mathrm{sys}}^{|\Traj{\VEC{\eta}}}$
[see Eq.~\eqref{eq:DeltaSsys|eta}].
Note that $\Delta S_{\mathrm{sys}}^{|\Traj{\VEC{\eta}}}$ is non-extensive with
$\tau$, and thus only a small correction to $\Delta\I_\pm$
which becomes negligible for long times.

In Fig.~\ref{fig:4}, we show the average conditional entropy
production $\langle \Delta S^\pm \rangle$ and the average
mutual information $\kB \langle \Delta \I_\pm \rangle-\langle \Delta S_{\mathrm{sys}}^{|\Traj{\VEC{\eta}}} \rangle$
for an active bath (Fig.~\ref{fig:4}a) and active self-propulsion (Fig.~\ref{fig:4}b).
The system parameters are the same as before in Figs.~\ref{fig:2}, \ref{fig:3},
in particular we again compare the a static trap $u=0$ with
a moving trap $u=1$.
We can see that now in all cases, both $\langle \Delta S^\pm \rangle$
and $\langle \Delta \I_\pm \rangle$ grow linearly with time for large $\tau$.
This conforms nicely with our previous findings:
For the conditional total entropy $\Delta S_{\mathrm{tot}}^\pm$,
the active fluctuations are treated like an external, time-dependent forcing.
Such a force is then naturally expected to produce entropy extensively.
For the ensemble- and time-averaged rate of total entropy production
$\langle \sigma_{\mathrm{tot}}^\pm \rangle=\lim_{t \to\infty} \langle\Delta S_{\mathrm{tot}}^\pm \rangle / \tau$,
we find
\begin{subequations}
\begin{align}
\langle \sigma_{\mathrm{tot}}^+ \rangle &=
\kB \left[
	\frac{\left( 1+\frac{\Da}{D} \right) u^2}{D+\Da} +
	\frac{\Da}{D} \left( \frac{1}{1 + \frac{k \ta}{\gamma}} \right) \frac{1}{\ta}
\right]
\, ,
\\
\langle \sigma_{\mathrm{tot}}^- \rangle &=
\kB \left[ 
	\frac{\left( 1+\frac{\Da}{D} \right) u^2}{D+\Da} +
	\frac{\Da}{D} \left( \frac{1}{1 + \frac{k \ta}{\gamma}} \right) \frac{k}{\gamma}
\right]
\, .
\end{align}
\end{subequations}
The first terms in these expressions contain the production rate
of $\langle \Delta\Sigma \rangle$ from \eqref{eq:SigmaRate}.
The second terms quantify the additional contributions from the
active fluctuations. They are balanced by the mutual
information production rates
$\langle \sigma_{\I}^\pm \rangle = \lim_{\tau\to\infty} \kB \langle\Delta\I_{\pm}\rangle / \tau$
(note that we include a factor of $\kB$ in this definition of the rates
so that they have units of entropy/time),
which explicitly read
\begin{subequations}
\begin{align}
\langle \sigma_{\I}^+ \rangle &=
\kB \frac{\Da}{D} \left[ \frac{u^2}{D+\Da} + \left( \frac{1}{1 + \frac{k \ta}{\gamma}} \right) \frac{1}{\ta} \right]
\, ,
\label{eq:DeltaIrate+}
\\
\langle \sigma_{\I}^- \rangle &=
\kB \frac{\Da}{D} \left[ \frac{u^2}{D+\Da} + \left( \frac{1}{1 + \frac{k \ta}{\gamma}} \right) \frac{k}{\gamma} \right]
\, .
\label{eq:DeltaIrate-}
\end{align}
\end{subequations}
In total, we therefore find in both cases that
$\langle \sigma \rangle = \langle \sigma_{\mathrm{tot}}^\pm \rangle - \langle \sigma_{\I}^\pm \rangle$
holds, where $\langle \sigma \rangle$ is given in \eqref{eq:SigmaRate}.

As discussed in Sec.~\ref{sec:EE}, the two interpretations of
the active fluctuations as active bath or as self-propulsion mechanism
correspond to measuring the mutual information
with respect to, respectively, even or odd time-reversal of
the active forcing, so that the rate of ``mutual information
production'' is different in the two cases, compare \eqref{eq:DeltaIrate+} and \eqref{eq:DeltaIrate-}.
Their difference reads
\begin{equation}
\langle \sigma_{\I}^+ \rangle - \langle \sigma_{\I}^- \rangle
	= \kB \frac{\Da}{D} \left( \frac{1}{1 + \frac{k \ta}{\gamma}} \right) \left( \frac{1}{\ta} - \frac{k}{\gamma} \right)
\, .
\end{equation}
We conclude that for the active bath the relevant time scale 
the trajectory duration $\tau$ is measured against is
the correlation time of fluctuations $\ta$,
whereas it is the system's relaxation time 
in the harmonic potential $\gamma/k$ for the self-propelled particle.

\section{Conclusions and Discussion}
\label{sec:CD}
Our present work contributes to assessing the out-of-equilibrium
character of active matter
\cite{Fodor:2016hff,marconi2017heat,mandal2017entropy,
Puglisi:2017crf,nardini2017entropy,shankar2018hidden,caprini2018comment,mandal2018mandal}
based on its observable dynamical behavior.
Having in mind that in a typical experiment the central observables
are particle trajectories, we quantify irreversibility in active matter systems
based on particle trajectories alone,
without resolving the microscopic mechanisms and associated dissipation
underlying the active fluctuations which drive particle motion.
For the thermodynamic-like features
emerging from the dynamical behavior, 
it is presumably of little relevance 
what kind of microscopic processes dissipate how much entropy 
in generating the active fluctuations
in the system (e.g.\ via self-propulsion),
as long as these processes
are not altered due to the motion of the particle.
We are therefore interested in how
(ir-)reversible a specific particle trajectory is, out of the set of all possible trajectories
which can be generated by the combined influence of thermal and active fluctuations,
but not in how (ir-)reversible the processes which underly the active fluctuations are.
In that spirit, we treat the active fluctuations as an active (non-thermal)
``bath'' the particle is in contact with in addition to the thermal bath,
with statistical properties which are completely independent of the internal
dynamical state of the particle.

We implement this active ``bath'' by following the common approach to include
stochastic ``active forces'' into the equations of motion, which emulate the
directional persistence and the non-equilibrium character of the active fluctuations
\cite{Note5}.
A particularly successful description of active fluctuations along these lines 
consists in a colored noise model,
more specifically a
Gaussian Ornstein-Uhlenbeck process
\cite{Maggi:2014gee,Argun:2016nbs,maggi2017memory,chaki2018entropy, 
Fily:2012aps,Farage:2015eii, 
Fodor:2016hff,marconi2017heat,mandal2017entropy,Puglisi:2017crf,
koumakis2014directed,szamel2014self,szamel2015glassy,maggi2015multidimensional,Flenner:2016tng,paoluzzi2016critical,Marconi:2016vdi,szamel2017evaluating,sandford2017pressure,caprini2018linear,fodor2018statistical, 
berthier2013non,Marconi:2015tas,shankar2018hidden}. 
For this model class,
we calculate the exact expression for the probability density of
particle trajectories by integrating over all possible realizations of
the active fluctuations [see Eq.~\eqref{eq:pp}].
Since the colored noise renders the particle's dynamics non-Markovian,
the standard Onsager-Machlup path integral
\cite{Onsager:1953fai,Machlup:1953fai,Chernyak:2006pia,cugliandolo2017rules}
cannot be applied directly to obtain the path probabilities.
While expressions for a single, colored noise source exist in
the literature \cite{McKane:1990pia,Hanggi:1989pis}, we here derive for the first
time the path weight for the superposition of colored Ornstein-Uhlenbeck noise
(the active fluctuations) and white thermal noise.

Building on this result, we then relate the probabilities of
time-forward and time-backward processes and establish an integral fluctuation theorem
\cite{Seifert:2008stp,Jarzynski:2011eai,Seifert:2012stf,van2015ensemble}
for their log-ratio $\Delta\Sigma$, a functional over the forward particle trajectory
[see Eqs.~\eqref{eq:IFT} and \eqref{eq:DeltaSigma}]
which quantifies the time (ir-)reversibility of the particle dynamics.
From the integral fluctuation theorem we directly obtain
a corresponding second law-like relation for $\Delta\Sigma$ [see Eq.~\eqref{eq:2ndLaw}].
With $\Delta\Sigma$ we therefore provide
an explicit expression to (exactly) evaluate the
fluctuation theorem for
a Brownian particle in contact not only with a thermal bath but at the same time
also with an active non-equilibrium bath.
In particular, it applies to trajectories of arbitrary, finite duration.
We expect that it can be tested experimentally
with state-of-the-art micro(-fluidic) technology for
biological or synthetic active matter systems, as used, e.g., in
\cite{Bechinger:2016api,Maggi:2014gee,maggi2017memory,
Argun:2016nbs,howse2007self,leptos2009dynamics,bohec2013probing,jepson2013enhanced,buttinoni2013dynamical}.
We point out again that our explicit results for the path weight
and the fluctuation theorem
are based on the assumption
that the active fluctuations are Gaussian with exponential correlations in time,
generated by an Ornstein-Uhlenbeck process.

Like in the case of usual Brownian motion in contact with a thermal bath only,
the path probability ratio \eqref{eq:ppratio}
is equal to the identity if there are no external forces acting
on the particle.
Therefore, for $\VEC{f}=0$ any time-backward trajectory is equally likely to occur as its time-forward twin,
so that the particle dynamics looks reversible and equilibrium-like.
Indeed on a coarse-grained time-scale (beyond $\ta$) they appear
very similar to free Brownian diffusion in thermal equilibrium.
Accordingly,
in the absence of external forces the probability ratio of particle trajectories
does not reveal the non-equilibrium nature of the system,
even though the whole system is out of equilibrium due to the active fluctuations.
This should be true for any stationary (and unbiased) non-equilibrium bath,
not just the Ornstein-Uhlenbeck implementation considered here.
In order to detect the irreversibility connected with the active fluctuations
we would have to resolve the corresponding degrees of freedom and
analyze their behavior under time-reversal \cite{gaspard2017communication,pietzonka2017entropy,speck2018active}.

However, as we can see from comparing \eqref{eq:DeltaSDa0} and \eqref{eq:DeltaSigma},
for non-vanishing forces $\VEC{f}$ the irreversibility measure $\Delta\Sigma$
is distinctively different from the entropy production of purely Brownian motion,
because it contains the non-local memory kernel $\Gamma_\tau(t,t')$.
Driving the particle by an external force thus reveals the
non-Markovian and
non-equilibrium character of the environment;
even in the limit $\ta \to 0$ of $\delta$-correlated active fluctuations
the kernel $\Gamma_\tau(t,t')$ yields a non-trivial contribution
(see also Appendix~\ref{app:limits}),
which renders $\Delta\Sigma$ different from the entropy production in
the actual thermal bath.
We leave for future exploration how this observation
that external forces may reveal the non-equilibrium character of the environment
may be used to probe properties of the active
bath. Preliminary results indicate
that the external force $\VEC{f}$ has to be non-linear or time-dependent, as a simple linear
$\VEC{f}$ leads to $\Delta\Sigma=0$ already on the level of individual trajectories
\cite{activeGyrator}, see also \cite{Fodor:2016hff}.

Our irreversibility measure $\Delta\Sigma$ from \eqref{eq:ppratioDeltaSigma}
quantifies the
combined ``dissipation'' into the thermal and the active bath, which occurs along
a (time-forward) particle trajectory $\Traj{\VEC{x}}$, but
cannot be interpreted easily as entropy production or dissipated heat.
However, if we keep track of a specific realization $\Traj{\VEC{\eta}}$ of
the active noise as a fluctuating force affecting the particle dynamics,
we find that $\Delta\Sigma$ can be split into two parts which have a
direct physical interpretation:
the usual entropy production in the thermal environment, and a
complementary dissipative component in the active bath, which is
expressed as the difference of path-wise
mutual information $\Delta\I$ accumulated along the time-forward
process $(\Traj{\VEC{x}},\Traj{\VEC{\eta}})$ versus its time-backward twin process
[see Eqs.~\eqref{eq:DeltaSigma+}, \eqref{eq:DeltaSigma-}].
This partition of $\Delta\Sigma$ is valid for any particle trajectory $\Traj{\VEC{x}}$
and any realization of the active noise fluctuation
$\Traj{\VEC{\eta}}$, but with process-dependent amounts of dissipation in the two baths.

All these general results and interpretations are independent of how
we choose the active fluctuations to behave under time reversal, even
[Secs.~\ref{sec:energeticsAB} and \ref{sec:MIAB}]
or odd [Secs.~\ref{sec:energeticsSP} and \ref{sec:MISP}].
The quantitative details, however, are different; compare, e.g.,
Eq.~\eqref{eq:DeltaI+} with Eq.~\eqref{eq:DeltaI-}.
These quantitative differences are a consequence of the different amounts of heat
exchanged with the thermal environment along a displacement $\mathrm{d}\VEC{x}$
in the two cases, see \eqref{eq:deltaQ+} and \eqref{eq:deltaQ-}.
When we interpret the active fluctuations $\VEC{\eta}(t)$
as an active environment the particle is moving through,
the viscous friction forces from the thermal bath which balance these active fluctuations
are included in the heat exchange.
In contrast, they are not counted as contributing to heat in the
case of self-propulsion, because self-propelled motion occurs without external forces,
i.e.\ on the coarse-grained level of description based on particle degrees of freedom
self-propulsion appears to be force- and thus dissipation-free.
We argue in Sec.~\ref{sec:EE}
that the interpretation of the active fluctuations $\VEC{\eta}(t)$
as an active environment requires $\VEC{\eta}(t)$ to be even under
time-reversal, while their interpretation as self-propulsion corresponds to $\VEC{\eta}(t)$ being odd
\cite{Note4}.

In any case, the corresponding stochastic energetics for active systems
related to displacements of the particles,
i.e.\ the definition of heat and work based on particle trajectories,
appears as a natural and consistent generalization of Sekimoto's stochastic energetics
of passive Brownian motion \cite{Sekimoto:StochasticEnergetics}
(see Section~\ref{sec:EE}), with a clear-cut connection to
the irreversibility measure $\Delta\Sigma$
as obtained from the path probability ratio.
The consistency and unambiguity of these results rely
on fully incorporating the thermal fluctuations from the
equilibrium heat bath into our description [cf.~Eqs.~\eqref{eq:LE}].
A number of previous works
\cite{Fodor:2016hff,marconi2017heat,mandal2017entropy,
Puglisi:2017crf,caprini2018comment,mandal2018mandal}
attempted to assess irreversibility in active particle systems and
its connection to heat and entropy production using
the model class \eqref{eq:LE}-\eqref{eq:OUP},
including, however, only viscous friction effects from the
heat bath and neglecting thermal fluctuations
($D=0$ in Eq.~\eqref{eq:LE} in our notation).
In this way they disregarded an essential part of the energy
exchange between system and heat bath \cite{Puglisi:2017crf}, leading to
ambiguities in the definition of heat and entropy production.
Likewise, no consensus could be reached about the thermodynamic interpretation
of the path probability ratio in terms of entropy production
(see \cite{Puglisi:2017crf} for a clear and enlightening summary,
as well as the Comment \cite{caprini2018comment}
and the corresponding Reply \cite{mandal2018mandal}).
We believe that we here resolve these problems in a unique and
consistent way by incorporating
the thermal fluctuations in our analysis [see model \eqref{eq:LE}-\eqref{eq:OUP}].
In particular this led to a clear interpretation of the path
probability ratio in terms of entropy production in the thermal bath
(connected to the actual heat dissipated in the bath) and mutual
information ``production'' with the active ``bath''
[see Eqs.~\eqref{eq:DeltaSigma+}, \eqref{eq:DeltaSigma-}].

In the present work, our focus is on deriving the path probability ratio
${\ppR{\trajR{\VEC{x}}|\tilde{\VEC{x}}_0}}/{\pp{\traj{\VEC{x}}|\VEC{x}_0}}$
as a measure of irreversibility in active matter systems and on establishing
its interpretation in thermodynamic terms.
As a fundamental concept in stochastic thermodynamics it contains essential
information about the system and entails many interesting implications.
With the fluctuations theorems and the second-law like relations we here derive
the most immediate ones.
Further potential applications of our results include
the exploration of features in $\Delta\Sigma$ and $\Delta\I$ which are characteristic
for the different phases of active matter \cite{Cates:2015mip},
the analysis of linear response under small external perturbations
and its potential for probing properties of the active bath \cite{gnesotto2018broken},
the connection between violations of the fluctuation-response relation
and irreversibility \cite{harada2005equality,gnesotto2018broken},
the characterization of universal statistics of infima, stopping times, and passage probabilities of
$\Delta\Sigma$ and $\Delta\I$, based on the property of
${\ppR{\trajR{\VEC{x}}|\tilde{\VEC{x}}_0}}/{\pp{\traj{\VEC{x}}|\VEC{x}_0}}$
\cite{neri2017statistics},
the connection of $\Delta\Sigma$ and $\Delta\I$ to the arrow of time
\cite{roldan2015decision,roldan2018arrow} in active matter systems,
and universal properties in the efficiency fluctuations of stochastic heat engines
operating between active baths \cite{Krishnamurthy:2016ams,Verley:2014tuc,Verley:2014uto}.

We here have analyzed the model \eqref{eq:LE}, \eqref{eq:vf}
in great detail from the viewpoint of a Brownian particle moving
under the influence of active fluctuations, which are represented
by the Ornstein-Uhlenbeck process \eqref{eq:OUP}.
However, most of our results, in particular the path weight
\eqref{eq:pp} and the associated integral fluctuation theorem \eqref{eq:IFT},
as well as its formulation in information-theoretic terms in Sec.~\ref{sec:MI},
are a direct consequence
of the mathematical structure of the model, and are therefore valid in a much
broader physical context.
In principle, our results can be applied to any Brownian dynamics that is driven by an
additional Gaussian Ornstein-Uhlenbeck process.
For instance, in \cite{crisanti2012nonequilibrium} an information-theoretic analysis similar to ours has been
conducted for a colored noise driven Brownian model.
Other examples are Brownian motion in a harmonic trap with fluctuating location
\cite{gomez2010steady,verley2014work,manikandan2017asymptotics,manikandan2018exact},
and thermodynamic nonequilibrium processes using external (artificial) colored noise sources
\cite{pal2013work,Mestres:2014ron}.

\begin{acknowledgments}
SB and RE acknowledge financial support from the Swedish Research Council
(Vetenskapsr{\aa}det) under the grants No.~621-2013-3956, No.~638-2013-9243 and No.~2016-05412.
LD acknowledges financial support by the Deutsche Forschungsgemeinschaft
under Grant No. RE 1344/10-1 and within the Research Unit FOR 2692 under
Grant No. RE 1344/12-1, as well as by the Stiftung der Deutschen Wirtschaft.
LD also thanks Nordita for the hospitality and support during an internship in 2016.
We moreover thank Erik Aurell, Aykut Argun, Giovanni Volpe and Jan Wehr
for illuminating discussions.
\end{acknowledgments}

\appendix

\section{Integrating over the active fluctuations $\VEC{\eta}(t)$}
\label{app:integrateEta}
Here we will show that the path integral~\eqref{eq:pp1} evaluates to expression~\eqref{eq:pp}.
Gaussian path integrals of this form are ubiquitous in field theories
\cite{Wiegel:IntroductionToPathIntegralMethods,Zinn-Justin:QFTAndCriticalPhenomena}
with well-established results.
Using the abbreviation $\VEC{w}_t = \frac{\sqrt{2\Da}}{2D} \left( \dot{\VEC{x}}_t - \VEC{v}_t \right)$,
the relevant terms in \eqref{eq:pp1} involving the active noise variable $\VEC{\eta}_t$ become
\begin{equation}
\label{eq:etaPathIntegral}
\int \D\Traj{\VEC{\eta}} \,
\exp \left[
	-\frac{1}{2} \int_0^\tau \!\! \mathrm{d}t \int_0^\tau \!\! \mathrm{d}t^\prime \,
			\VEC{\eta}_t\TRANS \hat{V}_\tau(t,t^\prime) \VEC{\eta}_{t^\prime}
	+ \int_0^\tau \!\! \mathrm{d}t \, \VEC{\eta}_t\TRANS \VEC{w}_t
\right] 
\, .
\end{equation}
Completing the square, we wish to write this as
\begin{multline}
\label{eq:completingTheSquare}
\int \D\Traj{\VEC\eta} \,
\exp \left\{
	\frac{1}{2} \int_0^\tau \!\! \mathrm{d}t \int_0^\tau \!\! \mathrm{d}t^\prime 
		\left[ \VEC{w}_t\TRANS \Gamma_\tau(t,t^\prime) \VEC{w}_{t^\prime}
\right.\right.
\\
\left.\left.
\mbox{} -(\VEC{\eta}_t + \VEC{\varepsilon}_t)\TRANS \hat{V}_\tau(t,t^\prime)
		(\VEC{\eta}_{t^\prime} + \VEC{\varepsilon}_{t^\prime}) \right]
\vphantom{\int_0^\tau }\right\}
\, ,
\end{multline}
where $\VEC{\varepsilon}_t$ and $\Gamma_\tau(t,t^\prime)$ are yet unknown functions. 
We can then shift the active noise histories and integrate over
$\VEC{\eta}^\prime_t = \VEC{\eta}_t + \VEC{\varepsilon}_t$ instead of $\VEC{\eta}_t$.
Since the path integral effectively integrates over all possible states of $\VEC{\eta}_t$
from $-\infty$ to $+\infty$ for any point in time $t$, this shift of trajectories does
not alter the domain of integration.
Moreover, the Jacobian associated with the transformation is the identity.
Performing the remaining functional integral over $\Traj{\VEC{\eta}}^\prime$,
expression~\eqref{eq:completingTheSquare} then reduces to
\begin{equation}
	(\operatorname{Det}\hat{V}_\tau)^{-1/2}
	\exp\left[ \frac{1}{2} \int_0^\tau \!\! \mathrm{d}t \int_0^\tau \!\! \mathrm{d}t^\prime
		\VEC{w}_t \Gamma_\tau(t,t^\prime) \VEC{w}_{t^\prime} \right]
\, ,
\end{equation}
assuming proper normalization of the integration measure $\D{\Traj{\VEC\eta}}^\prime$.
Absorbing the path-independent functional determinant $(\operatorname{Det}\hat{V}_\tau)^{-1/2}$
into the normalization,
we obtain the path weight stated in Eq.~\eqref{eq:pp}.

However, we still have to show that $\Gamma_\tau(t,t^\prime)$ is the
operator inverse of $\hat{V}_\tau(t,t^\prime)$, i.e.\ we have to verify~\eqref{eq:defGamma}.
Requiring equality of~\eqref{eq:etaPathIntegral} and~\eqref{eq:completingTheSquare},
we find that
\begin{multline}
\int_0^\tau \!\! \mathrm{d}t \, \VEC{\eta}_t\TRANS \VEC{w}_t
	\overset{!}{=} \frac{1}{2} \int_0^\tau \!\! \mathrm{d}t \int_0^\tau \!\! \mathrm{d}t^\prime
	\Big[ \VEC{w}_t\TRANS \Gamma_\tau(t,t^\prime) \VEC{w}_{t^\prime}
\\ \mbox{}
	- \VEC{\varepsilon}_t\TRANS \hat{V}_\tau(t,t^\prime) \VEC{\varepsilon}_{t^\prime}
	- 2 \VEC{\eta}_t\TRANS \hat{V}_\tau(t,t^\prime) \VEC{\varepsilon}_{t^\prime}
\Big]
\, .
\end{multline}
Note that we have used $\hat{V}_\tau(t,t^\prime) = \hat{V}_\tau(t^\prime,t)$.
We now observe that the term on the left-hand side as well as the last term on
the right-hand side are of first order in $\VEC{\eta}_t$, whereas
the remaining terms on the right-hand side do not contain $\VEC{\eta}_t$.
Therefore, these two types of expressions must cancel individually, i.e.\
\begin{subequations}
\begin{gather}
\label{eq:eta0}
\int_0^\tau \!\! \mathrm{d}t \int_0^\tau \!\! \mathrm{d}t^\prime \,
	\VEC{\varepsilon}_t\TRANS \hat{V}_\tau(t,t^\prime) \VEC{\varepsilon}_{t^\prime}
= \int_0^\tau \!\! \mathrm{d}t \int_0^\tau \!\! \mathrm{d}t^\prime \,
	\VEC{w}_t\TRANS \Gamma_\tau(t,t^\prime) \VEC{w}_{t^\prime}
\, , \\
\label{eq:eta1}
\int_0^\tau \!\! \mathrm{d}t \,
	\VEC{\eta}_t\TRANS \VEC{w}_t
= -\int_0^\tau \!\! \mathrm{d}t \, \VEC{\eta}_t\TRANS \int_0^\tau \!\! \mathrm{d}t^\prime \,
	\hat{V}_\tau(t,t^\prime) \VEC{\varepsilon}_{t^\prime}
\, .
\end{gather}
\end{subequations}
From~\eqref{eq:eta1} we immediately infer
\begin{equation}
\label{eq:w}
\VEC{w}_t = -\int_0^\tau \!\! \mathrm{d}t^\prime \, \hat{V}_\tau(t,t^\prime) \VEC{\varepsilon}_{t^\prime}
\, .
\end{equation}
Substituting this result into the left-hand side of
\eqref{eq:eta0}, we obtain
\begin{equation}
-\int_0^\tau \!\! \mathrm{d}t \, \VEC{\varepsilon}_t\TRANS \VEC{w}_t
= \int_0^\tau \!\! \mathrm{d}t \, \VEC{w}_t\TRANS \int_0^\tau \!\! \mathrm{d}t^\prime \,
		\Gamma_\tau(t,t^\prime) \VEC{w}_{t^\prime}
\, ,
\end{equation}
implying
\begin{equation}
\label{eq:epsilon}
\VEC{\varepsilon}_t = -\int_0^\tau \!\! \mathrm{d}t^\prime \, \Gamma_\tau(t,t^\prime) \VEC{w}_{t^\prime}
\, .
\end{equation}
This gives the shift $\VEC{\varepsilon}_t$ required
to complete the square in \eqref{eq:etaPathIntegral} (see also \eqref{eq:completingTheSquare})
as a function of $\Gamma_\tau(t,t^\prime)$ and $\VEC{w}_t$.
Moreover, substituting~\eqref{eq:epsilon} back
into~\eqref{eq:w}, we find
\begin{equation}
\VEC{w}_t = \int_0^\tau \!\! \mathrm{d}t^\prime \int_0^\tau \!\! \mathrm{d}t^{\prime\prime} \,
	\hat{V}_\tau(t,t^\prime) \Gamma_\tau(t^\prime,t^{\prime\prime}) \VEC{w}_{t^{\prime\prime}}
\, ,
\end{equation}
which implies~\eqref{eq:defGamma} and thus establishes
that $\Gamma_\tau(t,t')$ is indeed the Green's function of
the differential operator $\hat{V}_\tau(t,t')$.

\section{Construction of the Green's function $\Gamma_\tau(t,t')$}
\label{app:GreenFunctionConstruction}
In this appendix, we will construct the Green's function~\eqref{eq:Gamma}
by solving its defining equation~\eqref{eq:defGamma}
with the differential operator $\hat{V}_\tau(t,t')$ from \eqref{eq:Vfull}.
Exploiting the ``diagonal'' structure of $\hat{V}_\tau(t,t')$ we can
directly evaluate the integral and obtain
\begin{equation}
\label{eq:ODEGamma}
\left[ \hat{V}(t) + \hat{V}_0(t) + \hat{V}_\tau(t) \right] \Gamma_\tau(t,t') = \delta(t - t^\prime)
\, .
\end{equation}
This is a linear, second-order ordinary differential equation with a $\delta$-inhomogeneity.
We will calculate its solution in two steps.
First, we will compute the Green's function $\bar{\Gamma}(t,t')$ of the ordinary component $\hat{V}(t)$
with vanishing boundary conditions, i.e.\ $\bar{\Gamma}(0, t^\prime) = \bar{\Gamma}(\tau, t^\prime) = 0$.
Then we will add a solution $\Gamma_{0,\tau}(t,t')$ of the corresponding homogeneous problem,
$\hat{V}(t)\Gamma_{0,\tau}(t,t^\prime) = 0$, that fixes the two boundary terms.
Their sum $\Gamma_\tau(t,t^\prime) = \bar{\Gamma}(t,t^\prime) + \Gamma_{0,\tau}(t,t^\prime)$
satisfies \eqref{eq:ODEGamma}, and thus gives the desired solution.

We can construct both these parts, $\bar{\Gamma}(t,t^\prime)$ and $\Gamma_{0,\tau}(t,t^\prime)$,
from the homogeneous problem associated with $\hat{V}(t,t')$,
which reads (see \eqref{eq:V})
\begin{equation}
	\left[ -\tau_a^2 \partial_t^2 + (1 + D_a/D) \right] \Gamma(t) = 0
\, .
\end{equation}
We make an exponential ansatz $\Gamma(t) \sim e^{\lambda t}$ and
easily obtain
\begin{equation}
\label{eq:homogeneousSolution}
\Gamma(t) = \alpha^+ e^{\lambda t} + \alpha^- e^{-\lambda t}
\, , \quad
\lambda = \frac{1}{\ta} \sqrt{1 + \tfrac{\Da}{D}}
\, ,
\end{equation}
with constants $\alpha^\pm$ to be determined by the boundary conditions
or the $\delta$-inhomogeneity.

There exists a standard recipe \cite{BenderOrszag:AdvMathMethods,StakgoldHolst:GreensFunctionsAndBVPs}
for the construction of Green's functions of boundary value problems
for ordinary differential equations, which we will follow here to
calculate $\bar{\Gamma}(t,t')$.
Splitting the interval $[0, \tau]$ at $t = t^\prime$, we write
\begin{equation}
\label{eq:defGammabar}
\bar{\Gamma}(t,t^\prime)
= \Theta(t^\prime - t) \bar{\Gamma}_<(t,t^\prime) + \Theta(t - t^\prime) \bar{\Gamma}_>(t,t^\prime)
\, ,
\end{equation}
where $\Theta$ is the Heaviside step function and both,
$\bar{\Gamma}_<(t,t')$ and $\bar\Gamma_>(t,t')$, satisfy the homogenous problem,
i.e.\ they are of the form \eqref{eq:homogeneousSolution} with constants
$\alpha_<^\pm$ and $\alpha_>^\pm$, respectively, to be determined.
The constants are fixed by the boundary conditions,
\begin{subequations}
\begin{equation}
\label{eq:GammabarBC}
\bar{\Gamma}_<(0,t^\prime) = 0 \quad \text{and} \quad \bar{\Gamma}_>(\tau,t^\prime) = 0
\, ,
\end{equation}
and the requirements that $\bar{\Gamma}$ is continuous at $t = t^\prime$,
\begin{equation}
\label{eq:GammabarCont}
\bar{\Gamma}_<(t^\prime,t^\prime) = \bar{\Gamma}_>(t^\prime,t^\prime)
\, ,
\end{equation}
and has a jump of size $-1/\ta^2$ in its first derivative at $t = t^\prime$,
\begin{equation}
\label{eq:GammabarJump}
\left[ \partial_t\bar{\Gamma}_>(t,t^\prime) - \partial_t \bar{\Gamma}_<(t,t^\prime) \right]_{t=t'}
	= -\frac{1}{\ta^2}
\, .
\end{equation}
\end{subequations}
The last two conditions ensure the desired $\delta$-discontinuity at
$t=t^\prime$ in the second derivative.
Solving the resulting system of linear equations for
$\alpha_<^+$, $\alpha_<^-$, $\alpha_>^+$, and $\alpha_>^-$, we obtain
\begin{widetext}
\begin{equation}
\bar{\Gamma}_>(t,t^\prime)
= \left( \frac{1}{2 \ta^2 \lambda} \right)
\frac{e^{\lambda\tau} \left[ e^{-\lambda(t-t^\prime)} - e^{-\lambda (t+t^\prime)} \right]
			+ e^{-\lambda\tau} \left[ e^{\lambda (t-t^\prime)} - e^{\lambda(t + t^\prime)} \right]}
	 {e^{\lambda\tau} - e^{-\lambda\tau}}
\end{equation}
and $\bar{\Gamma}_<(t,t^\prime) = \bar{\Gamma}_>(t^\prime,t)$.
With \eqref{eq:defGammabar}, we can thus write the Green's function of the
ordinary component on the entire interval $[0,\tau]$ in the form
\begin{equation}
\label{eq:Gammabar}
\bar{\Gamma}(t,t^\prime)
= \left( \frac{1}{2 \ta^2 \lambda} \right)
\frac{e^{-\lambda |t-t^\prime|} - e^{-\lambda (t+t^\prime)} 
	+ e^{-\lambda (2\tau - |t-t^\prime|)} 
	- e^{-\lambda (2\tau - t - t^\prime)}}
	{ 1 - e^{-2 \lambda\tau} }
\, .
\end{equation}
\end{widetext}
We can immediately check that indeed
$\hat{V}(t) \bar{\Gamma}(t,t^\prime) = \delta(t-t^\prime)$
as well as $\bar{\Gamma}(0,t^\prime) = 0$ and $\bar{\Gamma}(\tau,t^\prime) = 0$.
Moreover, we note that $\bar{\Gamma}(t,t')$ is symmetric in its arguments,
$\bar{\Gamma}(t,t^\prime) = \bar{\Gamma}(t^\prime, t)$.

As announced earlier, we now add a homogeneous solution
$\Gamma_{0,\tau}(t,t^\prime) = \alpha^+ e^{\lambda t} + \alpha^- e^{-\lambda t}$
to the ordinary inverse $\bar{\Gamma}(t,t')$.
Since $\hat{V}(t) \Gamma_{0,\tau}(t,t^\prime) = 0$ and
$\bar{\Gamma}(t,t')$ vanishes at both boundaries of the interval $[0, \tau]$,
the differential equation \eqref{eq:ODEGamma} for the sum
$\Gamma_\tau(t,t')=\bar{\Gamma}(t,t')+\Gamma_{0,\tau}(t,t^\prime)$ becomes
\begin{multline}
\delta(t-\tau) \left[
	\ta^2 \, \left.\partial_t \bar{\Gamma}_>(t, t^\prime)\right|_{t=\tau} 
	+ \alpha^+ \ta \kappa_+ e^{\lambda\tau} + \alpha^- \tau_a \kappa_- e^{-\lambda\tau} 
\right]
\\ \mbox{}
+ \delta(t) \left[ 
	-\ta^2 \, \left.\partial_t\bar{\Gamma}_<(t,t^\prime)\right|_{t=0} 
	+ \alpha^+ \tau_a \kappa_- + \alpha^- \tau_a \kappa_+ 
\right] = 0
\, ,
\end{multline}
with $\kappa_\pm = 1 \pm \lambda \tau_a = 1 \pm \sqrt{1 + \frac{D}{D_a}}$.
We require the two terms in square brackets to vanish individually,
and thus obtain two linear equations for the two unknowns $\alpha^+$ and $\alpha^-$.
Solving them and combining the results for $\bar\Gamma(t,t')$ and $\Gamma_{0,\tau}(t,t')$
into $\Gamma_\tau(t,t')=\bar{\Gamma}(t,t')+\Gamma_{0,\tau}(t,t^\prime)$,
we finally find the Green's function~\eqref{eq:Gamma}.

\section{Limiting cases for the path weight}
\label{app:limits}

Here, we analyze three relevant limiting cases of the path weight
\eqref{eq:pp}, namely
$\Da \to 0$ (usual Brownian particle without active fluctuations),
$\ta \to 0$ (memory-less active fluctuations), and
$D \to 0$ (no thermal fluctuations).
All three limits reduce the Langevin equation \eqref{eq:LE} to
simpler set-ups, for which the path probabilities are already known in the literature.
We recover all these results when performing the respective limits in
our general expression \eqref{eq:pp}.

\subsection{Usual Brownian motion $\Da \to 0$}
Removing the effects of the active fluctuations from
the system amounts to setting $\VEC{\eta}(t) = 0$ or, equivalently, $\Da = 0$
in the equation of motion~\eqref{eq:LE}. Performing this limit for the path weight
is straightforward: In this case, $\lambda = 1/\ta$ is well-behaved, so that
the term containing the Green's function simply drops out in
\eqref{eq:pp}.
The remaining path probability is just the standard Onsager-Machlup expression
\eqref{eq:ppDa0} for a Brownian particle in a thermal bath.

\subsection{Vanishing correlation time $\ta \to 0$}
In the limit $\ta \to 0$, the correlator~\eqref{eq:corrEta} of the active noise
approaches a $\delta$-distribution, i.e.\
$\langle \eta_i(t) \eta_j(t^\prime) \rangle \to \delta_{ij} \delta(t - t^\prime)$.
This means that $\VEC{\eta}(t)$ becomes just another Gaussian white noise
with diffusion constant $D_a$. The equation of motion~\eqref{eq:LE} thus contains
two independent Gaussian white noises with zero mean and variance $2 D$ and $2 \Da$, respectively.
Their sum is itself a Gaussian white noise with zero mean,
but variance $2 (D+\Da)$. Therefore, we expect \eqref{eq:pp} to reduce to
\begin{equation}
\label{eq:ppta0}
\pp{\traj{\VEC{x}}|\VEC{x}_0}_{\ta \to 0}
\propto \exp \int_0^\tau \!\! \mathrm{d}t
\left[
	-\frac{\left( \dot{\VEC{x}}_t - \VEC{v}_t \right)^2}{4 (D + \Da)} 
	-\frac{\VEC{\nabla} \cdot \VEC{v}_t}{2}
\right]
\end{equation}
as $\ta \to 0$.

We first note that $\lambda$ diverges in this limit,
so that we have to take more care when analyzing the
Green's function~\eqref{eq:Gamma}.
Since the limit is expressed more compactly as $\lambda \to \infty$,
we rewrite all occurrences of $\ta$ in \eqref{eq:Gamma}
in terms of $\lambda$, $D$ and $\Da$.
By definition (see \eqref{eq:lambda} and \eqref{eq:homogeneousSolution}),
$\ta^2 = \left( 1 + \frac{\Da}{D} \right) / \lambda^2$.
Thus the leading order behavior of $\Gamma_\tau(t,t^\prime)$ is
\begin{equation}
\label{eq:Gammalambda1}
\Gamma_\tau(t,t^\prime) \sim
\left( \frac{\lambda}{2} \right)
\frac{ e^{-\lambda |t - t^\prime|}
		- \frac{\kappa_-}{\kappa_+}
		\left[ e^{-\lambda (t+t^\prime)} + e^{-\lambda (2 \tau - t - t^\prime)} \right]}
	 {1 + \frac{\Da}{D}}
\, .
\end{equation}
We observe that
\begin{subequations}
\begin{gather}
\lim_{\lambda\to\infty} \frac{\lambda}{2} e^{-\lambda |t - t^\prime |} = \delta(t - t^\prime)
\, , \\
\lim_{\lambda\to\infty} \frac{\lambda}{2} e^{-\lambda t} \Theta(t) = \frac{1}{2} \delta(t)
\, ,
\end{gather}
\end{subequations}
such that the exponentials become $\delta$-distributions as $\lambda \to \infty$:
\begin{equation*}
\Gamma_\tau(t,t^\prime) \sim
\frac{ \delta(t - t^\prime)
		- \frac{\kappa_-}{2 \kappa_+}
		\left[ \delta(t+t^\prime) + \delta(2 \tau - t - t^\prime) \right]}
	  {1 + \frac{\Da}{D}}
\, .
\end{equation*}
However, the last two $\delta$-distributions map either of the two times
$t$ and $t^\prime$ outside of the interval of integration when integrating over the other.
Therefore, they do not contribute when integrating over both $t$ and $t^\prime$.
We can see this also from partially integrating the corresponding term
in \eqref{eq:Gammalambda1} with two test functions $f$ and $g$,
\begin{align*}
&\int_0^\tau \!\! \mathrm{d}t \int_0^\tau \!\! \mathrm{d}t^\prime \;
	f(t) g(t^\prime) \frac{\lambda}{2}
		\left[ e^{-\lambda (t+t^\prime)} + e^{-\lambda (2 \tau - t - t^\prime)} \right]
\\
&\; =
\int_0^\tau \!\! \mathrm{d}t \;
	f(t) \left\{
	\left. g(t^\prime) \frac{1}{2}
		\left[ -e^{-\lambda (t+t^\prime)} + e^{-\lambda (2 \tau - t - t^\prime)} \right]
	\right|_{t^\prime=0}^\tau
	\right.
\\
&\;\qquad
	\left.
	\mbox{}- \int_0^\tau \!\! \mathrm{d}t^\prime \;
	\dot{g}(t^\prime) \frac{1}{2}
		\left[ -e^{-\lambda (t+t^\prime)} + e^{-\lambda (2 \tau - t - t^\prime)} \right]
	\right\}
\\
&\;
\to 0 \quad \text{as} \quad \lambda \to \infty
\, .
\end{align*}
Therefore, $\Gamma_\tau(t,t^\prime) \sim \delta(t - t^\prime) / (1 + \frac{D_a}{D})$ as
$\ta \to 0$. Substituting this finding into the path weight~\eqref{eq:pp}
leads precisely to the expected limit~\eqref{eq:ppta0}.

\subsection{No thermal fluctuations $D \to 0$}
The limit of vanishing thermal white noise is a little more involved.
If we let $D \to 0$ in the equation of motion \eqref{eq:LE},
we are left with a system driven by a single, colored noise source.
The resulting path probability density is known in the literature
\cite{McKane:1990pia,Hanggi:1989pis}
and given by Eq.~\eqref{eq:ppD0}.
However, it is not immediately obvious how we can obtain this result
from~\eqref{eq:pp}, because both the prefactor $1/(4D)$
and the exponent $\lambda$ diverge as $D \to 0$.

We first rewrite~\eqref{eq:ppD0}, so that the action
can be expressed in the form of~\eqref{eq:pp}.
To this end, we introduce the abbreviation $\VEC{w}_t = \dot{\VEC{x}}_t - \VEC{v}_t$
in \eqref{eq:ppD0},
and remember that we chose $p_{\mathrm{s}}$ to be the stationary
distribution \eqref{eq:pseta} of the colored noise.
After partial integration of the $\dot{\VEC{w}}_t^2$ term, we find
\begin{multline}
\label{eq:ppD0w}
\pp{\traj{\VEC{x}}|\VEC{x}_0}_{D \to 0} \propto
\exp \left\{
-\frac{1}{4\Da} \left[
	\int_0^\tau \!\! \mathrm{d}t \left( -\ta^2 \VEC{w}_t\TRANS \ddot{\VEC{w}}_t + \VEC{w}_t^2 \right)
	\right. \right. \\
	\left.  \left. \vphantom{\int_0^\tau} \mbox{}
	+ \ta^2 \dot{\VEC{w}}_t\TRANS \VEC{w}_t \big|_0^\tau
	+ \ta \VEC{w}_t^2 \big|_0^\tau
	+ 2 \ta^2 \VEC{w}_0^2
\right]
- \int_0^\tau \!\! \mathrm{d}t \, \frac{\VEC{\nabla} \cdot \VEC{v}_t}{2}
\right\}
\, .
\end{multline}

We will now show that the action
\begin{equation}
\mathcal{A}[\Traj{\VEC{x}}] =
\frac{1}{4D} \int_0^\tau \!\! \mathrm{d}t \int_0^\tau \!\! \mathrm{d}t^\prime \;
\VEC{w}_t\TRANS \left[ \delta(t-t^\prime) - \frac{\Da}{D} \Gamma_\tau(t,t^\prime) \right] \VEC{w}_t^\prime
\end{equation}
of the full path weight
$\pp{\traj{\VEC{x}}|\VEC{x}_0} \propto e^{-\mathcal{A}[\Traj{\VEC{x}}]-
\int_0^\tau \!\! \mathrm{d}t \, {\VEC{\nabla} \cdot \VEC{v}_t}/{2}}$
as given in \eqref{eq:pp},
reduces to the form
of the action in \eqref{eq:ppD0w} in the limit $D \to 0$.

We rewrite the limit $D \to 0$ again as $\lambda \to \infty$, i.e.\
we express all occurrences of $D$ in terms of $\lambda$, $\Da$ and $\ta$
according to $\frac{\Da}{D} = \lambda^2 \ta^2 - 1$
(see \eqref{eq:lambda} and \eqref{eq:homogeneousSolution}).
For the leading order behavior of the Green's function $\Gamma_\tau(t,t')$
we then find
\begin{widetext}
\begin{equation}
\Gamma_\tau(t,t^\prime) \sim
	\frac{ e^{-\lambda |t-t^\prime|}
		+ \left[ 1 - \frac{2}{\ta\lambda} + \mathcal{O}\left(\lambda^{-2}\right) \right]
		  \left[ e^{-\lambda (t+t^\prime)} + e^{-\lambda (2 \tau - t - t^\prime)} \right]}
		 {2 \ta^2 \lambda}
\, ,
\end{equation}
which implies
\begin{multline*}
\mathcal{A}[\Traj{\VEC{x}}]_{D \to 0} \sim
	\frac{1}{4\Da} \int_0^\tau \!\! \mathrm{d}t \int_0^\tau \!\! \mathrm{d}t^\prime \;
	\VEC{w}_t\TRANS \VEC{w}_{t^\prime}
	\left\{
		\left( \lambda^2 \ta^2 - 1 \right) \delta(t - t^\prime) 
	  - \left( \frac{\lambda^3 \ta^2}{2} - \lambda \right)
	  		\left[ e^{-\lambda |t-t^\prime|} + e^{-\lambda (t+t^\prime)} + e^{-\lambda (2 \tau - t - t^\prime)} \right]
	\right.		 
\\
	\left. \mbox{}
	+ \left[ \lambda^2 \ta + \mathcal{O}\left(\lambda\right) \right]
	  \left[ e^{-\lambda (t+t^\prime)} + e^{-\lambda (2 \tau - t - t^\prime)} \right]
    \vphantom{\frac{\lambda^3 \tau_a^2}{2} }
    \right\}
\, .
\end{multline*}
We divide our further analysis into three parts by writing
$\mathcal{A}[\Traj{\VEC{x}}] \sim \frac{1}{4\Da}
( \mathcal{B}_1[\Traj{\VEC{x}}] + \mathcal{B}_2[\Traj{\VEC{x}}] + \mathcal{B}_3[\Traj{\VEC{x}}] )$
with
\begin{subequations}
\begin{gather}
\mathcal{B}_1[\Traj{\VEC{x}}] = \int_0^\tau \!\! \mathrm{d}t \int_0^\tau \!\! \mathrm{d}t^\prime \;
	\VEC{w}_t\TRANS \VEC{w}_{t^\prime}
	\left[ \left( \lambda^2 \ta^2 - 1 \right) \delta(t - t^\prime)
		 - \left( \frac{\lambda^3 \ta^2}{2} - \lambda \right) e^{-\lambda |t-t^\prime|}
	\right]
\, , \\
\mathcal{B}_2[\Traj{\VEC{x}}] = - \int_0^\tau \!\! \mathrm{d}t \int_0^\tau \!\! \mathrm{d}t^\prime \;
	\VEC{w}_t\TRANS \VEC{w}_{t^\prime}
	\left( \frac{\lambda^3 \ta^2}{2} - \lambda \right)
	\left[ e^{-\lambda (t+t^\prime)} + e^{-\lambda (2 \tau - t - t^\prime)} \right]
\, , \\
\mathcal{B}_3[\Traj{\VEC{x}}] = \int_0^\tau \!\! \mathrm{d}t \int_0^\tau \!\! \mathrm{d}t^\prime \;
	\VEC{w}_t\TRANS \VEC{w}_{t^\prime}
	\left[ \lambda^2 \ta + \mathcal{O}(\lambda) \right]
	\left[ e^{-\lambda (t+t^\prime)} + e^{-\lambda (2 \tau - t - t^\prime)} \right]
\, .
\end{gather}
\end{subequations}
\end{widetext}
Upon repeated partial integration using
$\lambda e^{-\lambda t}=-\partial_t e^{-\lambda t}$
to remove powers of $\lambda$, and upon performing
the $\lambda \to \infty$ limiting procedure for well behaved terms,
we find for the first part
\begin{equation*}
\mathcal{B}_1[\Traj{\VEC{x}}] \sim \int_0^\tau \!\! \mathrm{d}t
	\left[ -\ta^2 \VEC{w}_t\TRANS \ddot{\VEC{w}}_t + \VEC{w}_t^2 \right]
  + \left( \VEC{w}_0^2 + \VEC{w}_\tau^2 \right) \frac{\lambda \ta^2}{2}
\, .
\end{equation*}
Similarly, the second part reduces to
\begin{equation*}
\mathcal{B}_2[\Traj{\VEC{x}}] \sim
	-\left( \VEC{w}_0^2 + \VEC{w}_\tau^2 \right) \frac{\lambda \ta^2}{2}
	+ \ta^2 \left( \VEC{w}_\tau\TRANS \dot{\VEC{w}}_\tau - \VEC{w}_0\TRANS \dot{\VEC{w}}_0 \right)
\, .
\end{equation*}
For the third part, we first remark that the terms of order $\lambda$
vanish here again under double time integrals because they become
$\delta$-distributions that map one of the times outside the
interval of integration, similarly to the case we had for the
limit of vanishing correlation time. Hence,
\begin{align*}
&\mathcal{B}_3[\Traj{\VEC{x}}] \sim
	\int_0^\tau \!\! \mathrm{d}t \int_0^\tau \!\! \mathrm{d}t^\prime \;
		\VEC{w}_t\TRANS \VEC{w}_{t^\prime} \lambda^2 \ta
			\left[ e^{-\lambda (t+t^\prime)} + e^{-\lambda(2\tau-t-t^\prime)} \right]
\displaybreak[0] \\
& \; =
	\int_0^\tau \!\! \mathrm{d}t \int_0^\tau \!\! \mathrm{d}t^\prime \;
		\VEC{w}_t\TRANS \VEC{w}_{t^\prime} \lambda \ta
			\partial_{t^\prime} \left[ -e^{-\lambda (t+t^\prime)} + e^{-\lambda(2\tau-t-t^\prime)} \right]
\displaybreak[0] \\
& \; =
\int_0^\tau \!\! \mathrm{d}t \left.
	\VEC{w}_t\TRANS \VEC{w}_{t^\prime} \lambda \ta
		\left[ -e^{-\lambda (t+t^\prime)} + e^{-\lambda(2\tau-t-t^\prime)} \right] \right|_{t^\prime=0}^\tau
\\
& \; \quad \mbox{}
	-\int_0^\tau \!\! \mathrm{d}t \int_0^\tau \!\! \mathrm{d}t^\prime \;
		\VEC{w}_t\TRANS \dot{\VEC{w}}_{t^\prime} \lambda \ta
			\left[ -e^{-\lambda (t+t^\prime)} + e^{-\lambda(2\tau-t-t^\prime)} \right]
\displaybreak[0] \\
& \; \sim
	\int_0^\tau \!\! \mathrm{d}t \;
		\VEC{w}_t\TRANS \left\{
			\VEC{w}_\tau \ta \lambda \left[ -e^{-\lambda(\tau + t)} + e^{-\lambda(\tau - t)} \right]
		\right\}
\\
& \; \quad \mbox{}
	-\int_0^\tau \!\! \mathrm{d}t \;
		\VEC{w}_t\TRANS \left\{
			 \VEC{w}_0 \ta \lambda \left[ -e^{-\lambda t} + e^{-\lambda (2 \tau - t)} \right]
		\right\}
\displaybreak[0] \\
& \; \sim
	\ta \left( \VEC{w}_\tau^2 - \VEC{w}_0^2 \right) + 2 \ta \VEC{w}_0^2
\, .
\end{align*}
Combining these three results, we finally find
\begin{multline}
\mathcal{A}[\Traj{\VEC{x}}]_{D \to 0} = \frac{1}{4\Da}
	\left[
		\int_0^\tau \!\! \mathrm{d}t \left( -\ta^2 \VEC{w}_t\TRANS \ddot{\VEC{w}}_t + \VEC{w}_t^2 \right)
	\right.
\\
	\left. \mbox{}
		+ \ta^2 \left( \VEC{w}_\tau\TRANS \dot{\VEC{w}}_\tau - \VEC{w}_0\TRANS \dot{\VEC{w}}_0 \right)
		+ \ta   \left( \VEC{w}_\tau^2 - \VEC{w}_0^2 \right) + 2 \ta \VEC{w}_0^2
	\vphantom{\int=0^\tau}
	\right]
\, ,
\end{multline}
which is identical to the path weight \eqref{eq:ppD0w}.

\section{Details for the harmonically trapped particle}
\label{app:HP}

In this appendix, we summarize some details and a few key steps
behind the calculations for the Brownian particle in a
harmonic potential from Sec.~\ref{sec:HP}.

\subsection{Dynamics}
The equations of motion for the joint Markovian system of particle and active fluctuations,
combined from~\eqref{eq:LEex} and~\eqref{eq:OUP} (for $d=1$), read
\begin{subequations}
\label{eq:LEexjoint}
\begin{equation}
\left(
	\begin{matrix} \dot{x}_t \\ \dot{\eta}_t \end{matrix}
\right)
= - \TENSOR{A}
	\left[ \left( \begin{matrix} x_t \\ \eta_t \end{matrix} \right)
		 - \left( \begin{matrix} u \\ 0 \end{matrix} \right) t
	\right]
+ \TENSOR{B} \left( \begin{matrix} \xi_t \\ \zeta_t \end{matrix} \right)
\end{equation}
with 
\begin{equation}
\TENSOR{A} =
\left(
	\begin{matrix}
		k/\gamma & -\sqrt{2 \Da} \\
		0        & 1/\ta
	\end{matrix}
\right)
\quad\text{and}\quad
\TENSOR{B} =
\left( 
	\begin{matrix}
		\sqrt{2 D} & 0 \\
		0        & 1/\tau_a
	\end{matrix}
\right) 
\, .
\end{equation}
\end{subequations}
Due to its Markovian character, all statistical properties of this joint
system are encoded in the propagator
$p(\VEC{q}, t |\VEC{q}_0, t_0)$,
where $\VEC{q} = (x, \eta)$, $\VEC{q}_0 = (x_0, \eta_0)$
are ``generalized coordinates'' summarizing particle position and
state of the active fluctuations.
This propagator represents the probability to be at ``position''
$\VEC{q}$ at time $t$ when having been at $\VEC{q}_0$ at an
earlier time $t_0<t$.
We obtain its explicit form by solving the
the Fokker-Planck equation associated with~\eqref{eq:LEexjoint},
\begin{subequations}
\label{eq:ExProp}
\begin{equation}
p(\VEC{q},t|\VEC{q}_0,t_0) =
	\frac{ e^{ -\frac{1}{2} \left[ \VEC{q} - \VEC{\mu}(t|\VEC{q}_0,t_0) \right]\TRANS \TENSOR{C}(t|t_0)^{-1}
			                \left[ \VEC{q} - \VEC{\mu}(t|\VEC{q}_0,t_0) \right]}}
		 {\sqrt{(2\pi)^2 \det \TENSOR{C}(t|t_0)}}
\, .
\end{equation}
Here, the expectation vector and covariance matrix
conditioned on the initial time point are given by
\begin{align}
\VEC{\mu}(t|\VEC{q}_0,t_0) &=
	(t-\TENSOR{A}^{-1}) \VEC{u} +
	e^{-(t-t_0)\TENSOR{A}} \left[ \VEC{q}_0 - (t_0-\TENSOR{A}^{-1}) \VEC{u} \right]
\, ,
\\
\TENSOR{C}(t|t_0) &=
	\TENSOR{C}(\infty) -
	e^{-(t-t_0)\TENSOR{A}} \, \TENSOR{C}(\infty) \,
	e^{-(t-t_0)\TENSOR{A}\TRANS}
\, .
\end{align}
with $\VEC{u} = (u, 0)$ and the stationary covariance matrix
\begin{equation}
\label{eq:Cinfty}
\TENSOR{C}(\infty) =
\left(
	\begin{matrix}
		\frac{\gamma}{k} \left[ D + \frac{\Da}{1 + \frac{k\ta}{\gamma}} \right] & \sqrt{\frac{\Da}{2}} \frac{1}{1 + \frac{k\ta}{\gamma}}  \\
		\sqrt{\frac{\Da}{2}} \frac{1}{1 + \frac{k\ta}{\gamma}}  & \frac{1}{2 \ta}
	\end{matrix}
\right)
\, .
\end{equation}
\end{subequations}

As described in the main text, we consider the situation
in which initially the distribution of
particle positions is independent of the active fluctuations (which are in their
stationary state), so that the
initial distribution for the joint system factorizes as
$p_0(x_0, \eta_0) = p_0(x_0) p_{\mathrm s}(\eta)$.
We further assume that $p_0(x_0)$ is Gaussian with zero mean and variance $c_0^2$,
whereas $p_{\mathrm s}(\eta_0)$ is given in~\eqref{eq:pseta}.
The initial probability density $p_0(x_0, \eta_0)$ is thus
also Gaussian with mean $\VEC{\mu}(0) = 0$ and covariance matrix
\begin{equation}
\TENSOR{C}(0) =
	\left(
		\begin{matrix}
			c_0^2 & 0 \\
			0     & 1/2\tau_a
		\end{matrix}
	\right)
\, .
\end{equation}
Due to the linearity of the system (see \eqref{eq:LEexjoint}),
the distribution remains Gaussian for all later times $t>t_0$,
with the expectation values and covariances evolving according to
\begin{subequations}
\label{eq:ExProbT}
\begin{align}
\VEC\mu(t) &=
	\left[ t - \left( 1 - e^{-t \TENSOR{A}} \right) \TENSOR{A}^{-1} \right] \VEC{u}
\, , \\
\TENSOR{C}(t) &=
	\TENSOR{C}(\infty) +
	e^{-t \TENSOR{A}} \left[ \TENSOR{C}(0) - \TENSOR{C}(\infty) \right] e^{-t \TENSOR{A}\TRANS}
\, .
\end{align}
\end{subequations}

In order to calculate averages of, e.g., $\Delta\Sigma$ or $\Delta\I_{\pm}$,
we have to evaluate correlators of particle positions and/or active fluctuations
at two different time points (see next section).
Using the propagator~\eqref{eq:ExProp} and the time-dependent probability
density~\eqref{eq:ExProbT}, we find
\begin{multline}
\label{eq:ExAutocorr}
	\left\langle
		\left[ \VEC{q}(t) - \VEC{\mu}(t) \right]_i \left[ \VEC{q}(t^\prime) - \VEC{\mu}(t^\prime) \right]_j
	\right\rangle
\\
= \begin{cases}
	\left[ e^{-(t - t^\prime) \TENSOR{A}}\,\TENSOR{C}(t^\prime) \right]_{ij} & \text{for } t \geq t^\prime \\
	\left[ e^{-(t^\prime - t) \TENSOR{A}}\,\TENSOR{C}(t) \right]_{ji}        & \text{for } t < t^\prime
\end{cases}
\, .
\end{multline}

\subsection{Evaluation of averages}
We are interested in the averages
$\langle\Delta S_{\mathrm{sys}}\rangle$, $\langle\Delta\Sigma\rangle$ and $\langle\Delta\I_\pm\rangle$,
where the general, trajectory-wise expressions of all these quantities are given
in \eqref{eq:DeltaSsysDa0}, \eqref{eq:DeltaSigma}, and \eqref{eq:DeltaI+}, \eqref{eq:DeltaI-},
respectively.
While calculating $\langle\Delta S_{\mathrm{sys}}\rangle$ is relatively straightforward
(and basically amounts to computing the variances of the Gaussians at initial and
final time), the evaluation of $\langle\Delta\Sigma\rangle$ and $\langle\Delta\I_\pm\rangle$
is more complicated. It involves the non-local
memory kernel $\Gamma_\tau(t,t')$ and the averages
$\langle\dot{x}(t)k[x(t^\prime)-ut^\prime]\rangle$
[for $\langle\Delta\Sigma\rangle$ and $\langle\Delta\I_\pm\rangle$, see
\eqref{eq:DeltaSigma} and \eqref{eq:DeltaI+}, \eqref{eq:DeltaI-}],
and $\langle\dot{x}(t)\eta(t^\prime)\rangle$ or $\langle\eta(t)k[x(t^\prime)-ut^\prime]\rangle$
[for $\langle\Delta\I_\pm\rangle$, see \eqref{eq:DeltaI+}, \eqref{eq:DeltaI-}].
Here, we used $f_{t^\prime}=f(x(t^\prime),t^\prime)=k[x(t^\prime)-ut^\prime]$, see \eqref{eq:LEex}.
It turns out to be convenient to move the time-derivative from $\dot{x}(t)$
over to $\Gamma_\tau(t,t')$ by partial integration. Then all the correlations reduce
to $\langle x(t)x(t^\prime) \rangle$ and $\langle x(t)\eta(t^\prime) \rangle$,
which we have already calculated in \eqref{eq:ExAutocorr}.

We exemplify the procedure in more detail for $\langle \Delta\Sigma \rangle$,
$\langle\Delta\I_\pm\rangle$ can be evaluated in an analogous way.
For the explicit calculation, it is useful to split~\eqref{eq:DeltaSigma}
into a white-noise and a colored-noise contribution,
\begin{subequations}
\begin{align}
\Delta\Sigma_{\mathrm{w}} &= 
	\frac{1}{T} \int_0^\tau \mathrm{d}x_t \, f_t
\, , \\
\Delta\Sigma_{\mathrm{c}} &=
	-\frac{1}{T} \left( \frac{\Da}{D} \right)
	\int_0^\tau \mathrm{d}t \int_0^\tau \mathrm{d}t^\prime \, \dot{x}_t \, f_{t^\prime} \, \Gamma_\tau(t, t^\prime)
\, ,
\end{align}
\end{subequations}
such that $\Delta\Sigma = \Delta\Sigma_{\mathrm{w}} + \Delta\Sigma_{\mathrm{c}}$.
The white-noise part $\Delta\Sigma_{\mathrm{w}}$ is an ordinary
stochastic integral and its average can be evaluated using standard techniques;
we recall that we use the Stratonovich convention throughout this work.
For evaluating $\Delta\Sigma_{\mathrm{c}}$ we observe
that $\Gamma_\tau(t, t^\prime)$ is continuously differentiable except at $t = t^\prime$
(see \eqref{eq:GammabarJump}).
Splitting the $t$-integral into the intervals $[0,t^\prime]$ and $[t^\prime,\tau]$,
we can transfer the time derivative from the trajectory $\dot{x}(t)$
to the memory kernel by partial integration, so that
\begin{widetext}
\begin{multline}
\Delta\Sigma_{\mathrm{c}} = 
	\frac{1}{T} \left( \frac{\Da}{D} \right)
	\int_0^\tau \mathrm{d}t^\prime \left[ 
		x_0 \, f_{t^\prime} \, \Gamma_<(0, t^\prime) - x_\tau \, f_{t^\prime} \, \Gamma_>(\tau, t^\prime) +
		\int_0^{t^\prime} \mathrm{d}t \; x_t \, f_{t^\prime} \, \partial_t \Gamma_<(t, t^\prime) +
		\int_{t^\prime}^\tau \mathrm{d}t \; x_t \, f_{t^\prime} \, \partial_t \Gamma_>(t, t^\prime)
	\right]
\, ,
\end{multline}
\end{widetext}
where $\Gamma_<(t,t^\prime)$ and $\Gamma_>(t,t^\prime)$ are defined in analogy to~\eqref{eq:defGammabar}.
Now, the average $\langle \Delta\Sigma_{\mathrm{c}} \rangle$ involves correlations
$\langle x_t f_{t^\prime} \rangle=k\langle x_t [x_{t^\prime}-ut^{\prime}]\rangle$
between different time points.
Using the autocorrelation functions~\eqref{eq:ExAutocorr} and the result~\eqref{eq:Gamma}
for $\Gamma_\tau(t, t^\prime)$, it
can thus be evaluated as an ordinary integral.

In its full general form the resulting expression for
$\langle\Delta\Sigma\rangle=\langle\Delta\Sigma_{\mathrm{w}}\rangle+\langle\Delta\Sigma_{\mathrm{c}}\rangle$
is rather lengthy, so that we omit it here.
A few relevant limiting cases are given in Sec.~\ref{sec:HP}.


%

\end{document}